\documentclass[twocolumn]{emulateapj}
\usepackage{lscape}

\def\etal{{et~al.}}

\shorttitle{The ACS Virgo Cluster Survey. VIII. The Nuclei of Early-Type Galaxies}
\shortauthors{C\^ot\'e \etal}

\begin{document}

\title{The ACS Virgo Cluster Survey. VIII. The Nuclei of Early-Type Galaxies\altaffilmark{1}}

\author{
Patrick C\^ot\'e\altaffilmark{2},
Slawomir Piatek\altaffilmark{3},
Laura Ferrarese\altaffilmark{2},
Andr\'es Jord\'an\altaffilmark{4,5},
David Merritt\altaffilmark{6},
Eric W. Peng\altaffilmark{2},
Monica Ha\c{s}egan\altaffilmark{7,8},
John P. Blakeslee\altaffilmark{9,10},
Simona Mei\altaffilmark{9},
Michael J. West\altaffilmark{11},
Milo\v s Milosavljevi\'c\altaffilmark{12,13},
John L. Tonry\altaffilmark{14}}

\altaffiltext{1}{Based on observations with the NASA/ESA {\it Hubble
Space Telescope} obtained at the Space Telescope Science Institute,
which is operated by the association of Universities for Research in
Astronomy, Inc., under NASA contract NAS 5-26555.}
\altaffiltext{2}{Herzberg Institute of Astrophysics, National
Research Council of Canada, 5071 West Saanich Road, Victoria, BC,
V8X 4M6, Canada; patrick.cote@nrc-cnrc.gc.ca;
laura.ferrarese@nrc-cnrc.gc.ca; eric.peng@nrc-cnrc.gc.ca}
\altaffiltext{3}{Department of Physics, New Jersey Institute of Technology,
Newark, NJ 07102; piatek@physics.rutgers.edu}
\altaffiltext{4}{European Southern Observatory, Karl-Schwarzschild-Str.
2, 85748 Garching, Germany; ajordan@eso.org}
\altaffiltext{5}{Astrophysics, Denys Wilkinson Building, University
of Oxford, 1 Keble Road, Oxford, OX1 3RH, UK}
\altaffiltext{6}{Department of Physics, Rochester Institute of
Technology, 84 Lomb Memorial Drive, Rochester, NY 14623;
merritt@astro.rit.edu}
\altaffiltext{7}{Department of Physics and Astronomy, Rutgers
University, New Brunswick, NJ 08854; mhasegan@physics.rutgers.edu}
\altaffiltext{8}{Institute for Space Sciences, P.O.Box MG-23,
Ro 77125, Bucharest-Magurele, Romania}
\altaffiltext{9}{Department of Physics and Astronomy, The Johns
Hopkins University, 3400 North Charles Street, Baltimore, MD
21218-2686; smei@pha.jhu.edu}
\altaffiltext{10}{Department of Physics, Washington State University,
Webster Hall 1245, Pullman, WA 99164-2814; jblakes@wsu.edu}
\altaffiltext{11}{Department of Physics and Astronomy, University of
Hawaii, Hilo, HI 96720; westm@hawaii.edu}
\altaffiltext{12}{Theoretical Astrophysics, California Institute of
Technology, Mail Stop 130-33, Pasadena, CA 91125;
milos@tapir.caltech.edu}
\altaffiltext{13}{Sherman M. Fairchild Fellow}
\altaffiltext{14}{Institute for Astronomy, University of Hawaii, 2680
Woodlawn Drive, Honolulu, HI 96822; jt@ifa.hawaii.edu}

\slugcomment{Accepted for publication in the {\it Astrophysical Journal Supplement Series.}}

\begin{abstract}
The ACS Virgo Cluster Survey is a {\it Hubble Space Telescope} program to
obtain high-resolution imaging, in widely separated bandpasses (F475W
$\approx g$ and F850LP $\approx$ $z$), for 100 early-type members of the
Virgo Cluster, spanning a range of $\approx$ 460 in blue luminosity.
We use this large, homogenous dataset to examine the innermost structure
of these galaxies and to characterize the properties of their compact
central nuclei.
We present a sharp upward revision
in the frequency of nucleation in early-type galaxies brighter than
$M_B \approx -15$ ($66 \lesssim f_n \lesssim 82$\%) and show that 
ground-based surveys underestimated the number of nuclei
due to surface brightness selection
effects, limited sensitivity and poor spatial resolution. We
speculate that previously reported claims that nucleated dwarfs
are more concentrated to the center of Virgo than their non-nucleated
counterparts may be an artifact of these selection effects. There is
no clear evidence from the properties of the nuclei, or from the overall
incidence of nucleation, for a change at $M_B \sim -17.6$, the traditional
dividing point between dwarf and giant galaxies. There {\it does}, however, appear to
be a fundamental transition at $M_B \sim -20.5$, in the sense that the
brighter, ``core-S\'{e}rsic" galaxies lack resolved (stellar) nuclei. 
A search for nuclei which may be offset from the photocenters of their
host galaxies reveals only five candidates with displacements of more
than 0\farcs5, all of which are in dwarf galaxies. In
each case, though, the evidence suggests that these ``nuclei"
are, in fact, globular clusters projected close to the galaxy photocenter. 
Working from a sample of 51 galaxies with
prominent nuclei, we find a median half-light radius of $\langle r_h \rangle = 4.2$~pc,
with the sizes of individual nuclei ranging from 62~pc down to $\le$~2~pc (i.e., 
unresolved in our images) in about a half dozen cases. Excluding these
unresolved objects, the nuclei sizes are found to depend 
on nuclear luminosity according to the relation
$r_h \propto {\cal L}^{0.50\pm0.03}$. Because the large majority of nuclei 
are resolved, we can rule out low-level AGN as an explanation for the
central luminosity excess in almost all cases. On average, the nuclei
are $\approx$ 3.5 mag brighter than a typical globular cluster. Based on
their broadband colors, the nuclei appear to have old to intermediate-age
stellar populations. The colors of the nuclei in galaxies fainter than
$M_B \approx -17.6$ are tightly correlated with their luminosities,
and less so with the luminosities of their host galaxies, suggesting
that their chemical enrichment histories were governed by local or internal
factors. Comparing the nuclei to the ``nuclear clusters" found
in late-type spiral galaxies reveals a close match in terms of
size, luminosity and overall frequency. A formation mechanism that
is rather insensitive to the detailed properties of the host
galaxy is required to explain this ubiquity and homogeneity.
The mean of the frequency function for the nucleus-to-galaxy luminosity ratio
in our nucleated galaxies, $\langle\log_{10}\eta\rangle = -2.49\pm0.09$~dex ($\sigma = 0.59\pm0.10$),
is indistinguishable from that of the SBH-to-bulge mass ratio,
$\langle \log_{10} {( {\cal M}_{\bullet} / {\cal M}_{gal} )} \rangle = -2.61\pm0.07$~dex
($\sigma = 0.45\pm0.09$),
calculated in 23 early-type galaxies with detected supermassive black holes (SBHs).
We argue that the compact stellar nuclei
found in many of our program galaxies are the low-mass counterparts of
the SBHs detected in the bright galaxies. If this interpretation is correct, then one should
think in terms of {\it Central Massive Objects} --- either SBHs or compact
stellar nuclei --- that accompany the formation of almost all early-type galaxies and
contain a mean fraction $\approx$ 0.3\% of the total bulge mass. In
this view, SBHs would be the dominant formation mode above $M_B \approx -20.5$.
\end{abstract}


\section{Introduction}
\label{sec:introduction}

Early-type galaxies are found in virtually all environments ---
from the field, to small groups, to rich clusters (Hubble \& Humason 1931;
Oemler 1974; Dressler 1980). In the highest density environments, 
ellipticals and lenticulars are known to dominate the overall fraction of
bright galaxies, $f_{\rm E+S0} \sim 0.4-0.9$, with the precise contribution
depending on local galaxy density and redshift (Smith \etal\ 2004;
Postman \etal\  2005). In the Virgo Cluster, the rich cluster nearest to
our own Galaxy, $f_{\rm E+S0} \approx 0.44$ for galaxies brighter than
$B\lesssim 13$ or $M_B \lesssim -18.1$ (Julian et~al. 1997).\footnote{Throughout
this paper, we adopt a Virgo distance modulus of $(m-M)_0 = 31.09$ mag
(Tonry \etal\ 2001; Mei \etal\ 2005b).} If one
considers not just giant galaxies, but also the much more common dwarfs,
then the dominance of early-type galaxies is even more pronounced: i.e., 
among the confirmed members of Virgo with unambiguous morphological 
classifications, the early-type fraction is $\approx$ 0.8 (Reaves 1983; 
Binggeli, Tammann \& Sandage 1987, hereafter BTS87).

It has long been recognized that early-type galaxies, both in Virgo and
elsewhere, often show compact nuclei near their centers. In their landmark study
of the Virgo Cluster, BTS87 carried out a visual search for nuclei
using wide-field, blue-sensitive photographic plates from the 2.5-m
du Pont telescope. Of the 1277 members and 574 probable members in their
Virgo Cluster Catalog (hereafter VCC), a total of 1192 were classified as non-nucleated dwarfs
(dEs or dS0s) while an additional 415 dwarfs (predominantly dE,Ns) were
found to be nucleated. Thus, roughly 25\% of the dwarf 
galaxies in Virgo were found by BTS87 to have a discernible nucleus, 
although the precise fraction was also found to depend on galaxy
luminosity and position within the cluster (see Figure~8 of Sandage, 
Binggeli \& Tammann 1985 and Figure~19 of BTS87, respectively). 
Unfortunately, progress toward understanding the nature of these nuclei
has been limited because of several factors: $e.g.$, ground-based studies must
contend with the contamination from the underlying galaxy light, it is
difficult to de-couple the brightness profiles of the
nuclei from those of their host galaxies, and the nuclei are 
sufficiently compact that they appear unresolved in even the sharpest
ground-based images (Caldwell 1983, 1987; Durrell 1997).

While the photographic survey of BTS87 remains a landmark study of
nucleated galaxies in the local universe, there are reasons to believe
that a modern survey of the nuclei belonging to early-type galaxies would
be advisable --- preferably one which capitalizes on the high angular
resolution afforded by the {\it Hubble Space Telescope} ({\it HST}).
First, {\it HST} imaging of late-type galaxies has revealed that
50--70\% of these systems have compact stellar clusters at or near
their photocenters (Phillips et~al. 1996; Carollo, Stiavelli \& Mack
1998; Matthews et~al 1999; B\"oker et~al. 2002; B\"oker et~al. 2004).
Second, if the early-type members of the Local Group are
any guide, then one may expect estimates for the fraction of nucleated galaxies to increase
as better imaging becomes available. For instance, in recent years
a number of Local Group dwarfs have been found to contain 
previously unrecognized central substructures and/or nuclei, including
Sagittarius (Layden \& Sarajedini 2000; Monaco et~al. 2005),
Ursa Minor (Kleyna et~al. 2003; Palma et~al. 2003),
Andromeda II (McConnachie \& Irwin 2006)
and Fornax (Coleman et~al. 2004). Third, in their WFPC2 
survey of dwarfs in the Virgo and Fornax Clusters, Lotz et~al. (2004) 
found that six of the 30 ``non-nucleated" dwarf ellipticals in their sample
actually contained nuclei which had gone unnoticed in the ground-based
surveys (Binggeli, Tammann \& Sandage 1985, hereafter BTS85; BTS87;
Ferguson 1989). Very recently, Grant, Kuipers \& Phillipps (2005) used
imaging from the Wide Field Camera on the Isaac Newton Telescope to
show that faint nuclei in Virgo dwarfs were frequently missed in
photographic surveys.

These results suggest that there may be significant incompleteness in our census of
nuclei in early-type galaxies.
Indeed, in their photographic study of Virgo, BTS87 cautioned that ``most nuclei in the
luminous E and S0 galaxies were probably missed due to [the] high surface
brightness [of the underlying galaxy.]" In addition to this surface-brightness
selection effect, BTS87 state explicitly that any nuclei with $B \gtrsim$ 23 
($M_B \gtrsim -8$) would fall below their plate detection limits and hence
be missing from their catalog.

The central regions of early-type galaxies have been favorite targets for
{\it HST} since its launch in 1990. For the most part, such surveys 
have tended to focus on the core structure of the galaxies. However, several
studies reported the discovery of compact nuclei in (predominantly bright)
samples of early-type galaxies, beginning with pre-refurbishment (WFPC) imaging 
(Crane \etal\ 1993; Lauer \etal\ 1995) and continuing with imaging from 
WFPC2 (Rest \etal\ 2001; Lauer \etal\ 2005) and NICMOS (Ravindranath \etal\ 2001).
These studies, which are primarily based on single-filter imaging of samples of
(33--77) galaxies with distances between 1 and 100~Mpc, have confirmed that some
early-type galaxies do contain compact nuclei, but there is 
disagreement over their overall frequency (with estimates ranging from 13\%
to $\approx$ 50\%), whether or not they are resolved structures, and their
classification as stellar or non-thermal (AGN) sources.
A better understanding of the physical properties of these nuclei is
important since they almost certainly hold clues to the violent processes
that have shaped the central regions of galaxies, which could include star
formation triggered by infalling gas, collisions and mergers of stars and 
star clusters, tidal disruption of clusters and the growth of stellar ``cusps"
by central black holes, and the mechanical and radiative feedback from 
accreting black holes or intense nuclear starbusts.

This paper presents a homogenous analysis of the nuclei belonging to
a sample of 100 early-type galaxies in the Virgo Cluster. Our images,
taken with the Advanced Camera for Surveys (ACS; Ford et~al. 1998),
form the basis of the ACS Virgo Cluster Survey (ACSVCS; C\^ot\'e \etal\ 2004; 
hereafter Paper~I). Other papers in this series have discussed the
data reduction pipeline (Jord\'an et al. 2004a = Paper II),
the connection between low-mass X-ray binaries in M87 (Jord\'an et al.
2004b = Paper III), the measurement and calibration of surface
brightness fluctuation magnitudes (Mei et~al. 2005ab = Papers IV and V),
the morphology, isophotal parameters and surface brightness
profiles for early-type galaxies (Ferrarese et~al. 2006a = Paper VI),
the connection between globular clusters and ultra-compact dwarf
galaxies (Ha\c{s}egan et al. 2005 = Paper VII), the color distributions
of globular clusters (Peng et~al. 2006a = Paper IX), the half light radii
of globular clusters and their use as a distance indicator (Jord\'an et al.
2006 = Paper X) and the discovery of diffuse star clusters in early-type
galaxies (Peng et~al. 2006b = Paper XI).

There are several features of the ACS Virgo Cluster Survey which make it uniquely
suited to the study of nuclei in early-type
galaxies. First, the survey itself targets a large sample of 100 early-type
galaxies lying at a common distance of about 16.5~Mpc so that the
$\approx 0\farcs1$ FWHM
of the ACS point-spread function (PSF) corresponds to a small, and nearly 
constant, physical scale of $\approx 8$~pc. This excellent spatial
resolution, coupled with the fine plate scale of 0\farcs049~pix$^{-1}$,
makes it possible to measure
structural parameters for any nuclei larger than a few parsecs in size.
Second, with blue magnitudes in the range $9.31 \lesssim B_T \lesssim 15.97$
($-21.88 \lesssim M_B \lesssim -15.21$),
our program galaxies span a wide range in luminosity
so it is possible to study the 
phenomenon of nucleation in giant and dwarf galaxies simultaneously.
Third, the images are sufficiently deep that they reveal not only the nuclei,
but also the many globular clusters belonging to 
our program galaxies; thus, the same images which provide information on
the nucleus and host galaxy can also be used to study the associated
globular cluster systems and to explore possible evolutionary links between the
clusters and nuclei.  And finally, because multi-band imaging is available
in two widely-separated bandpasses (F475W and F850LP)
for each object in the survey, it is possible to use broadband colors to
place rough constraints on the star formation and chemical enrichment histories
of the nuclei and their host galaxies.

The organization of this article is as follows. \S2 gives a brief summary of
the observational material used in our analysis.  A description of the galaxy
brightness profiles and the method of their analysis is presented in \S3.
\S4 contains a discussion of the empirical properties of the
nuclei in our survey, such as their overall numbers, possible displacements
from the galaxy photocenters, luminosities, colors, surface brightnesses and half-light radii.
In \S5 we discuss the implications of our findings for various formation
scenarios. The article concludes with a summary of the main results in \S6.
A future paper in this series will discuss the broader implications of our
findings for models of nucleus formation in early-type galaxies (Merritt \etal\ 
2006).

\section{Observations and Data Reductions}
\label{sec:obdata}

Our analysis is based on \textit{HST} imaging for 100 early-type galaxies
having morphological types E, S0, dE, dE,N and dS0. All are confirmed members
of the Virgo Cluster based on radial velocity measurements. Images were taken
with the ACS instrument used in Wide Field Channel (WFC) mode
with the F475W and F850LP filter combination, which are roughly equivalent
to the $g$ and $z$ bands, respectively, in the Sloan Digital Sky Survey
photometric system.  These images form the basis
of the ACS Virgo Cluster Survey, a complete description of which may be
found in Paper~I. Note that the 26 brightest galaxies in this survey constitute
a complete sample of early-type members of Virgo with $B_T \le 12.15$,
and that the full sample represents 44\% of all early-type members of
Virgo spanning the magnitude range $9.3 \lesssim B_T \lesssim 16$. A
customized data pipeline (described in detail in Paper II) produces
geometrically-corrected, flux-calibrated, cosmic-ray-free images in the
F475W and F850LP bandpasses.

Table~\ref{tab:data} gives some basic information about the target galaxies, tabulated in order
of increasing blue magnitude (decreasing luminosity).  An identification number
for each galaxy is given in the first column, followed by the identification from the 
VCC (BST85) and other names for the galaxy in the Messier, NGC, UGC or IC catlogs.
Blue magnitudes, $B_T$, from BST85
are presented in column 4, while the fifth column records the 
adopted Galactic reddening from Schlegel, Finkbeiner  \& Davis (1998). Columns
6 and 7 record the surface brightness of each galaxy, in both the $g$
and $z$ bandpasses, measured via spline interpolation at a geometric mean
radius\footnote{The geometric mean radius is defined as $r \equiv a\sqrt{(1-\epsilon)}$
where $a$ is the semi-major axis and $\epsilon$ is the ellipticity.} of
$r =$~1\arcsec~($\approx$ 80~pc).
This model-independent choice of surface brightness should closely approximate
the galaxy's {\it central} surface brightness, but is measured at a radius 
large enough to ensure negligible contamination from any central nucleus. 

The remaining columns of Table~\ref{tab:data} will be described below. Coordinates,
morphological classifications and other information on the program galaxies
may be found in Papers I and VI.

\section{Analysis}
\label{sec:analysis}

Our goals in this paper include the measurement of the structural and
photometric properties of the nuclei in our program galaxies,
and an investigation into the relationship between these nuclei
and their host galaxies. Additionally, we
wish to compare the properties of the nuclei to
those of the globular clusters in the program galaxies and,
more generally, to the Virgo ultra-compact dwarf
(UCD) galaxies (e.g., Drinkwater et~al. 1999; Hilker et~al. 1999; 
Drinkwater et~al. 2000; Phillipps et~al. 2001) 
identified in the course of this
survey and described in Ha\c{s}egan \etal\ (2005; 
hereafter Paper~VII) and Ha\c{s}egan \etal\ (2006).  A companion
paper in this series (Paper~VI) presents
an analysis of the surface brightness profiles of the
program galaxies along with a tabulation of the best-fit structural
parameters, while two other articles examine the photometric (Paper~IX)
and structural (Paper~X) parameters of the globular clusters.
As we make use of several results from these
studies, the reader is referred to these papers for complete details.

\subsection{Parameterization of the Surface Brightness Profiles}

Because the nuclei are always superimposed on the light of the underlying
galaxy, measuring their photometric and structural properties requires
a model for the galaxy surface brightness profile. For each galaxy
in our survey, $g$- and $z$-band azimuthally-averaged radial surface 
brightness profiles are available from Paper~VI.  These profiles were
derived by fitting the isophotes with the ELLIPSE task in IRAF which,
in turn, is based on the algorithm of Jedrzejewski (1987). The $g$- and $z$-band
brightness profiles were parameterized with a standard S\'{e}rsic (1968) model,
\begin{equation}
\begin{array}{rrrrr}
$$I_{g}(r) & = & I_0\exp[-b_n(r/r_e)^{1/n})],$$
\end{array}
\label{eq1}
\end{equation}
where $I_0$ is the central intensity and 
$n$ is a shape parameter which yields an $R^{1/4}$-law profile for $n =4$
(de Vaucouleurs 1948) and an exponential profile for $n = 1$.
The parameter $b_n$ is defined such that
$\Gamma(2n) = 2\Gamma_1(2n,b_n)$, where $\Gamma$ and $\Gamma_1$ are 
the complete and incomplete gamma functions, respectively (e.g., 
Graham \& Driver 2005). 
As shown by Caon, Capaccioli \& D'Onofrio (1993), a convenient
approximation relating $b_n$ to the shape parameter $n$ is
$b_n \approx 1.9992n - 0.3271$ for $1 \lesssim n \lesssim 10$. Given
this definition of $b_n$, $r_e$ is the effective radius of the galaxy.

The $g$- and $z$-band brightness profiles for each galaxy 
were also fit with a ``core-S\'{e}rsic'' model,
\begin{equation}
\begin{array}{rrrrr}
I(r) = I^{\prime} \biggl [1 + \biggl ( {r_b \over r} \biggr ) ^{\alpha} \biggl ]^{\gamma / \alpha} \exp \biggl [-b_n  \biggl ( {r^{\alpha} + r_b^{\alpha} \over r_e^{\alpha}} \biggr ) ^{1/(\alpha n)} \biggr ],
\end{array}
\label{eq2}
\end{equation}
where
\begin{equation}
\begin{array}{rrrrr}
I^{\prime} = I_b2^{-\gamma / \alpha} \exp \biggl [b_n \biggl (2^{1/\alpha}r_b/r_e \biggl )^{1/n} \biggr ]
\label{eq3}
\end{array}
\end{equation}
This model, which was first proposed by Graham \etal\ (2003),
consists of a power-law
component in the inner region of a galaxy, which ``breaks'' to a
traditional S\`{e}rsic profile beyond some radius, $r_{b}$.
The model has a total of six free parameters: the
logarithmic slope of the inner power-law ($\gamma$); 
the shape of the S\'{e}rsic function ($n$); the break radius ($r_b$);
the effective half-light radius of the S\'{e}rsic profile ($r_e$); 
the intensity at the break radius ($I_b$) and
a parameter ($\alpha$) which governs the sharpness of the
transition between the inner power law and the outer S\'{e}rsic function. After
some experimentation, it was decided to use the modified parametrization of
Trujillo \etal\ (2004), 

$$
I_{g}(r) = I_b \biggl [(r_b/r)^{\gamma}u(r_b-r) + 
$$
\begin{equation}
+e^{b_n(r_b/r_e)^{1/n}}e^{-b_n(r/r_e)^{1/n}}u(r-r_b) \biggr ]
\label{eq4}
\end{equation}
in which $\alpha \rightarrow \infty$ and $u(x-z)$ is the Heaviside step function.
This
model produced more stable fits, with better consistency between
the five remaining parameters ($I_b$, $\gamma$, $n$, $r_e$ and $r_b$)
measured in the $g$ and $z$ bandpasses.

Equations (1) and (4) are intended to describe the profiles of
galaxies which have no central nucleus. However, it is obvious that many
galaxies in our sample do indeed have compact sources at or near their centers.
For such nucleated galaxies, a single-component King model (Michie 1963;
King 1966) was used to represent this central component. This introduces three 
additional parameters to the fit: the total intensity of the nucleus ($I$);
the projected half-light radius ($r_h$); and the King concentration index
($c$). In other words,
for {\it nucleated
galaxies}, the fitted model, $I(r)$, takes the form
\begin{equation}
\begin{array}{rrrrr}
$$I(r) & = & I_{g}(r) + I_k(r),$$
\end{array}
\label{eq5}
\end{equation}
where $I_{g}(r)$ is either a pure S\'ersic model (Equation 1)
or a core-S\'ersic model (Equation 4), depending on the galaxy in question,
and $I_k(r)$ is the central King model component.
For non-nucleated galaxies, the profiles are fit simply with models
of the form of Equations (1) or (4).
A detailed justification for the choice of galaxy model (i.e.,
S\'{e}rsic vs. core-S\'{e}rsic) is given on a case-by-case basis
in Paper~VI. We adopt these classifications verbatim,
with the exception of three intermediate-luminosity galaxies: VCC543
(UGC7436), VCC1528 (IC3501) and VCC1695 (IC3586).\footnote{Note that
$r_b >> r_h$ for all nucleated (Type~Ia) core-S\'{e}rsic galaxies;
$\langle r_b/r_h\rangle = 74$ for the four galaxies in this category.} 
While the {\it global} brightness profiles of
these galaxies are adequately represented by S\'ersic models, such
models overpredict the amount of galaxy light on subarcsecond
scales. For the purposes of measuring photometric and structural 
properties for the nuclei in these galaxies, we parameterize the
galaxy profiles with core-S\'ersic models in all three cases.

We note that the definition of a ``nucleus" invariably 
hinges on some assumption --- explicit or otherwise --- about the intrinsic
brightness profile of the host galaxy. Our study is no exception in
this regard. Choices for the galaxy profiles made by previous workers have
included King models (Binggeli \& Cameron 1993), pure exponentials (Binggeli
\& Cameron 1993; Stiavelli \etal\ 2001), Nuker laws (Rest \etal\ 2001;
Ravindranath \etal\ 2001; Lauer \etal\ 2005) and Sersic profiles (Durrell
1997; Stiavelli \etal\ 2001). After considerable experimentation (Paper IV),
we opted to use the family of models represented by Equations~(1) and
(2) because they have the great advantage they are flexible
enough to provide accurate fits to the brightness profiles of both
giant and dwarf galaxies (see Paper~VI).
The use of a single
(S\'{e}rsic) model to describe the full sample of galaxies also seems
advisable in light of mounting evidence that, at least in terms of their
{\it structural parameters}, the longstanding perception of a
fundamental dichotomy between giant and dwarf ellipticals
(e.g., Kormendy 1985) may be incorrect (see, e.g., Jerjen \&
Binggeli 1997; Graham \& Guzman 2003; Paper~VI). From a theoretical perspective, the
choice of Equations~(1) and (4) also seems reasonable given recent
findings that the S\'{e}rsic law provides an accurate representation
of the spatial and surface density profiles of dark matter halos
in high-resolution $\Lambda$CDM simulations (Navarro et~al. 2004;
Merritt et~al. 2005).

At the same time, our decision to parameterize the central nuclei with King models is motivated
by high-resolution observations of the nuclei in nearby galaxies. In nucleated
Local Group galaxies such as NGC205 and Sagittarius,
King models are found to provide accurate representations of the central
components (e.g., Djorgovski \etal\ 1992; Butler \& Mart\'inez-Delgado 2005;
Monaco \etal\ 2005). Nevertheless, for galaxies at the distance of the Virgo
Cluster, we are working close to the limits of resolvability, so we caution that
our choice of King models to parameterize the central components may not be unique,
particularly for faint nuclei in the highest surface brightness galaxies.
Alternative parameterizations of the central brightness ``cusps" in our sample
galaxies will be explored in a future paper in this series (Merritt \etal\ 2006).

\subsection{Choice of Drizzling Kernel, PSF Determination and Fitting Procedure}

As described in Paper~II, our analysis of the nuclei, brightness profiles,
and isophotal structure of the galaxies is based on F475W and F850LP images in which
a {\it Gaussian} kernel is used to
distribute flux onto the output (drizzled) images. This choice of kernel
has the advantage that, relative to {\it Lanczos3} kernel, bad pixels can be repaired 
more effectively, albeit with the penalty of a slight reduction in angular
resolution.\footnote{Using the {\it Lanczos3} kernel produces images with
better noise characteristics and a somewhat sharper PSF (0\farcs09 versus 0\farcs1),
so this kernel was used for both the surface brightness fluctuation measurements 
and the determination of the globular cluster photometric and structural parameters.}
Due to the compact
nature of the nuclei (even the most extended objects have effective radii 
$\lesssim 1$\arcsec), it is important that the effects of the PSF
are taken into account when fitting models to the observed brightness profiles.

\begin{figure}
\figurenum{1}
\plotone{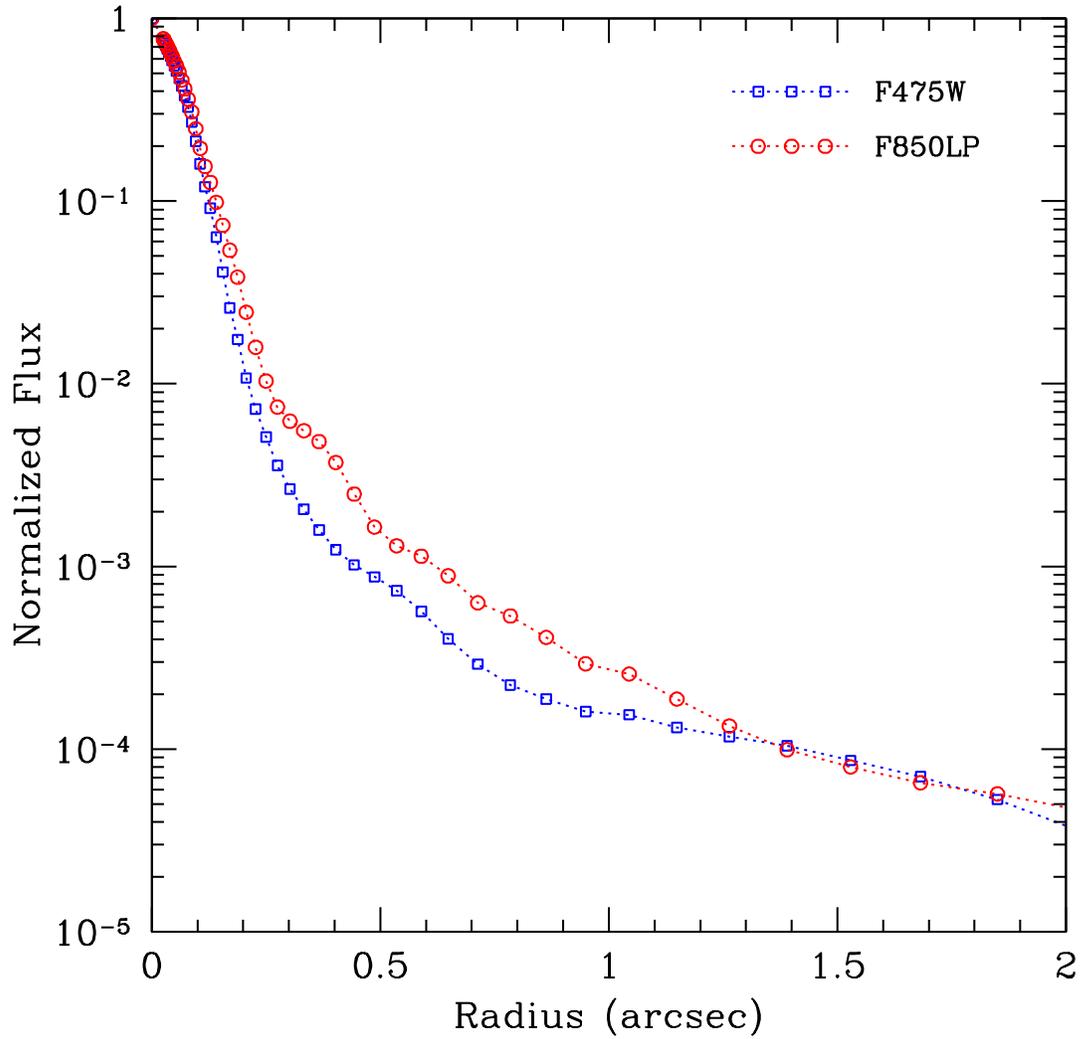}
\caption{Azimuthally-averaged point spread functions for the nucleus of 
VCC1303, the galaxy closest to the mean position of the centers for the
full sample of galaxies. The profiles for F475W and F850LP are shown as
squares and circles, respectively. The profiles have been normalized
to the same central intensities.
\label{fig01}}
\end{figure}

PSFs in the F475W and F850LP filters, varying quadratically
with CCD position, were derived using DAOPHOT II as described in Paper~II. Briefly,
archival images of the Galactic globular cluster NGC104 (47~Tucanae) taken during
programs G0-9656 and GO-9018 were used to construct empirical PSFs in the two
bandpasses. These archival images were drizzled in the same manner as the
images for the program galaxies. A total of $\approx$~200 stars in each filter were used
to construct the PSFs, which extend to a radius of 0\farcs5 in both bandpasses. To
follow the behavior of the PSFs to still larger radii, we matched our
empirical PSFs at a radius of 0\farcs3 to those measured for high-S/N composite stars 
by Sirianni \etal\ (2005). These latter PSFs extend to radii of 3\arcsec, and
were constructed from images of 47~Tuc fields taken as part of the
photometric calibration of ACS. Figure~\ref{fig01} shows azimuthally averaged
PSFs for the F475W and F850LP filters measured at the position of the nucleus in
VCC1303 (NGC4483) --- the program galaxy whose center is nearest to the mean
position for the full sample of program galaxies.

A $\chi^2$ minimization scheme was used to find the models which best fit the 
azimuthally-averaged, one-dimensional intensity profiles for each galaxy. 
Minimizations were carried out using the {\tt Minuit} package in the CERN program 
library; initial determinations of the minima, obtained using
a Simplex minimization algorithm (Nelder \& Mead 1965), were later refined using a variable
metric method with inexact line search (MIGRAD).
Following Byun et al. (1996), all points in the profile were assigned equal weight.
For both nucleated and non-nucleated galaxies, the
PSFs at the location of the galaxy's center were convolved with the models
before fitting to the intensity profiles.
Customized PSFs were created for each galaxy in the survey, centered at the exact
(sub-pixel) location of the nucleus.
While, in practice, the PSF convolution has little impact on the fitted S\'ersic or
core-S\'{e}rsic model parameters, with the exception of $\gamma$, this step
is critically important when evaluating accurate structural parameters
for the central nucleus. 

Profile fits are carried out independently in the two
bandpasses, with the exception of the 11 nucleated galaxies brighter
than $B_T = 13.5$ (i.e., Type~Ia galaxies; see \S\ref{sec:results}). Our
numerical experiments suggest that in this
high surface brightness regime, the profile of the underlying galaxy 
makes the measurement of nuclei half-light radii and total magnitudes
extremely challenging (see Appendix~A). For these
galaxies, the composite $g$- and $z$-band profiles were first fitted
simultaneously and the individual
fits constrained so that the galaxy shape index parameter,
nucleus concentration index and half-light radius were the same in
the two bandpasses. When dust is present (see below), the models
are fitted to the dust-corrected surface brightness profiles if
$\ge$ 50\% of the points along a given isophote are affected;
otherwise the dust affected regions are masked. More details on the
correction for dust obscuration are given in Paper~VI.

Sufficiently compact nuclei will appear unresolved even in our ACS images. To estimate
the resolution limit of our observations, we constructed brightness profiles 
for a number of likely stars which appear in our images. These candidate stars
were classified as unresolved
in the object catalogs produced by KINGPHOT, the reduction package used to measure 
structural and photometric parameters for the globular clusters in these 
fields (see Papers II and X). Fitting King models to the brightness profiles
of these objects gives median half-light radii of 0\farcs011$\pm$0\farcs004 and
0\farcs018$\pm$0\farcs005 in the F475W and F850LP bandpasses, respectively.
As an additional test, we may make use of the fact that VCC1316 (M87 = NGC4486), one of 
the AGN galaxies in our survey (see below), contains a prominent non-thermal
central point source. Although this source is saturated in our F475W images,
a King model fitted to the central source in the $z$-band brightness profile
gives $r_h = 0\farcs021$. In what follows, we
adopt a conservative upper limit of 0\farcs025 $\approx$ 2~pc for 
the resolution limit in both bandpasses. 

\begin{figure}
\figurenum{2}
\plotone{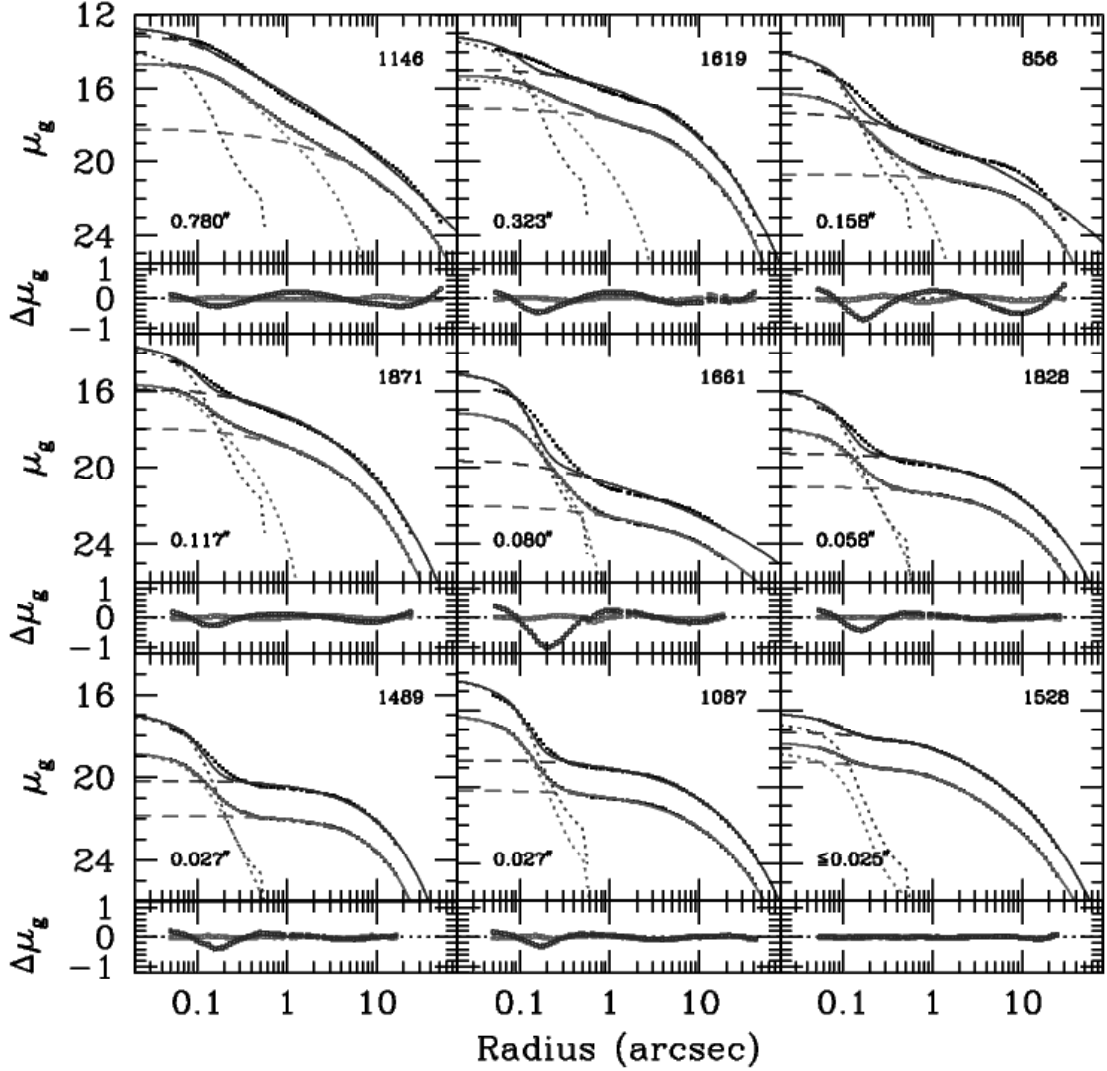}
\caption{Surface brightness profiles for nine Type Ia
galaxies whose nuclei span the full range in half-light radius. In
each panel, we show the azimuthally-averaged brightness profile in the
$g$ band (small squares) along with the fitted models. The respective
components for the nucleus and galaxy (a Sersic model in all cases except
for VCC1528, where a core-S\'{e}sic model was used) are shown as the
dotted and short dashed curves. Two different assumptions for the nucleus component
are shown for each galaxy --- red curves show the results obtained for a King model
while blue curves show the results obtained by fitting
a central point source. In the latter case, both the
data and model have been shifted upwards by 1.5 mag for clarity.
Residuals about the fitted
relations, ${\Delta}\mu_g$, are shown below in each panel.
For VCC1528, the best-fit half-light radius of
$r_{h,g}$ = 0\farcs015 falls below the resolution limit of 0\farcs025, meaning
that the nucleus is formally unresolved in this case. In the eight remaining cases, a
point-source nucleus provides an inadequate representation of the measured brightness
profiles. The median
half-light radii measured for the sample of Type Ia galaxies (see \S4.1) are
0\farcs051 ($g$ band) and 0\farcs048 ($z$ band).
\label{fig02}}
\end{figure}

Before proceeding, we pause to demonstrate that the vast majority of the nuclei
belonging to our program galaxies are indeed more extended than point sources. In
Figure~\ref{fig02}, we show $g$-band surface brightness profiles
for a representative sample of nine nucleated galaxies, chosen to span the full
range in fitted half-light radius (with $\langle{r_h}\rangle$ decreasing from left
to right and from top to bottom). In each panel, the red curves show the results of
fitting the nuclear component with a King model, while the blue curves show the
results of fitting a central point source; residuals from both fits are shown in the 
lower panel. With the exception of VCC1528 (IC3501), the central nucleus
is resolved for all of the galaxies in Figure~\ref{fig02}. In total, six
galaxies in our sample --- VCC1883 (NGC4612), VCC140 (IC3065), VCC1528,
VCC1695 (IC3586), VCC1895 (UGC7854) and VCC1826 (IC3633) --- have
best-fit half-light radii, measured in at least one bandpasses, that fall below
our nominal resolution limit of 0\farcs025. These half-light radii are given
in parantheses in Table~\ref{tab:data}. They have been included in the
following analysis, but we caution that they are formally unresolved in
our ACS images. We shall return to the issue of these compact nuclei in
\S\ref{sec:discussion}.
Additional tests on the resolution
limits, possible biases in the derived photometric and structural 
parameters, and a discussion of measurement errors, are given in 
\S4.1 and Appendix~A.

\clearpage

\section{Results}
\label{sec:results}

As many as eighteen of the 100 galaxies in Table~\ref{tab:data} show evidence for
dust --- either as isolated patches and filaments, or in the form of disks having
varying degrees of regularity (see Paper~VI). For the most part, this
dust has no impact on the identification of possible nuclei. However,
for four galaxies in our sample (i.e., VCC1535 = NGC4526, VCC1030 =
NGC4435, VCC685 = NGC4350 and VCC571) the central dust obscuration is
severe enough to make a reliable  classification of these galaxies as
nucleated or non-nucleated impossible.  Moreover, for VCC1535 and VCC1030, both
of which harbour massive, kpc-scale dust disks, the surface brightness
profiles are themselves so limited that it is not even possible to
place the galaxies in the appropriate S\'ersic or core-S\'ersic
categories.

In general, the census of active galactic nuclei (AGN) in intrinsically
faint galaxies --- and in the ACS Virgo Cluster Survey galaxies
in particular --- is far from complete. However, two of the brighter galaxies in our sample
(VCC1316 = NGC4486, M87, 3C 274 and VCC763 = NGC4374, M84, 3C 272.1) are
known to host AGNs with strongly non-thermal spectral energy distributions
(e.g., Wrobel 1991; Ho 1999; Chiaberge, Capetti \& Celotti 1999).
In both cases, the
unresolved non-thermal nucleus is clearly seen in the ACS images;
in neither instance, however, does there appear to be a resolved 
stellar nucleus. A third galaxy (VCC1619 = NGC4550), is classified
as a LINER by Ho \etal\ (1997). This galaxy contains some dust within the
central $\sim$ 25\arcsec, but there is clear evidence for a resolved
stellar nucleus.

Wrobel (1991) detected nuclear radio emission in three other galaxies
in our survey (VCC1226 = NGC4472, M49; VCC1632 = NGC4552, M89; and VCC1978 =
NGC4649, M60). In both VCC1226
and VCC1632, the innermost $\sim$ 1\arcsec~are slightly obscured by dust 
(see Paper VI), but once a correction for dust obscuration is performed, 
there is no evidence of a stellar nucleus in either case.  We see no sign
of a nucleus in VCC1978.

A search for low-level AGN in
our program galaxies is now underway using low- to intermediate-resolution
ground-based optical spectra, the results of which will be presented in a
future paper in this series. These spectroscopic data will be useful in
establishing the extent to which non-thermal sources are responsible for,
or contribute to, the central luminosity excesses observed in a number of
these galaxies.  For the time being,
Table~\ref{tab:class} summarizes our classifications for the program galaxies, as
discussed in the next section.  We begin by defining a class of galaxies
(Type 0) in which dust obscuration (four galaxies) or AGN emission (two galaxies)
renders a reliable classification as nucleated or non-nucleated impossible. In what
follows, we shall limit our analysis to the remaining 94 galaxies.

\subsection{Identification and Classification of the Nuclei}

As a first step in the identification of nuclei in our program galaxies,
the $g$- and $z$-band surface brightness profiles were each fitted with the
appropriate galaxy model (i.e., either a pure S\'ersic or core-S\'ersic model)
outside a geometric mean radius of 0\farcs5. Those galaxies with brightness profiles
which lay systematically above the inward extrapolation of fitted model for
$\lesssim 0\farcs5$ were considered to be nucleated. Because
many of the nuclei are somewhat bluer than the underlying galaxies, a 
central excess was often more apparent in the $g$-band profile than in
the redder bandpass. In addition to classifying the galaxies on the
basis of their brightness profiles, the F475W and F850LP images for each
galaxy were carefully inspected for the presence of a distinct central excess. Using
these two criteria, a total of 62 galaxies were found to show clear
evidence for a central nucleus; such galaxies are classified
as Type Ia or Ib.

Unfortunately, for 11 of these 62 galaxies, although the presence of a
faint central component could be established from the images themselves or
from the brightness profiles, the nucleus itself was too faint 
to allow us to recover trustworthy photometric or structural parameters
from the surface brightness profiles. Such galaxies are
referred to as Type Ib in Tables~\ref{tab:data} and \ref{tab:class}. Our analysis of the structural and
photometric properties of the nuclei is therefore based on the subset of
51 nucleated galaxies for which it was possible to obtain a reliable
fit to the central brightness profiles: i.e., Type~Ia galaxies.
The Type~Ib galaxies are classified as
nucleated for the purposes of computing the overall frequency of
nucleation, but their nuclei are omitted from the analysis in \S4.4 to 4.8.

Of the remaining 94--62 = 32 galaxies, five may have nuclei which
are offset by $\approx 0\farcs5$ or more from the centers of the
isophotes (Type~Ie galaxies). We consider these five galaxies in
more detail in \S4.3. The remaining 94--62--5 = 27 objects consist of
galaxies which are either unquestionably non-nucleated, or galaxies with
uncertain classifications. 
As described in Appendix~A, we have carried out a series of experiments
in which simulated nuclei having sizes and
luminosities that obey the empirical scaling relations found in
\S\ref{sec:results}, are added to --- and removed from --- the observed
brightness profiles. By re-fitting the brightness profiles obtained
in this way, we aim to refine the nuclear classifications of these galaxies.
To summarize our conclusions from these simulations, we 
classify 12 of these 27 galaxies as
{\it certainly non-nucleated} (Type~II),
11 as {\it possibly nucleated} (Type~Id)
and four others as {\it likely nucleated} (Type~Ic).
These classifications are reported on a case-by-case basis in Table~\ref{tab:data},
and summarized for the entire sample in Table~\ref{tab:class}.

\begin{figure}
\figurenum{3}
\plotone{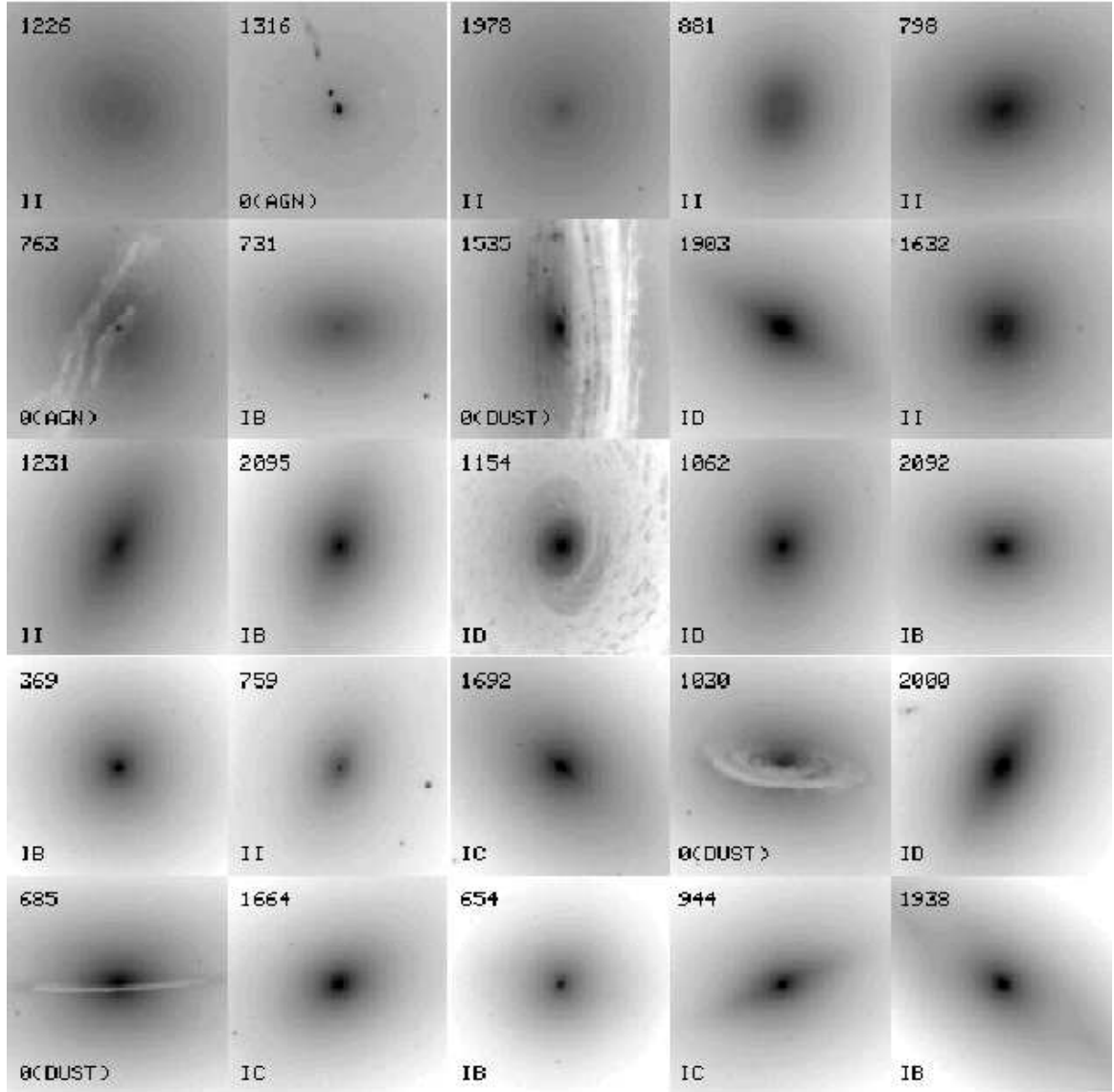}
\caption{F475W ($g$) images for galaxies from the ACS Virgo Cluster Survey.
Panels show the inner $10\arcsec\times10\arcsec$ ($\approx$ 800$\times$800~pc) of each
galaxy. Galaxies are rank ordered by blue magnitude, which increases from left to
right and from top to bottom (i.e., decreasing luminosity). Classifications from 
Table~\ref{tab:data} are reported in the lower left corner of each panel. The classification 
scheme itself is summarized in Table~\ref{tab:class}. For those galaxies with possible offset
nuclei (Type Ie galaxies), the arrow shows the presumed nucleus while the cross
indicates the galaxy photocenter. Only the brightest 25 galaxies are included
in this submission; images for the remaining galaxies can be found in
the version posted on the ACSVCS webpage: http://www.cadc.hia.nrc.gc.ca/community/ACSVCS/publications.html\#acsvcs8.
\label{fig03}}
\end{figure}

\begin{figure}
\figurenum{4}
\plotone{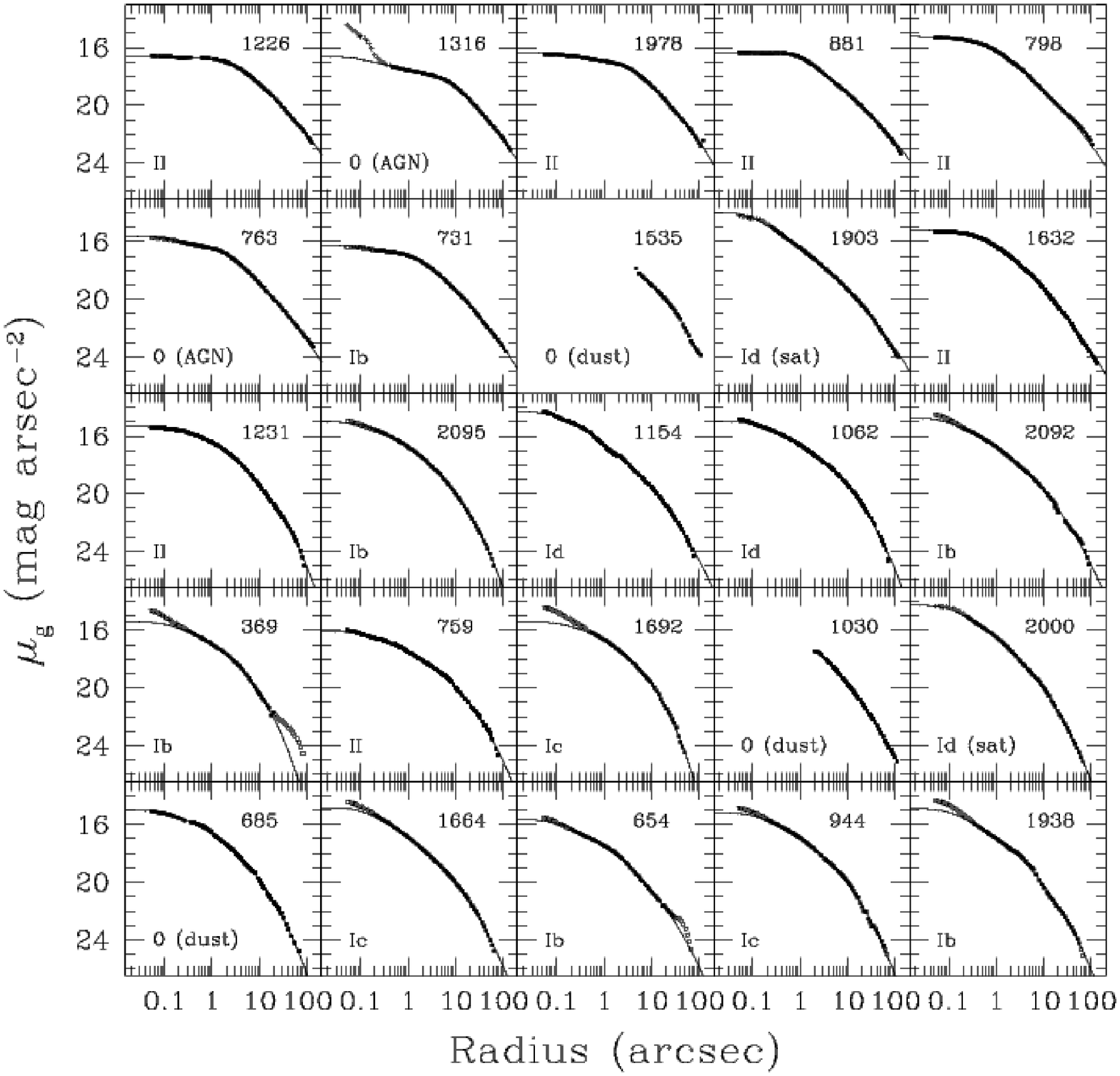}
\caption{Azimuthally-averaged $g$-band surface brightness profiles for the
galaxies from the ACS Virgo Cluster Survey (see Figure~\ref{fig03}).
The galaxies are rank ordered by blue magnitude, which decreases from left to
right and from top to bottom. Classifications from Table~\ref{tab:data} are reported in
the lower left corner of each panel. The classification scheme itself is
summarized in Table~\ref{tab:class}. In each panel, the solid curve shows the best-fit
galaxy model; open symbols show datapoints that were excluded during the
fitting of the models. For Type Ia galaxies, dashed and dotted curves show
the respective models for the galaxy and the nucleus. The fitted models for
galaxies with possible offset nuclei (the Type~Ie galaxies) do not include
components for the nuclei.
For VCC1316, VCC1903 and VCC2000, the crosses show regions affected by
saturation of the F475W images; these regions were excluded when fitting
models to the brightness profiles. 
The brightness profiles for two galaxies (VCC1192 and VCC1199)
flatten outside of $\approx$ 10$^{\prime}$ due to contamination from the
halo of VCC1226 (M49), which is $\lesssim 5^{\prime}$ away in both cases.
These outer points have been excluded in the fits.
\label{fig04}}
\end{figure}

Figure~\ref{fig03} shows F475W images for the central
10\arcsec$\times$10\arcsec~regions ($\approx$ 800$\times$800~pc) 
of all 100 galaxies in the survey. Each galaxy is labelled according
to the classifications from Table~\ref{tab:data}. Azimuthally-averaged surface
brightness profiles in the $g$-band for all 100 galaxies are
shown in Figure~\ref{fig04}.  For the 51 Type Ia galaxies shown
in this figure, the dashed
and dotted curves indicate the best-fit models for the nucleus and galaxy,
while the combined profile is shown by the solid curve. For all remaining
galaxies, the solid curve simply shows the best-fit S\'ersic or
core-S\'ersic model. Note that no fit was possible for either VCC1535
or VCC1030, the two galaxies with the most severe dust obscuration.
Open symbols in Figure~\ref{fig04} denote datapoints that were omitted
when fitting the galaxy profile (e.g., the innermost datapoints for galaxies
which contain nuclei too faint to be fit reliably,
outer datapoints for the close companions of
luminous giant galaxies, and, occasionally, datapoints corresponding to
pronounced rings, shells, or other morphological peculiarities). 

\subsection{Errors on Fitted Parameters}

Given that independent fits of the $g$- and $z$-band brightness profiles 
are performed for the Type~Ia galaxies, it is natural to
ask how well the photometric and structural parameters of the nuclei measured in the
two bands agree. The first two panels of Figure~\ref{fig05} compare
the King model half-light radii, $r_h$, and total magnitudes, 
$g_{\rm AB}$ and $z_{\rm AB}$, measured from the separate profiles (filled
circles). Note that for 11 of these 51 galaxies (i.e., those objects
with $B_T \lesssim 13.5$), the King concentration index and
half-light radii of the nuclei were constrained to be the same in the
two bandpasses; these nuclei are plotted as open stars in the first
panel of Figure~\ref{fig05}. In addition, we include
in this figure the five galaxies with possible offset nuclei, 
bearing in mind that in these cases, the $r_h$, $g_{\rm AB}$ and
$z_{\rm AB}$ measurements were carried out in a rather
different way (see \S4.3 for details).
The open circles show the nuclei of these five galaxies.

The third panel of Figure~\ref{fig05} compares two estimates for
the color of the nuclei: i.e., that obtained by integrating the 
best-fit $g$- and $z$-band King models, ${\it (g-z)_{AB}}$, and
an aperture color, ${\it (g-z)^a_{AB}}$, obtained using a circular 
aperture of radius 4 pixels (0\farcs20 $\approx$ 16~pc) applied to
the nucleus of the galaxy-subtracted image. 
The mean difference between the total and aperture colors is 0.018~mag, in
the sense that the aperture colors are slightly redder.
The $rms$ scatter in the measured radii and
colors is found to be $\langle\sigma(r_h)\rangle \sim$ 0\farcs007 and
$\langle\sigma(g-z)\rangle \sim$ 0.059~mag, respectively. Assuming
the latter uncertainty arises equally from errors in the $g$ and $z$ bands,
we find $\langle\sigma(g)\rangle = \langle\sigma(z)\rangle \sim$ 0.041~mag
for the nuclei magnitudes.  We adopt these
values for the typical uncertainties on the fitted radii, colors and magnitudes,
bearing in mind that additional systematic errors (e.g., in the photometric
zeropoints or in the construction of the PSFs) may affect the measurements.
In any case, we conclude from Figure~\ref{fig05} that there is 
excellent internal agreement between the measured sizes, colors and magnitudes.

\subsection{Frequency of Nucleation}

VCC classifications for our program galaxies are given in Table~\ref{tab:data} where
column (8) reports the classification from BST85: {\tt Y} means nucleated,
{\tt N} means non-nucleated. Our new classifications are given in
column (9). Column 10 indicates which type of model was used to represent the
galaxy: ``{\tt S}" = S\'ersic or ``{\tt cS}" = core-S\'ersic.
 
The most basic property of the nuclei which might serve as a constraint on
theories for their origin is the overall frequency, $f_n$,  with which they
are found in our program galaxies. 
Among the 94 galaxies which can be reliably classified as either 
nucleated or non-nucleated, we find 62 galaxies, or $f_n = 62/94 \approx 66$\% 
of the sample to show clear evidence for a central nucleus (Types~Ia and Ib).
However, we believe
this estimate should be considered a firm lower limit on the frequency. Including
the Type~Ic galaxies (which are very likely to be nucleated but could not be classified
as such unambiguously), gives $f_n = 66/94 \approx 70$\%. If one also includes
the Type~Id galaxies, which {\it may} be nucleated, one then finds 
$f_n = 77/94 \approx 82$\%.  Finally, if all
five galaxies with possible offset nuclei are included (although we caution
in \S4.3 that the weight of evidence argues against doing so), the
percentage of nucleated galaxies could be as high as $f_n = 82/94 \approx 87$\%.
While the true frequency probably lies between these extremes, it is
nevertheless striking to think that, among our sample of 94 classifiable
galaxies, in only 12 cases can the {\it absence} of a nucleus be established
with any degree of certainty.

\subsubsection{4.3.1 Comparison with Ground-Based Studies}
\label{sec:ground}

Among the 100 elliptical, lenticular and dwarf galaxies in the ACS Virgo
Cluster Survey, 24 dwarf galaxies (dE,Ns) and one E galaxy were classified
as nucleated in the original VCC (BST85; see also Table~\ref{tab:data} of
Paper~I).\footnote{The lone elliptical in our sample which was classified
as nucleated by BST85 is VCC1422 = IC3468 (E1,N:). However, Binggeli \& Cameron (1991) argue
that this galaxy is in fact a misclassified {\it dwarf}. In what
follows, we take the total number of nucleated dwarf galaxies
in our sample, estimated from the BTS87 classifications, to be 25.}
The frequency of nucleation which we derive
here, $f_n \approx$ 66--87\%, is much
higher than the value of $f_n \approx 25\%$ found using the
classifications of BST85, and represents a sharp upward
revision of the nucleation frequency for early-type galaxies
in this luminosity range.

There are several reasons why 
such a discrepancy should come as no surprise. 
To the best of our knowledge, ours is the first systematic
census of nuclei in early-type galaxies that includes both dwarf and
giant galaxies (spanning a factor of $\sim$ 460 in blue luminosity).
More importantly, the studies of BST85 and BTS87, along with most of the
major subsequent studies of dwarf galaxies and their nuclei (e.g.,
Binggeli \& Cameron 1991; 1993), were based on visual inspection of
photographic plates. As pointed out in \S1 and stressed by
BST85 themselves, the VCC classifications are known to be incomplete
fainter than $B \gtrsim$ 23 ($M_B \gtrsim -8$) and to suffer from
surface brightness selection effects for the luminous E and S0
galaxies. Clearly, selection effects of this sort are less of an
issue for our survey, where the identification of the nuclei is
relatively straightforward thanks to the depth and high spatial 
resolution of the ACS images.

In any case, care must be taken when comparing our measurement to
previous estimates since the frequency of nucleation is known to
depend on the luminosities of the galaxies under consideration 
(e.g., Sandage, Binggeli \& Tammann 1985).
Figure~\ref{fig06} shows the luminosity functions for our
sample of 62 nucleated galaxies (Types~Ia and Ib) as the double-hatched
histogram; the hatched histogram shows this same sample plus the
15 likely or possibly nucleated galaxies of Types Ic and Id 
(i.e., 77 galaxies in total).  For comparison,
the 25 galaxies classified as nucleated by BST85 are shown by the filled
histogram, and the open histogram shows the distribution of the 94
classifiable galaxies from the ACS Virgo Cluster
Survey. As expected, the disagreement between our classifications
and that of BST85 is quite dramatic for galaxies brighter than
$B_T \approx 13.7$.  This happens to be the approximate dividing
point between dwarf and giant galaxies in the VCC, which strongly 
suggests that the disagreement is the result
of selection effects that made it difficult or
impossible for BST85 to identify nuclei in bright, high-surface-brightness
galaxies. For $B_T \gtrsim 13.7$, there is better
agreement although we still find significantly more
nuclei even among these faint galaxies: i.e., we classify 46 of 53 galaxies,
or $87\%$, of this subsample as nucleated, compared to just
25/56 ($\approx$ 47\%) using the BST85 classifications. The
luminosity dependence of $f_n$ is shown explicitly in the lower
panel of Figure~\ref{fig06}. 

A vivid demonstration of the importance of surface brightness selection 
effects when classifying nuclei
is shown in Figures~\ref{fig07} and \ref{fig08}. The first
of these figures compares the distribution of galaxy surface 
brightnesses, measured at a geometric mean radius of 1\arcsec, for the 
same four samples shown in the Figure~\ref{fig06}. By contrast,
Figure~\ref{fig08} shows nucleus magnitude as a function of
galaxy surface brightness measured at a geometric mean radius of 1\arcsec. 
Filled symbols show the 51 Type~Ia galaxies in our sample, while the
open squares show the 25 galaxies classified as nucleated by BST85.
Open circles in this figure denote the five galaxies with possible offset
nuclei.

Figures~\ref{fig07} and \ref{fig08} leave little doubt that the survey of BST85
preferentially missed nuclei in the bright, high-surface-brightness
galaxies. We further note that a recent survey of 156 Virgo dwarfs with the
Wide Field Camera on the Isaac Newton Telescope uncovered faint nuclei
in 50 galaxies previously classified as non-nucleated, consistent with
our upward revision for frequency of nucleation (Grant, Kuipers \&
Phillipps 2005). Of course, it is conceivable we {\it too} may
be missing faint nuclei in the highest surface brightness
galaxies; it is certainly true that the 11 galaxies for
which we are unable to measure reliable photometric or structural parameters
for the nuclei are among the highest surface brightness galaxies in
our survey. Accordingly, we
stress once again that the estimate of $f_n \approx 66$\% from
\S4.1, which is based on galaxies with unambiguous nuclei, {\it is
certainly a lower limit to the true frequency of nucleation among
our sample of early-type galaxies}.
We shall return to this point in \S5.2 (see also Appendix~A).

Figures~\ref{fig09} and \ref{fig10} illustrate the
importance of {\it HST} imaging for
the identification of nuclei in these galaxies. In 
Figure~\ref{fig09} we show a comparison between
the co-added F475W image for VCC2048 (IC3773) --- a Type~Ia galaxy --- with three simulated
ground-based images for this same galaxy. In these three
cases, the co-added F475W frame has been binned $4\times4$ and 
convolved with Gaussians having dispersions of 1, 2 and 3 pixels, corresponding
to FWHM of 0\farcs5, 0\farcs9 and 1\farcs4. It is clear that seeing effects
alone make the detection of faint, compact nuclei challenging under normal
conditions of ground-based seeing. This finding is all the more sobering
when one considers that VCC2048, classified as dS0(9) in the
VCC, was thought on the basis of the original BST85 classifications
to be the brightest non-nucleated dwarf galaxy in our sample.

The first two panels of Figure~\ref{fig10} compare the F475W
image for VCC784 (NGC4379) with a $V$-band image taken with the 2.4m Hiltner
MDM telescope on 21 April, 1993 in conditions of 1\farcs14 seeing.
This galaxy, one of the brightest Type Ia galaxies in our survey, was
also classified as non-nucleated in the study of BST85. As the
third panel of Figure~\ref{fig10} demonstrates, there is no hint of
a central nucleus in the ground-based surface brightness profile,
despite the fact that the nucleus, which is clearly visible in the
ACS brightness profile, is among the brightest and largest in our sample.

\subsubsection{4.3.2 Comparison with Previous HST Studies}
\label{sec:previous}

As noted in \S1, a few {\it HST} studies had previously revealed the
presence of compact nuclei in bright early-type galaxies (e.g., Rest \etal\ 2001; 
Ravindranath \etal\ 2001; Lauer \etal\ 2005).  While these programs
preferentially focussed on distant, high-luminosity ellipticals and
lenticulars --- with 80\% of the galaxies in these respective surveys having
absolute magnitudes brighter than $M_V \sim -20, -20$ and $-20.8$, compared
to $M_V \sim -16$ for the present survey --- there is nevertheless
some overlap with our sample at the bright end due to the large number
of luminous E and S0 galaxies in the Virgo Cluster. In this section, we
compare our nuclear classifications with those reported in these previous
surveys, limiting the comparison to those galaxies in our survey which
have unambiguous classifications (i.e., Types~Ia, Ib and II). For 
completeness, we also compare our classifications for three faint galaxies
to those of Lotz \etal\ (2004) who carried out a WFPC2 snapshot survey of
early-type dwarf galaxies in the Virgo and Fornax Clusters.
Table~\ref{tab:comp} summarizes the nuclear classifications for galaxies
in common with the surveys of Rest \etal\ (2001), Ravindranath \etal\ (2001),
Lauer \etal\ (2005) and Lotz \etal\ (2004).

The Rest \etal\ (2001) study presented WFPC2 (F702W $\approx R$) imaging for
67 early-type galaxies between 6 and 54 Mpc, with a mean distance of 
$\langle d\rangle = 28\pm$9
Mpc. To minimize spurious detections, Rest \etal\ (2001) adopted rather
conservative criteria in their search for nuclei, identifying nucleated
galaxies as those objects which showed a central excess, along both the major
and minor axes, over the best-fit ``Nuker" model inside a radius of
0\farcs15. Based on these criteria, they identified nuclei in 
9 of their 67 galaxies (13\%). No structural and photometric parameters
were measured for the nuclei.
There are six galaxies in common between their survey and ours. We find
reasonable agreement between the two studies, with the exception
of VCC731 (NGC4365): Rest \etal\ (2001) report no nucleus in this galaxy, 
whereas we find a small, but definite, central brightness excess.
Accordingly, we classify this galaxy as Type~Ib (i.e., certainly nucleated). 

The NICMOS study of Ravindranath \etal\ (2001) was carried out using F160W 
($\approx H$) images from the NIC2 (FWHM = 0\farcs17, scale =
0\farcs076) and NIC3 (FWHM = 0\farcs22, scale =
0\farcs2) cameras. For 33 galaxies with distances
in the range 7 to 69~Mpc and $\langle d\rangle = 21\pm$14~Mpc, these authors fitted
two-dimensional, PSF-convolved ``Nuker" models to their NICMOS images. Compact
sources --- consisting of narrow, PSF-convolved Gaussians --- were then included
for those galaxies whose one-dimensional surface brightness profiles
showed evidence for a central excess (14 of their 33 galaxies).
FWHMs and magnitudes for the nuclei were then obtained by $\chi^2$ minimization. We
find good agreement with the Ravindranath \etal\ (2001) classifications.
Specifically, we confirm the absence of nuclei in VCC1226 and VCC881 
(M86 = NGC4406). For VCC763 (M84 = NGC4374), which is classified as
nucleated by these authors, we confirm the presence of a
central point source, although the galaxy is classified as Type~0
in Table~\ref{tab:data} due to the presence of strong AGN activity. All of the
nuclei in the study of Ravindranath \etal\ (2001) were found to be
unresolved point sources, although this is probably a consequence of the
relatively poor resolution of their images: i.e., at the mean distance of
their sample galaxies, the NICMOS FWHM corresponds to $\sim$ 20~pc.

The WFPC2 study of Lauer \etal\ (2005) was based on F555W or F606W ($\approx V$)
imaging for 77 galaxies; a little more than half of their galaxies (45) were also
imaged in the F814W ($\approx I$) bandpass. The galaxies have distances
in the range 10 to 97~Mpc, with mean $\langle d\rangle = 33\pm21$, so the
0\farcs07 FWHM for the PC1 CCD corresponds to a physical
scale of 11~pc for the typical galaxy.
Magnitudes and colors for the nuclei in their sample --- identified as an
excess above the ``Nuker" model which best fits the observed brightness
profile --- were derived by direct integration of the model residuals.
In only two of their 25 nucleated galaxies did the nucleus appear resolved.
There are seven ACSVCS galaxies having unambiguous
nuclear classifications which are in common with Lauer \etal\ (2005). The
classifications are in agreement in three cases: VCC1978, VCC731 and
VCC1903 (M59 = NGC4621). For VCC1146 (NGC4458),  which Lauer \etal\ (2005) 
classify as non-nucleated, we believe the discrepancy may be due to
the highly extended nature of the nucleus. With
$r_h \approx$ 0\farcs8 = 62~pc, it is largest nucleus in our sample, and
would be difficult to distinguish from the underlying galaxy profile
in brightness profiles of limited radial extent; the Lauer \etal\ (2005) 
brightness profile for this galaxy covers just the inner 5\arcsec.

The three remaining galaxies --- VCC1226, VCC881 and VCC1632 --- are
listed as nucleated in Lauer \etal\ (2005), but we classify each of
these galaxies as Type~II (non-nucleated). We speculate that the
detection of nuclei in VCC1226 and VCC1632 is
an artifact resulting from the presence of dust in both galaxies,
which partly obscures the innermost $\sim$ 1\arcsec.
Lauer \etal\ (2005) 
do not correct their images for dust obscuration; once
such a correction is performed, we find no indication of a central
nucleus in either galaxy (see also Paper~VI).

In the case of VCC881, which is classified as nucleated by Lauer \etal\ (2005),
a faint continuum enhancement is indeed detected in both the $g$ and $z$ bands
at the central location. This feature would certainly be enhanced by the
deconvolution procedure applied by Lauer \etal\ (2005) to their data.
However, it is unclear whether this
corresponds to a stellar nucleus. If one assumes that VCC881 follows the scaling
relation between nucleus and galaxy luminosity obeyed by
the rest of the sample, then the putative nucleus would be
underluminous by a factor of $\sim$ 250. Furthermore, starting around 0\farcs4, the
surface brightness profile of VCC881 {\it decreases} towards the center
(Carollo \etal\ 1997). The origin of this central surface brightness depression 
is unclear (Lauer \etal\ 2002; Paper~VI): e.g., an intrinsic decrease in the
luminosity (or mass) density, or perhaps obscuration by gray dust, might be
responsible. Since either processes could produce a modest and localized 
continuum enhancement such as the one seen at the nuclear location, we
believe this galaxy is best classifed as non-nucleated (Type~II). We further
note that there is no evidence of nuclear activity in VCC881 from its
optical, radio and X-ray properties (Ho \etal\ 1997; Rangarajan \etal\ 1995;
Fabbiano \etal\ 1989).

Lotz \etal\ (2004) carried out WFPC2 (F555W $\approx V$ and F814W $\approx I$)
imaging for 69 dEs and dE,Ns, mostly belonging to the Virgo and Fornax Clusters.
In their analysis, a nucleus was identified as a bright compact object within
1\farcs5 of the galaxy photocenter. While the Lotz \etal\ (2004) survey 
tended to focus on fainter galaxies than does our survey (i.e., their program galaxies have
absolute blue magnitudes in the range $-17 \lesssim M_V \lesssim -11.7$ mag, 
with mean $\langle M_V\rangle = -14.2$~mag), there are three galaxies which
appear in both studies: VCC9 (IC3019), VCC543 and VCC1948. In the case of VCC9 and
VCC1948, the two studies agree in finding no evidence of a nucleus at the
position of the galaxy photocenter. However, we have identified both of these
galaxies as possible examples of galaxies with offset nuclei (see \S\ref{sec:offset}). 
Although Lotz \etal\ (2004) do not comment on a possible offset nucleus in the
case of VCC1948, they state that: ``VCC9 was originally classified
as nucleated by Binggeli \etal\ (1985), but its brightest globular cluster
candidate is 1\farcs8 from its center". For comparison, we measure an
offset of 1\farcs91$\pm$0\farcs07 for this object and, like Lotz \etal\ (2004), 
conclude that it is probably a star cluster projected close to the galaxy
photocenter, rather than a {\it bonafide} nucleus. The remaining galaxy, VCC543,
appears in the list of non-nucleated galaxies in Table~3 of Lotz \etal\ (2004), 
although we find unmistakable evidence for a nucleus in this object (see
Figures~\ref{fig03} and \ref{fig04}) that is offset by no more than
0\farcs07$\pm$0\farcs12 from the galaxy photocenter.

Finally, we note that two recent papers (de Propris \etal\ 2005; Strader \etal\ 2006)
have examined the properties of nuclei belonging to subsets of the galaxies from
the ACS Virgo Cluster Survey, based on the same observational material used in
this paper. A detailed comparison of the sizes, magnitudes and colors we measure
for the nuclei with those reported by de Propris \etal\ (2005) and 
Strader \etal\ (2006) is given in Appendix~B.

\subsection{Possible Offset Nuclei}
\label{sec:offset}

Before proceeding, we pause to consider those galaxies that may have
offset nuclei. Nuclei displaced from the photocenters of their host 
galaxies are potentially interesting since they may hold clues to the general processes
which trigger and/or regulate the formation of nuclei in general. For instance, offsets
may arise through the ongoing merging of globular clusters through
dynamical friction (Tremaine, Ostriker \& Spitzer 1975; Miller \& Smith 1992), 
the fading of stellar populations in dwarf irregular or blue compact dwarf 
galaxies as they evolve into dwarf ellipticals (e.g., Davies \& Phillips 1988),
recoil events following the ejection of a supermassive black hole from the nucleus
(Merritt \etal\ 2004) or counter-streaming instabilities that develop in flat 
and/or non-rotating systems (Zang \& Hohl 1978; De Rijcke \& Debattista 2004).

From an observational perspective, the identification of such nuclei is
a complicated problem. They are prone to confusion with globular clusters, foreground
stars or background galaxies --- difficulties that are particularly serious in ground-based
imaging, where the nuclei and contaminants will appear unresolved. The most
ambitious study of offset nuclei undertaken to date is that of Binggeli,
Barazza \& Jerjen (2000), who measured offsets for a sample of
78 nucleated dwarf galaxies in the Virgo Cluster using digitized images of
blue photographic plates obtained in conditions of FWHM $\approx$ 1\farcs2 seeing. 
They found offset nuclei to be commonplace, with 
$\delta{r_n} \gtrsim 0\farcs5$ in 45 (58\%) and 
$\delta{r_n} \gtrsim 1$\arcsec~ in 14 of the objects (18\%).
It is of interest to check these results given the small sizes of the measured
offsets relative to the ground-based seeing disk, the absence of color information
that might be used to identify contaminants, 
and the possibility of confusion with Galactic stars, globular
clusters and, to a lesser extent, background galaxies.

We have used our ACS images to measure offsets for the nuclei of the 62 Type~Ia
and Ib galaxies
in our sample. In both the F475W and F850LP images for each galaxy,
we first calculate the centroid of the nucleus and its corresponding uncertainty.
The location of the galaxy photocenter is then found by averaging the centers
of ellipses fitted to the galaxy isophotes over the range 
1\arcsec~$\le r \le r_e$ (Paper~VI). The uncertainty on the
position of the photocenter is taken to be the standard deviation 
about the mean ellipse center. Adding in quadrature
the uncertainties for the position of the nucleus and photocenter 
then yields the uncertainties for the offset. The
results reported in column 17 of Table~\ref{tab:data} are
averages of the offsets measured from the F475W and F850LP images.

Figure~\ref{fig11} shows the measured offsets for the 62 galaxies.
Offsets are shown both in arcseconds (upper panel)
and in units of the effective radius of the galaxy, $\langle r_e\rangle$,
taken from Paper~VI (lower panel). In only three galaxies do we see
evidence for an offset as large as 1\arcsec. Using a less restrictive criterion
of $\delta{r_n} \gtrsim$ 0\farcs5, we find
only five galaxies that may have offset nuclei (i.e., Type~Ie galaxies).
These galaxies, which are shown as the open circles in
Figure~\ref{fig11}, are:

\begin{itemize}

\item[] {\it VCC9}. This very low surface brightness galaxy has
multiple bright sources near its photocenter; it may be a dIrr/dE
transition object and seems to contain a rich population of ``diffuse
star clusters" (Paper~XI). In addition to the
presumed nucleus, there is a second source about two magnitudes
fainter which is located $\approx$ 1\farcs5 from the photocenter (and a
similar distance from the presumed nucleus). Both the color and
the half-light radius of the presumed nucleus are similar to those of 
metal-poor globular clusters in our dwarf galaxies. Thus, it is
conceivable that this galaxy has no nucleus at all. 

\item[] {\it VCC21 (IC3025)}. More than a dozen bright sources are found in the
inner regions of this very low surface brightness galaxy. Based on its
mottled appearance, this galaxy too should be re-classified as a dIrr/dE 
transition object. The presumed nucleus is 
located $\approx$ $0\farcs76\pm0\farcs07$ from the galaxy photocenter, the
smallest offset in our sample of five candidates. There are, however,
two fainter sources close by, and given the large number of sources 
in this galaxy, its transitional morphology, and the fact that the
presumed nucleus has a very blue color of $(g-z) \approx 0.3$, 
we believe the evidence indicates that the ``nucleus" in VCC21 is 
probably a young star cluster.

\item[] {\it VCC1779 (IC3612)}. This highly flattened galaxy ($\epsilon \simeq 0.5$)
is noteworthy in that it contains dust filaments --- unusual for 
low- and intermediate-luminosity galaxies in our sample (see Paper~VI).
Like VCC9 and VCC21, this galaxy may be a dIrr/dE transition object. The ACS images
reveal four bright sources, all of which may be globular clusters, near 
the galaxy center. We identify the brightest
of these sources, which is {\it not} the nearest to the center, as the
putative nucleus. If the nearest (and second brightest) source is instead
identified as the nucleus, then the offset would be $\approx$ 0\farcs4
rather than $\approx$ 0\farcs5. 

\item[] {\it VCC1857 (IC3647)}. This galaxy, another very low surface brightness
object, has a very bright source located $\approx$~7\arcsec from its
center. This is by far the largest offset for any galaxy in our
sample, so the identification of this source as a nucleus should
be viewed with some caution. The color and half-light 
radius of the presumed nucleus are consistent with those expected
for an otherwise normal (metal-poor) globular cluster.

\item[] {\it VCC1948}. The presumed nucleus in this galaxy, another
very low surface brightness object, is located $\approx 1\farcs4$ 
from the galaxy photocenter. It is the brightest of several sources in 
the inner few arcseconds. There also appears to be a 
very faint surface brightness ``excess" that is nearly coincident
with the galaxy photocenter. It is therefore possible that this galaxy 
may be a normal (Type Ib) nucleated galaxy, albeit one with an 
unusually faint nucleus. If so, then the source identified as the
possible nucleus may be a globular cluster.

\end{itemize}

We conclude that in every case there is considerable uncertainty
regarding the nature of the presumed offset nucleus. It is possible ---
and we consider it likely --- that the ``offset nuclei" in all five of
these galaxies are merely globular clusters residing in non- or weakly-nucleated galaxies. 

Are there nuclei with even smaller offsets?  The nuclei of four other galaxies
(VCC1539, VCC2019, VCC1895 and VCC1695) have offsets $0\farcs1 \lesssim \delta_{r_n} \lesssim 0\farcs5$ and, in 
fractional terms, there are four other galaxies that have nuclei offset by more than 1\% of 
the galaxy effective radius (VCC2019, VCC1199, VCC1695 and VCC1895).
We remind the reader, however, that these offsets correspond to just two ACS/WFC
pixels, and should be confirmed using deeper, higher resolution imaging. 
Our only secure conclusion is that, {\it at most}, only five of the
nucleated galaxies in our survey,
or $\approx$ 7\% of the sample, have nuclei that are offset
by more than 0\farcs5~from their photocenters --- at least three times
smaller than the value of $\sim$~20\% found by Binggeli \etal\ (2001).
Moreover, we believe that most --- and perhaps {\it all} --- of the 
``nuclei" in the Type~Ie galaxies
are probably globular clusters, so this should be considered a
firm upper limit on the percentage of galaxies
with nuclei offset by more than 0\farcs5. The actual percentage may
in fact be zero.

\subsection{The Spatial Distribution of Nucleated and Non-Nucleated Galaxies}
\label{sec:spat}

A key result to emerge from the survey of BTS87 was the discovery of a
spatial segregration between nucleated and non-nucleated dwarf galaxies in
Virgo: the 
dE,Ns are more strongly concentrated to the cluster center than the dEs (see, 
e.g., Figure~9 of BTS87). A similar trend was
later reported for the Fornax Cluster by Ferguson \& Sandage (1989).
Our discovery of nuclei in many of the galaxies classified as non-nucleated by
BST85 suggests that it is worth reconsidering this important issue.

To facilitate comparison with the BTS87 and BST85 results, we limit our analysis 
in this section to those galaxies fainter than $B_T \gtrsim 13.7$, the approximate
dividing point between dwarfs and giants in the VCC. As it happens, this magnitude
also divides the ACS Virgo Cluster Survey equally into two samples of 50 galaxies.

In the left panel of Figure~\ref{fig12}, the heavy solid curve shows the 
cumulative distribution of projected distances from the center of the Virgo Cluster
for the 50 galaxies with $B_T \gtrsim 13.7.$\footnote{The cluster center is taken to
be the position of M87: $\alpha$(J2000) = 12:30:49.4 and $\delta$(J2000) = 12:23:28.} 
Using the VCC classifications, one finds this sample to be composed predominantly
of dwarfs (i.e., 33 of 50 galaxies according to Table~1 of Paper~I).
The dotted and dashed curves show the corresponding distributions for the nucleated
(23) and non-nucleated (27) galaxies in this sample, once again using the VCC
classifications. A KS test confirms the visual impression from this figure
that the nucleated galaxies in our survey exhibit the same trend noted by BTS87 for 
the full sample of nucleated dwarfs in the VCC: i.e., the dE,N galaxies are more
centrally concentrated than their non-nucleated counterparts.

In the right panel of this figure, we show the sample of 49 galaxies
with $B_T \gtrsim 13.7$ which we are able to classify as nucleated or
non-nucleated from our ACS images (heavy solid
curve).\footnote{One galaxy in this magnitude range, VCC571, is
excluded because of an irregular dust lane which obscures 
the nucleus and makes a definitive classification impossible.} Excluding
for the moment the five galaxies with possible offset nuclei, whose true
nature is uncertain, we find
40 of 44 remaining galaxies to be nucleated (dotted curve). Given
the preponderance of nucleated galaxies, it is not surprising to see that
a systematic difference in central concentration between the two populations
is no longer apparent. Of course, with just four non-nucleated galaxies in
this regime, the sample falls below the minimum
size needed for statistically reliable results using a KS test, but
our point in showing this comparison is to stress again that the
overwhelming majority of program galaxies with $B_T \le 13.7$ contain nuclei.

A definitive investigation into the spatial distribution of nucleated and
non-nucleated galaxies in the Virgo Cluster would require ACS imaging for
many hundreds of galaxies. Nevertheless, we can speculate on
the origin of the trend noted by BTS87. It has long been known that galaxies 
in Virgo show some segregation in terms of luminosity and morphology. Ichikawa
et~al. (1988) noted that the dwarf elliptical galaxies in the central regions of
Virgo appear to be larger and brighter than those in the cluster outskirts. At the
same time, BTS87 showed that the bright early-type galaxies (E+S0) are less
strongly concentrated to the cluster center than the faint (dE) early-type 
galaxies (see their Figures~7 and 8). Since central surface brightness
is proportional to total luminosity for early-type galaxies, the BTS87 finding implies
that bright, high-surface-brightness dwarfs (HSBDs) in Virgo are more 
spatially extended than low-surface-brightness dwarfs (LSBDs). Because
the original BTS87 classifications suffer from a serious surface brightness
selection effect --- in the sense that nuclei belonging to galaxies with central 
surface brightnesses $\mu_g(1\arcsec) \lesssim 20$~mag~arcsec$^{-2}$
will go undetected; see Figure~\ref{fig08} --- the observed
trend may simply be a consequence of this surface brightness selection effect.

To put this claim on a more quantitative footing, we have calculated the density 
profiles for HSBD and LSBD early-type galaxies in Virgo, using a surface brightness
of $\mu_g(1\arcsec) \approx 20$~mag~arcsec$^{-2}$ as a dividing point
between the two subgroups. As shown in Figure~\ref{fig08}, BTS87 would 
have tended to classify LSBDs as nucleated, while the HSBDs would have
been preferentially classified as non-nucleated.
A least-squares fit to
our sample galaxies gives $\mu_g(1\arcsec) = 1.139B_T + 3.44$, so that 
$\mu_g(1\arcsec) \lesssim 20$~mag~arcsec$^{-2}$
corresponds to a total galaxy magnitude of $B_T \approx 14.55$. Restricting
ourselves to the
early-type members of Virgo with $13.7 \lesssim B_T \lesssim 18$, this
leaves us with a total of 448 galaxies. The upper limit of $B_T = 13.7$ represents
the approximate transition between dwarfs and giants in Virgo, while the lower
limit reflects the completeness limit of the BTS87 survey. Among this sample
of 448 galaxies, there are 42 HSBDs with $13.7 \lesssim B_T \lesssim 14.55$,
and 406 LSBDs with $14.55 \lesssim B_T \lesssim 18$.

Figure~\ref{fig13} shows the density profiles, $\Sigma(r)$, for these two
populations. In calculating the profiles, we have excluded galaxies belonging
to the M and W Clouds, and discarded galaxies with declinations less 
than 9$^{\circ}$ to guard against contamination from the Virgo B subgroup
centered on VCC1226 (M49). Fitting exponentials of 
the form $\Sigma(R) \propto e^{-\alpha R}$ gives scalelengths of 
$\alpha = 0.49\pm0.06$ and $0.36\pm0.06$~deg$^{-1}$ for the LSBD
and HSBD populations, respectively. In other words, there is a statistically
significant tendency for the HSBD early-type galaxies --- the same galaxies
which would preferentially be misclassified as non-nucleated in the VCC
because of their high central surface brightnesses --- to be more spatially
extended. This is consistent with the interpretation
that the differing spatial distributions for dE and dE,N galaxies noted by
BTS87 is, in fact, a consequence of their survey's surface brightness limit.

\subsection{The Nucleus-to-Galaxy Luminosity Ratio} 
\label{sec:lum}

Lotz \etal\ (2004) and Grant, Kuipers \& Phillipps (2005) found that brighter
galaxies tend to contain brighter nuclei.
The upper panels of Figure~\ref{fig14} plot the magnitudes of the nuclei against
those of the host galaxies; results for the $g$ and $z$ bandpasses are shown in the
left and right panels, respectively. Filled and open symbols show the results for
51 Type~Ia and five Type~Ie galaxies.
The dashed lines in these panels show the least-squares lines of best fit:
$g_n^{\prime} = (0.90\pm0.18)g_g^{\prime} + (7.59\pm2.50)$ and
$z_n^{\prime} = (1.05\pm0.18)z_g^{\prime} + (5.77\pm2.19)$. For comparison,
the solid lines show the best-fit
relations, with (fixed) unity slope:
\begin{equation}
\begin{array}{rrrrrr}
g_n^{\prime} & = & g_g^{\prime} + (6.25\pm0.21) \\
z_n^{\prime} & = & z_g^{\prime} + (6.37\pm0.22) \\
\end{array}
\label{eq6}
\end{equation}

The lower panels of Figure~\ref{fig14} show these same data in a slightly
different form. In these panels, we plot the ratio of nucleus luminosity, ${\cal L}_n$,
to host galaxy luminosity, ${\cal L}_g$,
\begin{equation}
\begin{array}{rrrrr}
$$\eta & = & {\cal L}_n / {\cal L}_g,$$
\end{array}
\label{eq7}
\end{equation}
as a function of galaxy magnitude. Total luminosities for the nuclei were obtained by integrating the brightness
profiles of the best-fit King model components (see \S3). These
magnitudes are recorded in columns (11) and
(12) of Table~\ref{tab:data}. Galaxy luminosities are taken from Paper~VI, in which the
best-fit galaxy model --- either S\'ersic or core-S\'ersic, as specified in
column (10) of Table~\ref{tab:data} --- was integrated over all radii to obtain the total luminosity.
The contribution of the nucleus itself was excluded in the calculation
of~${\cal L}_g$.

The primary conclusion to be drawn from Figure~\ref{fig14} is that the nucleus-to-galaxy
luminosity ratio does not vary with galaxy luminosity, although there is considerable
scatter about the mean value.
In terms of $\eta$, the relations in equation~6 are equivalent to
\begin{equation}
\begin{array}{rrrrr}
\langle \eta_g\rangle & = 0.32\pm0.06~\% \\
\langle \eta_z\rangle & = 0.28\pm0.06~\% \\
\end{array}
\label{eq8}
\end{equation}
for the two bands, where the quoted uncertainties refer to the mean errors.
Our best estimate for the mean nucleus-to-galaxy luminosity ratio is then
\begin{equation}
\begin{array}{rrrrr}
\langle \eta \rangle & = 0.30\pm0.04~\%. \\
\end{array}
\label{eq9}
\end{equation}
This is well below previous
estimates: only 5 of the 51 nucleated galaxies in Figure~9
of Binggeli, Barazza \& Jerjen (2000) have nuclei with
fractional luminosities smaller than this. While the discrepancy
may partly be the result of different choices for the models used
to parameterize the galaxy brightness profiles (e.g., Binggeli et~al. 2000
use King models for the galaxy when calculating the luminosity of the
central excess), it is also
true that the greater depth and sensitivity of the ACS imaging allows
us to identify fainter nuclei than is possible from the ground, while
the high spatial resolution allows a more
accurate subtraction of the underlying galaxy light.

\subsection{Luminosity Functions}
\label{sec:lf}

The luminosity function of nuclei is one of the most 
powerful observational constraints on models for their formation.
For instance, one theory involves
the growth of a central nucleus through mergers of globular clusters whose
orbits have decayed because of dynamical friction (Tremaine, Ostriker \&
Spitzer 1975; Tremaine 1976; Lotz \etal\ 2001). While this scenario is consistent
with the well known fact that the brightest nuclei have luminosities
that exceed those of the brightest globular clusters (e.g., Durrell et~al.
1996; Durrell 1997), a reliable measurement for the luminosity function of the nuclei has
been hard to come by due to the lack of high-resolution CCD imaging for
large, homogenous samples of early-type galaxies. The need for {\it HST}
imaging in this instance is clear, since subtle differences in the
subtraction of the galaxy light (particularly the choice of model to 
represent the galaxy) can lead to large differences in the inferred
luminosities of the nuclei (see, e.g., section 5 of Binggeli \&
Cameron 1991).

In Figure~\ref{fig15}, we plot the luminosity functions, in $g$ and $z$, for the
sample of 51 Type~Ia nuclei given in Table~\ref{tab:data}. A Gaussian distribution,
\begin{equation}
\begin{array}{rrrrr}
\Phi(m_n^0) & \propto A_ne^{-(m_n^0-\overline{m}_n^0)^2 / 2\sigma_n^2} \\
\end{array}
\label{eq10}
\end{equation}
provides an adequate representation of the luminosity functions,
although there is no physical justification for this particular choice of fitting 
function (and it is likely that the luminosity function suffers from some degree 
of incompleteness at both the bright and faint ends). Moreover, if the mean luminosity
of nuclei in early-type galaxies is indeed a roughly constant fraction,
$\eta \approx 0.3\%$, of that of their host (\S\ref{sec:lum}), then the distribution
shown in Figure~\ref{fig14} may largely be a reflection of the luminosities of
the program galaxies. With these caveats in mind, we
overlay the best-fitting Gaussian distributions in 
Figure~\ref{fig15} as the dotted curves. Fitted parameters and their 
uncertainties are recorded in Table~\ref{tab:lf}. 

A core objective of the ACS Virgo Cluster Survey is a study of the
globular cluster populations associated with early-type galaxies. Since
many thousands of Virgo globular cluster candidates have been identified 
in the course of the survey (e.g., Papers IX, X and XI), it
is possible to compare directly the luminosity
functions of the nuclei with those of the globular clusters.
Figure~\ref{fig15} shows the $g$- and $z$-band luminosity
functions for $\approx$ 11,000 high-probability globular 
cluster candidates from the survey. These objects were chosen to
have globular cluster probability indices, ${\cal P}_{\rm gc}$,
in the range $0.5 \le {\cal P}_{\rm gc} \le 1$ (see Jord\'an \etal\ 2006
for details). A complete discussion of the globular cluster luminosity
function is beyond the scope of this paper and will be presented in a
future article.  For the time being, we simply note that,
brighter than the 90\% completeness limits of $g_{\rm lim} \sim 26.1$~mag and
$z_{\rm lim} \sim 25.1$~mag, the luminosity functions of the globular
clusters in Virgo (which are dominated by the contributions from the brightest
galaxies) are well
described by Gaussian distributions with $\sigma = 1.3$~mag and 
reddening-corrected turnover magnitudes of $g_{\rm to} \approx 23.9$~mag and 
$z_{\rm to} \approx 22.8$~mag.  These Gaussians are shown as the
upper dotted curves in each panel of Figure~\ref{fig15}.

It is apparent that the luminosity function of the nuclei extends
to higher luminosities than that of the globular clusters and that,
irrespective of the functional form used to parameterize the luminosity
function of the nuclei, their distribution is significantly
broader than that of the globular clusters. In addition, their distribution is
displaced to higher luminosities than that of the globular clusters. We
measure this
offset to be $\Delta{g} = 3.52$~mag and $\Delta{z} = 3.63$~mag
in the two bands. Thus, on average, the nuclei are $\sim$ 25
times brighter than a typical globular cluster. 
We shall return to this point in \S5.2.

Also shown in Figure~\ref{fig15} are seven probable UCD galaxies in the
Virgo Cluster, drawn from Paper~VII and from Ha\c{s}egan \etal\ (2006). These
objects were identified on the basis of magnitude and half-light
radius from the same images used to study the nuclei and globular
clusters. Although the UCD sample is limited in size, membership in Virgo 
has been established for each object through radial velocity
measurements, surface brightness fluctuation distances, or both. 
Furthermore, the mass-to-light ratios presented in Paper VII demonstrate
that at least some of these objects appear genuinely distinct
from globular clusters. Several explanations for their origin
have been proposed; according to
what is probably the leading formation scenario, they
are the surviving nuclei of dwarf galaxies which have been extensively
stripped by gravitational tidal fields in the host cluster (e.g., Bassino
\etal\ 1994; Bekki \etal\ 2001). 
The UCDs shown in Figure~\ref{fig15} have luminosities
which coincide with the peak of the nuclei luminosity function, which is
certainly consistent with this ``threshing" scenario. However, it
is important to bear in mind that the luminosities of the UCDs shown in
Figure~\ref{fig15} are entirely due to the construction of the sample: i.e., candidates
from Paper~VII and Ha\c{s}egan \etal\ (2006) were {\it selected} to
have $18 \le g \le 21$~mag and $17 \le z \le 20$~mag.

\subsection{Structural Properties}
\label{sec:structural}

Prior to the launch of {\it HST}, virtually nothing was known about the sizes
of the nuclei in dwarf galaxies. A notable exception was the compact, 
low-luminosity nucleus of the Local Group dwarf elliptical galaxy NGC205,
which was measured to have $r_h \sim$~0\farcs4 = 1.4~pc by Djorgovski et~al.
(1992). This early estimate, which was based on deconvolved ground-based
images, is in good agreement with more recent values obtained using
ACS surface brightness profiles (e.g., Merritt \etal\ 2006). However, 
measuring half-light radii for nuclei at the distance of Virgo
using ground-based images is impossible (e.g., Caldwell 1983;
Sandage, Binggeli \& Tammann 1985; Caldwell \& Bothun 1987). For instance, using
high-resolution CFHT images for ten Virgo dwarfs, Durrell (1997) was only able
to place an upper limit of $r_h \lesssim$ 0\farcs4--0\farcs5 (30--40~pc) on 
the sizes of the nuclei.

Even with the excellent angular resolution and spatial sampling afforded by {\it HST}/ACS, 
the measurement of structural parameters for the nuclei is challenging --- more so
than for a typical Virgo globular cluster because the nuclei are observed
on a bright background which is varying rapidly in both the radial and
azimuthal spatial directions.
In their WFPC2 snapshot survey of dwarf galaxies in the Virgo
and Fornax clusters, Stiavelli \etal\ (2001) and Lotz~et~al. (2004) did not
attempt to measure half-light radii for the nuclei, but they noted that
these nuclei have ``sizes" less than 0\farcs13 (10~pc). Working from the same
WFPC2 data for a subset of five nucleated dwarfs, Geha,
Guhathakurta \& van der Marel (2002) derived half-light radii in the
range 9--14~pc (0\farcs11--0\farcs18).

With their greater depth and superior
sampling of the instrumental PSF, our
ACS images are better suited to the measurement of half-light
radii than any previous dataset, including the WFPC2 imaging.
Moreover, images are available in two filters, so it is also possible to carry out
independent size measurements and identify possible systematic errors
arising from uncertainties in the F475W and F850LP PSFs. As shown in
Figure~\ref{fig05}, we have made such a comparison and find 
good agreement between the half-light radii
measured in the different bandpasses, with a typical {\it random} measurement
error of $\sigma(r_h) \sim$ 0\farcs007. We note that half-light radii 
measured for the nuclei of approximately two dozen of our program galaxies have
recently been reported by de Propris \etal\ (2005) and Strader \etal\ (2006).
Appendix~B presents a comparison of our structural and photometric parameters
with those measured in these studies.

Figure~\ref{fig16} shows the distribution of half-light radii for the nuclei 
of Type~Ia galaxies from Table~\ref{tab:data}.
The distribution is evidently quite broad, 
with a peak at $r_h \lesssim0\farcs05$ (4~pc) and an extended tail to much
larger radii ($0\farcs83$ $\approx$ 62~pc). The dashed line at 0\farcs025
($\approx$ 2~pc) in each panel shows our estimate for the resolution limit
of the images used to characterize the properties of the nuclei.\footnote{Note 
that this resolution limit applies only to those images which were drizzled
with a {\it Gaussian} kernel. In Paper~X, we estimated from numerical simulations that
the half-light radii of globular clusters --- which are measured using the KINGPHOT
software package directly from 2D images
generated with a {\it Lanczos3} kernel --- are largely unbiased for
$r_h \gtrsim$ 0\farcs0125 $\approx$ 1~pc. However, the negative lobes of this 
kernel makes it difficult to repair bad pixels, so the {\it Gaussian}
kernel is preferred for the analysis
of the galaxy/nucleus surface brightness profiles.}
In both bandpasses, the median
half-light of the nuclei in our sample is found to be $0\farcs05$ (4~pc).
Clearly, the nuclei have a size distribution that is different from
that of the globular clusters. In the latter case, the distribution
is sharply peaked, with a typical (and nearly constant; Paper~X) half-light radius of 
$\langle{r_h}\rangle = 0\farcs033 \approx 2.7$~pc (i.e., $\sim$ 30\% larger than the
resolution limit for the nuclei). 
It is clear that
the nuclei are not only {\sl brighter} than typical globular clusters
(\S4.5) but they are also, on average, larger. There is, however,
considerable overlap between the two distributions, and the most 
compact nuclei have half-light radii that are indistinguishable from 
those of globular clusters. The UCDs, on the other hand, 
have half-light radii which resemble those of the nuclei. As 
with the luminosity functions, this agreement may be a consequence of the
selection process: i.e., UCD candidates were identified from the sample of 
sources with sizes in the range 0\farcs17--0\farcs5 (14--40~pc).

Figure~\ref{fig17} shows that there is a clear correlation between size
and luminosity for the nuclei, in the sense that the brighter objects have 
larger half-light radii.
We have fit relations of the form $r_h \propto {\cal L}^{\beta}$
to the data in Figure~\ref{fig17}, excluding both the offset nuclei and
the 5--6 nuclei which fall below
the nominal resolution limit of 0\farcs025 (shown by the dashed lines in the
two panels). The solid lines in the two panels show the relations:
\begin{equation}
\begin{array}{rrrrr}
r_{h,g} & \propto & {\cal L}_g^{0.505\pm0.042} \\
r_{h,z} & \propto & {\cal L}_z^{0.503\pm0.039} \\
\end{array}
\label{eq11}
\end{equation}
This luminosity dependence constitutes another clear point of distinction
between nuclei and globular clusters: the latter, both in the Milky Way
(van den Bergh et~al. 1991; Paper~VII) and in our program galaxies
(Paper~X), have a near-constant size of $\langle r_h\rangle = 2.7$~pc. 
This value is indicated by the arrows in Figure~\ref{fig17}. Nuclei
fainter than $g \sim 19$~mag and $z \sim 18$~mag have typical half-light
radii of 0\farcs04 (3.2~pc), or about 20\% larger than a
typical globular cluster; the brightest nuclei are an order
of magnitude larger still. Given their uncertain nature, it is 
worth noting that all five of the candidate offset nuclei from \S4.3 
have half-light radii close to the mean of the globular clusters.

It is interesting to see that the UCDs --- which in Figure~\ref{fig16} were
found to have half-light radii comparable to those of the largest nuclei ---
are outliers in this size-luminosity plane.
Compared to nuclei of comparable luminosity, the UCDs are nearly
three times larger, with $r_h \approx$ 0\farcs2--0\farcs5.
Alternatively, one might consider the UCDs to be
$\sim$~2~mag underluminous for their size. In any event, the
fact that UCD candidates from Paper VII were chosen to lie within a specific 
range of magnitude and half-light radius suggests that a general conclusion 
about systematic size differences between the two populations would be premature.

Figure~\ref{fig18} shows that there also exists a difference in surface
brightness between the globular clusters and nuclei. This figure plots
the average surface brightness within the half-light radius,
\begin{equation}
\begin{array}{rrrrr}
\langle \mu_h^{\prime}\rangle & = m^{\prime} + 0.7526 + 2.5\log{(\pi{r_h}^2}),\\
\end{array}
\label{eq12}
\end{equation}
for these two populations. 
Because of their near-constant size, the globular clusters fall along 
a diagonal swath in this diagram. Note that the dashed line in Figure~\ref{fig18}
is {\it not} a fit to the globular clusters, but simply the expected relation for 
clusters which obey Equation~11 and have a constant half-light radius
of $r_h \equiv 2.7$~pc. Although there is sizeable scatter,
The nuclei have a mean surface brightness of 
$\langle{\mu_h}\rangle = 16.5$ mag~arcsec$^{-2}$ in $g$ and 
$15.2$ mag~arcsec$^{-2}$ in $z$, although there is considerable scatter
($\sigma \approx 1.5$~mag) about these values.
The basic properties for the UCDs and nuclei in Virgo are compared in Table~\ref{tab:global}.

By virtue of their larger radii at fixed luminosity,
the Virgo UCDs have surface brightnesses that are lower than those of comparably
bright nuclei.  This is opposite to the claim of de~Propris et~al. (2005)
who argued that Fornax UCDs have {\it higher} surface brightness than the
Virgo nuclei. However, these authors seem to base this conclusion on a
visual comparison of the nuclei brightness profiles with that for
their ``mean UCD". We have calculated the average surface brightness
within the half-light radius for the four Fornax UCDs which have
absolute magnitudes and half-light radii reported in their Table~2.
In doing so, we have transformed their $V$-band magnitudes 
into the $g$ and $z$ bandpasses using assumed colors of $(g-V) = 0.48$ and
$(V-z) = 0.76$, which are appropriate for old, intermediate-metallicity
populations (see Table~3
of Paper~III). Their radii have been converted from parsecs to arcseconds
using their adopted Fornax distance modulus of $(m-M)_0$ = 31.39. The
resulting surface brightnesses for these four UCDs are shown as the
open squares in Figure~\ref{fig18}. We find that the Fornax and
Virgo UCDs occupy similar locations in the diagram, and that both populations
have {\it lower} surface brightness (by $\sim$
2.5~mag~arcsec$^{-2}$ in both bandpasses) than comparably bright nuclei,
contrary to the claims of de~Propris et~al. (2005).

Finally, we note that the five candidate offset nuclei are observed to fall along
the diagonal swath defined by the globular clusters in Figure~\ref{fig18}. This
strengthens the conclusion from \S4.3 that these objects are globular clusters,
rather than {\it bonafide} nuclei.

\subsection{Nuclei Colors}
\label{sec:col}

The first comprehensive investigation into the colors of nuclei in dwarf galaxies
was the series of papers by Caldwell (1983; 1987) and Caldwell \& Bothun
(1987). Based on imaging of 30 dwarfs in the Fornax Cluster, Caldwell \& Bothun 
(1987) found no evidence for a color difference between the nuclei and their
host galaxies. They did, however, find a correlation between
nuclei luminosity and galaxy color, in the sense that the reddest galaxies 
tended to harbor the brightest nuclei. At a given luminosity
the nucleated galaxies were also found to be slightly redder than 
their non-nucleated counterparts. A decade later, high-resolution CFHT imaging for
two Virgo dwarfs (Durrell 1997) hinted at an apparent diversity in nuclei
colors: in one galaxy (VCC1254), the nucleus was found to be significantly
bluer than the galaxy, while in the case (VCC1386), the colors 
were indistinguishable.

Recently, Lotz et~al. (2004) have carried out aperture photometry for the nuclei
of 45 dE,N galaxies in the Virgo and Fornax Clusters using $VI$ images from
three WFPC2 snapshot programs. They find that: (1) the nuclei are 
consistently bluer than the underlying galaxy light, with offset ${\Delta}$($V-I$) = 0.1--0.15~mag;
(2) the nuclei colors correlate with galaxy colors and luminosities, in
the sense that the redder nuclei are found in the redder and brighter galaxies;
(3) and the nuclei are slightly redder 
than the globular clusters associated with the host galaxy. 

Our examination of  the nuclei colors begins with Figure~\ref{fig19}. The left
panel of this figure shows
the color-magnitude diagram for the nuclei of the 51 Type~Ia galaxies
(filled circles) and the five galaxies with possible offset nuclei (open circles).
For the Type~Ia galaxies, the symbol size is proportional to the blue luminosity of the
host galaxy. For reference, 11 galaxies with nuclei redder than 
$(g-z)^{\prime}_{AB} = 1.35$ have been labeled.\footnote{These galaxies are
VCC1146, VCC1619, VCC1630, VCC1913, VCC784,
VCC1720, VCC828, VCC1627, VCC1250, VCC1242 and VCC1283.}
Note that one other  galaxy, VCC21, has a very blue nucleus with $(g-z)^{\prime}_{AB} \approx 0.30$.
Although it is listed in Table~\ref{tab:data} as a possible example of a galaxy with an offset
nucleus, we have argued in \S4.3, \S4.6 and \S4.7 that such offset ``nuclei"
are likely to be misclassified star clusters. In the case of VCC21, the blue color of the
object points to a young age (i.e., $\le 1$~Gyr for virtually any choice of metallicity;
see Figure~6 of Paper~I). This interpretation is consistent with the galaxy's dIrr/dE 
transitional morphology.

There are a number of noteworthy features in the color-magnitude diagram shown in 
Figure~\ref{fig19}.
First, both the colors and luminosities of the nuclei are seen to correlate with
host galaxy luminosity, in the sense that the nuclei belonging to the
most luminous galaxies are the brightest and reddest objects in our sample. This finding 
is consistent with the trend noted by Lotz et~al. (2004). Even more striking,
though, is the tendency for {\it the nuclei themselves to follow a clear
color-luminosity relation}. To the best of our knowledge, this is the first time
such a trend has been detected. The dashed line in Figure~\ref{fig19} shows the
relation
\begin{equation}
\begin{array}{rrlrr}
(g-z)^{\prime}_{AB} & = & -0.095(\pm0.015)g^{\prime}_{AB} & + &  2.98(\pm0.30), \\
\end{array}
\label{eq13}
\end{equation}
obtained from a least-squares fit to the 37 nuclei belonging to galaxies with $B_T \le 13.5$.
While this relation provides an excellent description of the color-magnitude 
relation for the nuclei in faint galaxies, it appears to break down
for brighter galaxies: in this regime, the nuclei not only show considerable scatter,
but they lie systematically to the faint/red side of the extrapolated relation.
These red nuclei cause the histogram of nuclei colors to have secondary peak at
$(g-z)^{\prime}_{AB} \approx$ 1.5 (see the right panel of Figure~\ref{fig19}).
They are found exclusively in high-surface-brightness environments,
which raises the possibility of a bias in the measured colors. However, 
the simulations described in Appendix~A --- in which artificial nuclei of known size, 
magnitude and color are added to the non-nucleated galaxy VCC1833 and
their properties measured in the same way as the actual nuclei --- show no
evidence for a significant color bias for such bright nuclei. In addition, the
colors for most of these nuclei are actually
redder than the galaxies themselves, by $\sim$~0.1~mag, so it seems unlikely
that contamination from the host galaxy can entirely explain their red colors.

For comparison, the open stars in Figure~\ref{fig19} show the sample of Virgo
UCDs from Paper~VII and Ha\c{s}egan \etal\ (2006).
The agreement with the nuclei is striking: i.e., with a
mean color of $\langle(g-z)^{\prime}_{AB}\rangle$ = 1.03$\pm$0.06~mag, the UCDs
have colors that are virtually identical to those of comparably bright nuclei.
This constitutes yet another piece of evidence for a link between UCDs and the nuclei
of early-type galaxies.

To better visualize how the properties of the host galaxy may affect
the relationship between nuclei color and magnitude, 
Figure~\ref{fig20} divides the sample by host galaxy magnitude
into four subgroups. These subsamples are shown in the four separate panels, with the
color-magnitude relation given by Equation (13) repeated in each case
(the dashed line).  Also included in each panel are the
globular clusters (small points) belonging to the
galaxies in each of these magnitude intervals; to reduce contamination from
stars and compact galaxies, we plot only those sources
with ``globular cluster probabilities" (see Paper~IX for details) in the
range $0.5 \le {\cal P}_{\rm gc} \le 1$. Note the clear bimodality in the
colors of globular clusters belonging to
these galaxies (Paper~IX). With the exception of the very red nuclei
noted above, we conclude that the nuclei in our Type~Ia
galaxies have colors which fall within the range spanned by the bulk of
the globular clusters in these same galaxies:
$0.7 \lesssim (g-z)^{\prime}_{AB} \lesssim 1.4$~mag.
Comparing the globular cluster colors to those of the UCDs 
from Figure~\ref{fig19}, we see that the UCDs
are $\approx$ 0.1--0.2 mag redder than the population of blue globular clusters,
but $\approx$ 0.2--0.3 mag bluer than the red clusters. This may be a point of distinction
with the UCDs in Fornax, which Mieske \etal\ (2006) find to have
colors similar to the red globular clusters.

Figure~\ref{fig21} shows how the colors of the galaxies, nuclei and globular
clusters depend on the galaxy luminosity. Results are shown for the $g$ and $z$
bands in the left and right panels, respectively. A common distance
of 16.5~Mpc has been assumed for all galaxies (Mei \etal\ 2005). Galaxy
colors are taken from Paper~VI and represent the average color in the range
1\arcsec~$\le r \le r_e$, subject to the ACS/WFC field view and
excluding those regions with surface brightnesses 1 mag~arcsec$^{-2}$
or more below the sky.
The majority of our galaxies show no evidence for strong color gradients,
so these colors should accurately reflect their integrated colors.
For the globular clusters, we plots colors for the red and blue subpopulations,
as determined in Paper~IX, along with that of the composite cluster system.
To highlight the subtle trends exhibited by these various samples,
we show mean colors for the nuclei, globular clusters
and galaxies in four broad bins of approximately
equal width in galaxy magnitude ($\sim$~2~mag). Results for the nuclei
are shown for three bins containing an equal number of objects.

This figure reveals a correlation between nucleus color and 
galaxy luminosity that is broadly consistent with the finding of Lotz \etal\
(2004) for fainter galaxies. However, the trend is relatively weak and is
in fact due mainly to the $\sim$ one dozen galaxies noted above that have very
red nuclei.
These galaxies make up most of the objects in the bins at $M_g \approx -18.6$,
$(g-z)_{AB}^{\prime} \approx 1.4$ and $M_z \approx -19.6$, $(g-z)_{AB}^{\prime} \approx 1.35$.
In the fainter galaxies, the nuclei colors show
essentially no correlation with galaxy luminosity.
For galaxies fainter than $M_g = -17$, the nuclei have
$\langle(g-z)_{AB}\rangle \approx 1.02$ ---
intermediate in color to the globular clusters and stars in galaxies
in this luminosity regime.

\section{Discussion}
\label{sec:dis}

In the preceding sections we have focussed on the observed properties
of the nuclei found in our program galaxies. We now turn to the broader question of
what these observations may be telling us about the origin and evolution of 
galactic nuclei. Before doing so, we pause to briefly review some of the
scenarios that have been proposed as possible explanations for stellar nuclei
in early-type galaxies. A more complete discussion of the theoretical implications
of our findings will be given in Merritt \etal\ (2006).

\subsection{A Review of Formation Models}
\label{sec:review}

Tremaine, Ostriker \& Spitzer (1975) were the first to suggest that
galactic nuclei may be the remains of merged globular clusters, which
were driven inward to the galactic center by dynamical friction 
(Chandrasekhar 1943).
According to this ``merger
model", the metallicity and luminosity of the nucleus should be a superposition
of the metallicity and luminosity of the progenitor clusters. 
Because dynamical friction causes the orbits of most massive globular
clusters to decay most rapidly, a nucleated galaxy would be expect to show a 
selective depletion of {\it bright} globular clusters, at least in the inner
regions where the dynamical friction timescale is short compared to a
Hubble Time. The contribution of globular clusters mergers
to the growth of central black holes and galactic nuclei has
been explored in a series of papers by Capuzzo-Dolcetta and coworkers
(e.g., Capuzzo-Dolcetta 1993; Capuzzo-Dolcetta \& Tesseri 1999).
Recently, Bekki \etal\ (2004) have used numerical simulations
to examine the physical properties (e.g., half-light radii, central 
velocity dispersion, mean density) of nuclei that form in such mergers.

Motivated by the discovery that the dE galaxies in Virgo are less centrally
concentrated than the dE,N galaxies (c.f. \S4.4), Oh \& Lin (2000)
revisited the question of how the tidal field from the Virgo cluster
affects the evolution of globular cluster orbits within dE galaxies.
They found that tidal perturbations acting on galaxies near the
center of the cluster tend to be compressive, and have little net
effect on the rate of decay of the globular cluster orbits.
In the outer regions of  the Virgo cluster, tidal forces tend
to disrupt galaxies, and the resulting decrease in density
leads to longer time scales for dynamical friction. Thus,
tidal forces favor the formation of nuclei in galaxies which are
located in the cluster core, and suppress the formation in more
distant galaxies. Clearly, this will cause the relative number of
nucleated and non-nucleated galaxies to vary within the cluster,
with the highest fraction of nucleated galaxies in the core.

A second, broad category of models focuses on a dissipational origin for the
nuclei. Noting that galaxies with nuclei are
typically rounder than those without, van den Bergh (1986) speculated
that nuclei form from the gas which collects more easily in the centers
of slowly-rotating galaxies. 
Silk, Wyse \& Shields (1987) argued that
dwarf galaxies experience late accretion of cool gas from the 
intergalactic medium, leading to star formation and the growth of
compact central nuclei. In a similar vein,
Davies \& Phillips (1988) proposed that early-type dwarfs result from 
the fading of stellar populations in dwarf irregular or blue compact
dwarf galaxies. In this scenario, intermittent bursts of central star
formation --- driven by the infalling gas --- continue until the
gas reservoir is depleted. According to this scenario, the final
star-forming event gives rise to the nucleation observed today.

Babul \& Rees (1992) examined the impact of the local intergalactic medium
on the evolution of a low-mass galaxy.  They argued that the pressure of the
intergalactic medium acts as a confining agent: in a high-pressure
environment, early-type dwarfs are able to retain more gas and produce
a nucleus from the gas that has been prevented from escaping by the
intergalactic medium.  Since the external pressure acting on galaxies
decreases with increasing distance from the cluster center, 
some properties of the nuclei (such as their luminosity and color)
should also depend on position within the cluster, with the highest
frequency of nucleation in the central regions of a cluster.

Gas inflow models have also been explored within the context of disk
galaxy mergers. Mihos \& Hernquist (1994) used N-body/hydrodynamical
simulations to show that such mergers are accompanied by gas dissipation
and central star formation which may result in the formation of
a dense stellar core, or the fueling of a pre-existing AGN. Following Weedman
(1983), Mihos \& Hernquist (1994) further note that the dense stellar 
core may itself collapse to form a supermassive black hole (SBH). The
observational evidence for a possible link between such SBHs and the
stellar nuclei of early-type galaxies is examined in \S\ref{sec:bh}.

\subsection{Implications for Nucleus Formation}
\label{sec:discussion}

\subsubsection{5.2.1 Connection to Nuclear Star Clusters in Late-Type Galaxies}
\label{sec:boker}

High-resolution {\it HST} imaging for Sa-Sd galaxies has shown that these
objects frequently contain compact nuclear clusters near their photocenters
(e.g. Phillips et~al. 1996; Carollo, Stiavelli \& Mack
1998; Matthews et~al 1999; B\"oker et~al. 2002; B\"oker et~al. 2004).
Figure~\ref{fig22} compares the sample of nuclear clusters from
B\"oker \etal\ (2004) to our sample of early-type nuclei. In the upper panel,
we plot the physical sizes for both samples, where we have assumed a common
distance of 16.5~Mpc for the Virgo galaxies (Tonry \etal\ 2001; Paper V).
It is clear that the nuclear clusters of B\"oker \etal\ 
(2004) have sizes similar to the early-type nuclei.

The lower panel of Figure~\ref{fig22} compares the absolute magnitudes
of the two samples. Note that the observations of B\"oker \etal\ (2004)
were carried out in the F814W ($I$) bandpass. Comparing the means of the
samples, we find the two populations to be comparably bright, with 
$\langle M_g\rangle = -10.9$ and $\langle M_z\rangle = -12.0$
for the early-type nuclei, and $\langle M_I\rangle = -11.7$ for the
nuclear clusters.
B\"oker \etal\ (2002) further report that
59 of 77 late-type spirals in their survey contain a nuclear
cluster close to the galaxy photocenter, giving an overall frequency
of nucleation of $f_n \approx77\%$. For comparison, in \S4.2 we estimated
$66 \lesssim f_n \lesssim 82\%$ for our sample of 
early-type galaxies, counting galaxies with possible offset nuclei
as non-nucleated. Thus, in all
these respects, the nuclear clusters found in late-type galaxies are nearly
identical to the nuclei studied here. The lone point of 
distinction between the nuclear clusters and the early-type nuclei seems
to be one of age: the majority of the nuclear clusters appear to have
$\tau \lesssim 10^8$ yr (Walcher \etal\ 2005 and references therein),
while the broadband colors rule out such young ages for {\it all} of the 
Type~Ia nuclei, irrespective of metallicity (see \S\ref{sec:stellpop} and
Figure~6 of Paper~I). This difference notwithstanding, the similar properties
of the nuclei and nuclear clusters --- and their appearance in galaxies of such
disparate morphology --- clearly points to a rather generic formation
mechanism: e.g., one which is largely independent of local or global environmental
factors, such as the gas content and present-day morphology of the
host galaxy, or the density of neighboring galaxies.

\subsubsection{5.2.2 A Fundamental Division Between S\'{e}rsic and core-S\'{e}rsic Galaxies}
\label{sec:fund}

The above conclusion applies equally well to the luminosity of the host
galaxy: i.e., the nuclei are not confined to just the dwarfs, but are also
found with regularity in many of the giants. In fact,
half (21/42) of the galaxies brighter than $B_T = 13.6$ or $M_B \approx -17.6$ 
(the approximate division between dwarfs and giants in the VCC) have
classifications of Type~Ia or Ib.\footnote{Excluding
the five Type 0 galaxies in this luminosity range.} The fact that nuclei are
common above and below the canonical dwarf-giant boundary suggests that, at least in terms
of their {\it nuclear} properties, there is no evidence for a fundamental
change in galaxies at this magnitude. This is consistent with
the mounting evidence from photometric scaling relations that the
``dichotomy" between normal and dwarf ellipticals, as originally envisioned
by Kormendy (1985) and others, may not be real (e.g., Jerjen \& 
Binggeli 1997; Graham \& Guzm\'an 2003; Paper~VI).

On the other hand, there {\it does} appear to be a fundamental transition at
$M_B \approx -20.5$ in terms of nuclear properties. Brighter than
this, we measure $f_n \sim 0$ and, in almost all cases, the presence
of a nucleus can be ruled out with some confidence (see Appendix A). 
Fainter than $M_B \approx -20.5$, the fraction of nucleated galaxies
rises sharply, as shown in the lower panel of Figure~\ref{fig06}. 
It has been argued (Graham \&
Guzm\'an 2003; Trujillo \etal\ 2004; Graham 2004; Paper~VI)
that this
magnitude marks a transition from faint, S\'ersic-law
galaxies to bright, core-S\'ersic-law galaxies, whose flat
``cores" are presumed to result from core depletion 
by coalescing of supermassive black holes (Ebisuzaki \etal\ 1991; Quinlan \&
Hernquist 1997; Faber \etal\ 1997; Milosavljevi\'c \& Merritt 2001). The absence
of nuclei in galaxies brighter than $M_B \approx -20.5$ is 
consistent with this scenario. Of course, it is equally possible that
these ``missing" nuclei are absent in the bright galaxies not because 
of the disruptive effects of mergers and black hole coalescence,
but because they failed to form in such galaxies in the first place.
Discriminating between these competing scenarios should prove to be
a fruitful area for future theoretical study.

\subsubsection{5.2.3 Coincidence of Nuclei with Galaxy Photocenters}

In almost all cases, the nuclei are found to be coincident with 
the photocenters of their host galaxies. In only five cases does there appear
to be a statistically significant offset of ${\delta}r_n / \langle r_e\rangle \ge 0.04$.
The bulk of the evidence, however, favors the view that these ``nuclei"
are, in actuality, star clusters projected close to the galaxy photocenters. 
That is to say, the sizes, surface brightnesses and colors of the five possible
offset nuclei more closely resemble those
of globular clusters than those of the other nuclei in our sample. Interestingly,
all five of the host galaxies show some characteristics of dIrr/dE transition objects,
including blue colors, low central surface brightnesses, the presence of dust,
and a mottled or irregular appearance.
This suggests that if dwarf ellipticals represent an evolutionary
stage that follows gas exhaustion and stellar fading (Davies \& Phillips 1988),
ram pressure stripping (Mori \& Burkert 2000) or harrassment (Moore, Lake \& 
Katz 1998) of gas-rich dIrr/disk galaxies, then the formation of a
central nucleus is not an immediate or inevitable consequence. Additional
time seems to be required to ``grow" a central nucleus.

\subsubsection{5.2.4 Nucleus Formation through Globular Clusters Mergers?}

Because the theoretical framework for the globular cluster merger model is at
a more mature stage than for any other model, we now turn our attention to the
question of whether this scenario is consistent with our new observations for
the nuclei. We note that Lotz \etal\ (2001) have previously examined the
viability of the merger hypothesis on the basis of data collected for
nuclei and globular clusters in their WFPC2 snaphot survey of dwarf galaxies. Apart from
identifying a possible depletion of bright clusters in the innermost regions 
of the galaxies, they could find no strong evidence for a merger origin
of the nuclei, either from the spatial distribution of the clusters or 
from the measured luminosities of the nuclei.

As pointed out in \S4.6, a comparison of the luminosity functions of
nuclei and globular clusters in these galaxies shows that the {\it typical} nucleus
is $\approx$ 3.5 magnitudes brighter than a typical globular cluster. If 
cluster mergers are responsible for the formation of a central nucleus, then
one might expect an average of $\sim$ 25 mergers would be needed to assemble
a nucleus from typical clusters. Of course, as Figure~\ref{fig20} shows, a
single number does not tell the whole story. The four panels
of this figure plot the distribution of nuclei and globular clusters in
the color-magnitude diagram. For the brightest galaxies (shown in the first panel),
the nuclei have a median luminosity $\approx$ 125$\times$ that of globular 
clusters at the peak of the cluster luminosity function. For the fainter galaxies
(shown in the three remaining panels), the nuclei are brighter than a typical
globular clusters by factors of 29, 15 and 17, respectively. 

Are these numbers feasible? In Figure~\ref{fig23} we attempt to answer
this question by plotting the integrated luminosity in globular clusters
against the luminosity of the nucleus for Type~Ia galaxies in our survey. Results are
shown in the upper panels, with measurements made in the $g$ and $z$ bands
given in the left and right panels, respectively. As in Figures~\ref{fig19} and
\ref{fig20}, symbol type indicates the magnitude of the host galaxy. In calculating the
total luminosity in globular clusters for these galaxies, we have simply
summed the luminosities of globular cluster candidates with probabilities in the
range ${\cal P_{\rm gc}} \ge 0.5$. Although this approach will obviously miss
any globular clusters located outside the ACS field, the correction should be
small for the Type~Ia galaxies in our survey which, with $M_B \lesssim -19$, have
$\langle{R_e}\rangle \sim$ 15\arcsec~or less (Paper~VI). The correlations
apparent in these panels are a consequence of the fact that both the total number
of globular clusters, and the luminosity of the nucleus, scale with host
galaxy luminosity.

The lower panels of Figure~\ref{fig23} plot the ratio of globular-to-nucleus 
luminosities, $\kappa$, in the two bandpasses. In both cases, the ratio is 
near unity. This should perhaps come as no surprise since the mean nucleus-to-galaxy
luminosity ratio, $\eta = 0.30\pm0.04$\%, found in \S4.5 is nearly 
identical to the globular
cluster formation efficiency of $\epsilon = 0.26\pm0.05$\% measured by
McLaughlin (1999) for early-type galaxies. This latter measurement is in
turn based on observations of 97 early-type galaxies
and represents the total mass in globular clusters normalized
to the total baryonic (stellar + gas) mass.
While the agreement may be purely coincidental, it is 
a remarkable empirical result that the formation of early-type galaxies
results in a nearly constant fraction of the initial
baryonic mass, $\sim$ 0.3\%, being deposited into both globular clusters and, in many
cases, a central nucleus. Of course, this conclusion appears {\it not} to
apply to galaxies brighter than $M_B \approx -20.5$, which lack nuclei
either because they did not form in the first place or because they were
subsequently destroyed by some as-yet-unidentified process.

In any case, galaxies which lie below the dashed line at $\kappa = 1$ in
Figure~\ref{fig23} pose a clear challenge to the merger model for the
obvious reason that they simply have too few clusters
to explain the luminosity of the nucleus. The difficulty is most severe for the
dozen or so red nuclei associated with the brightest Type~Ia galaxies. Of course,
this argument is based on the number of clusters contained by the host galaxy
{\it at the present time}. If the observed clusters are the rare ``survivors"
descended from a much larger initial cluster population, then it may be possible
to circumvent this problem. 

An additional test of the merger model is possible. If the mergers were
dissipationless so that star formation and chemical enrichment can be
ignored, then we can use the observed colors of globular
clusters to predict colors for the nuclei. Since both globular cluster
color and the fraction of red
globular clusters are increasing functions of host galaxy
luminosity (Paper~IX), we expect the nuclei in this model to
become progressively redder in brighter galaxies. The heavy 
solid curve shown in the four panels of Figure~\ref{fig20} shows the
predicted color magnitude relation for nuclei which grow through
globular cluster mergers. This curve is based on Monte Carlo
experiments in which the color evolution of the nuclei is followed
using the observed colors of the globular clusters in these galaxies.
The thin curves show the 95\% confidence limits on the relation.
Although these simulations do indeed predict redder colors for
the brighter nuclei (which are found preferentially in the brighter
galaxies containing a larger proportion of red clusters),
the predicted scaling is much milder than what is observed.

Bekki \etal\ (2004) have used numerical simulations to investigate the physical
properties of nuclei which form through repeated mergers of globular clusters.
Their predicted scaling relation between half-light radius and luminosity,
$r_h \propto {\cal L}^{0.38}$, is shown by the dotted line in Figure~\ref{fig17}.
The relation has a somewhat milder luminosity dependence than the 
observed relation, $r_h \propto {\cal L}^{0.50\pm0.03}$, but is nevertheless
in reasonable agreement. A similar conclusion applies to the luminosity 
dependence of surface brightness averaged within the half-light radius. We
find no strong correlation between $\langle\mu_h\rangle$ and $\cal L$, but
the predicted relation, $r_h \propto {\cal L}^{0.23}$ (shown as the dotted line in
Figure~\ref{fig18}), is reasonably consistent with the observations, having 
only a weak luminosity dependence. 
In the future, it will be of interest to compare the predicted and observed 
relationship between luminosity and central velocity dispersion. Spectra for most
of our program galaxies are now in hand and such a comparison will be the subject of a
future paper in this series. 

\subsubsection{5.2.5 Stellar Populations in the Nuclei: Clues from Colors}
\label{sec:stellpop}

Ground-based spectroscopy will also be useful for investigating
the history of star formation and chemical enrichment in the nuclei,
although care must be exercised when decoupling the contributions
from the galaxy and nucleus. This separation is more straightforward
in the {\it ACS} imaging, although in this case we are limited in
our ability to measure ages and metallicities because of the well known
age-metallicity degeneracy of broadband colors. The upper
panel of Figure~\ref{fig24} shows linear interpolations of
the [Fe/H]-$(g-z)_{AB}^{\prime}$ relation for simple
stellar populations from the models of Bruzual \& Charlot (2003). The four
relations show color-metallicity relation for ages of $\tau$ = 1, 2, 5 and
10 Gyr, although it is, needless to say, quite unlikely that a single age is
appropriate for all of the nuclei in our sample. For comparison, 
the heavy dashed curve in black shows the 
color-metallicity relation derived from globular clusters in the Milky Way,
M49 and M87 (Paper~IX). If it is assumed that the nuclei have ages 
similar to the globular clusters, then this empirical relation may be
used to derive metallicities for the nuclei.

Converting from colors to metallicities with these relations, we find 
the five metallicity distributions shown in the lower panel of
Figure~\ref{fig24}. The results are summarized in Table~\ref{tab:met}.
Not suprisingly, the metallicity distributions derived from the models
depend sensitively on
the assumed age. For $\tau$ = 10 Gyr, the colors of the bluest nuclei, with
$(g-z)_{AB} \sim 0.8$~mag, would require very low 
metallicities: i.e., [Fe/H] $\sim -2$ or less. By the same token, 
the reddest nuclei in our sample would require
metallicities 1-100$\times$ solar for an assumed age of 1 Gyr. 
For ages as young as $\tau \lesssim 10^8$ yr, which is appropriate 
for many of the nuclear clusters in late-type galaxies (see \S\ref{sec:boker})
no reasonable choice of metallicity can explain the colors
of $\approx$ 0.8--1.5 that are observed.  Thus, to the
extent that the nuclei can be characterized by a single formation epoch, they
show evidence for an old to intermediate age: i.e., $\tau > 1$~Gyr. Using 
the globular cluster color metallicity relation gives a mean metallicity of
$\langle{\rm [Fe/H]}\rangle$ = $-0.88\pm0.79$~dex. Firmer conclusions
on the ages and metallicities of the nuclei must await the spectroscopic
analysis.

Spectroscopic constraints on the mix of stellar populations in the nuclei should also
shed some light on what may be the most serious challenge facing the merger model:
the existence of a tight correlation between nucleus luminosity
and color (Figures~\ref{fig19}-\ref{fig20}). Such a correlation
is generally thought to be a signature of self enrichment in stellar systems, and
is reminiscent of the color-magnitude relation for dwarf and giant galaxies
(e.g., Bower, Lucey \& Ellis 1992; Caldwell \etal\ 1992). That the
colors correlate tightly with the luminosities of the nuclei, and less so
with those of the host galaxies, suggests that the chemical enrichment
process was governed primarily by local/internal factors. The existence of a
tight color-magnitude relation for the nuclei is a difficulty for the
merger model as envisioned by Tremaine \etal\ (1975) since the
clusters from which the nuclei are assembled show
no color-magnitude relation themselves, and our Monte-Carlo experiments reveal
the slope of the observed color-magnitude relation is steeper than
that predicted in dissipationless cluster mergers.

We speculate that the merger model in its original form (i.e., ``dry"
mergers of stellar clusters) is an oversimplication of a process that almost
certainly involves some gas dissipation. In fact, if nuclei do indeed 
have stellar ages of a few Gyr old or more, then they were
assembled during the earliest, most gas-rich stage of galaxy evolution. It would be 
interesting to revisit the merger model with the benefit of numerical simulations that
include the effects of not just dark matter and stars, but also gas, to see
if star formation and chemical enrichment caused by mergers/inflows are
capable of producing a color-magnitude relation consistent with 
that shown in Figure~\ref{fig19}. In a number of respects, the dozen or
so bright nuclei labelled in Figure~\ref{fig19} appear to form a 
population distinct from their faint counterparts, most notably in
their integrated colors (which appear redder than the galaxies themselves).
These nuclei may be candidates for the ``dense stellar cores"
which form in numerical simulations (Mihos \& Hernquist 1994)
when (chemically-enriched) gas is driven inward, perhaps as a result of mergers.

\subsubsection{5.2.6 Ultra-Compact Dwarfs, Nuclei and Galaxy Threshing}

Our ACS observations may also provide some clues to the origin of
UCD galaxies. In terms of color, luminosity and size, the UCDs from Paper~VII
bear a strong resemblance to many of the nuclei studied
here, leading credence to the galaxy threshing scenario
(Bassino \etal\ 1994; Bekki \etal\ 2001).
However, these UCDs were selected for study on the basis of luminosity
and half-light radius, so it is unclear to what extent these conclusions
may apply to the population of UCDs as a whole. An unbiased survey of the
UCD population in Virgo should be undertaken to clarify this issue,
although this will be a difficult and time-consuming task as it requires
high-resolution spectroscopy and HST imaging to distinguish true UCDs from bright
globular clusters (see \S7 of Paper~VII). Jones \etal\ (2006) have recently
reported the first results from a program to search for UCDs
in Virgo using radial velocities for $\sim$ 1300 faint,
star-like sources in the direction of the cluster. However, lacking structural
and internal dynamical information for UCD candidates found in this way,
it is impossible to know to what extent their sample has been ``polluted" by 
globular clusters: either those associated
with galaxies or intergalactic in nature (e.g., West \etal\ 1995).

For the time being, we may use the existing sample of Virgo UCDs from Paper~VII
and Ha\c{s}egan \etal\ (2006) to re-examine
the threshing model in light of our findings for the Virgo nuclei.
Specifically, we may estimate the luminosities of the
UCD progenitor galaxies within the context of the threshing  model:
for the
typical ratio of nucleus-to-galaxy luminosity found in \S4.5,
$\langle\eta\rangle \approx 0.3$\%, we expect the progenitors
to have $-18.2 \lesssim M_B \lesssim -16.6$, with a
value mean of $\langle M_B\rangle = -17.3\pm0.6$.
If the threshing
scenario is correct, then we might expect the surviving analogs of the UCD
progenitors to resemble galaxies \#40--69 in Table~\ref{tab:data}. It is
interesting to note that only about half (16/30) of these galaxies were
classified as dwarfs by BST85, meaning that a significant fraction of 
{\it bright} galaxies may need to have been ``threshed" to explain the UCD
luminosity function within the context of this model. Photometric, dynamical
and structural parameters
for these candidate UCD progenitor galaxies may serve as useful constraints
for self-consistent numerical simulations of galaxy threshing and UCD formation.

\subsubsection{5.2.7 Connection to Supermassive Black Holes}
\label{sec:bh}

A large body of literature now exists on the SBHs that reside in the 
centers of many galaxies (see, e.g., the review of Ferrarese \& Ford 2005). 
While it had been known for some time that SBH masses, $\cal M_{\bullet}$, 
correlate with the bulge masses, ${\cal M}_{gal}$, of their host galaxies
(Kormendy \& Richstone 1995), it was only after the discovery of a tight relation 
between $\cal M_{\bullet}$ and bulge velocity dispersion (Ferrarese \&
Merritt 2001; Gebhardt \etal\ 2001) that 
Merritt \& Ferrarese (2001) were able to show that the frequency function for
galaxies with SBHs
has a roughly normal distribution in $\log_{10} {( {\cal M}_{\bullet} / {\cal M}_{gal} )}$. Fitting to the data available at that time,
Merritt \& Ferrarese (2001) found a mean of value of $-2.90$ (0.13\%)
and a standard deviation of 0.45~dex.

Remarkably, this mean value is, to within
a factor of $\approx$ two, identical to the
mean fractional luminosity contributed by nuclei to their host galaxies (\S\ref{sec:lum}).
In fact, the nuclei and SBHs share a number of other key similarities that are
highly suggestive of a direct connection: e.g., they
share a common location at the bottom of the gravitational potential wells
defined by their parent galaxies and dark matter halos, and both are probably
old components that formed during the earliest stages of galaxy evolution
(\S\ref{sec:stellpop}; Haehnelt \etal\ 1998; Silk \& Rees 1998; Wyithe \&
Loeb 2002). Could it be that the compact nuclei which are found
so frequently in the low- and intermediate-luminosity early-type galaxies are
related in some way to
SBHs dectected in the brighter galaxies?

Figure~\ref{fig25} examines the connection between nuclei and SBHs in more
detail. In the upper panel, we show the distribution of absolute blue
magnitudes, $M_B$, for the 51 galaxies in our survey that contain Type~Ia nuclei
(solid histogram).
This distribution should be compared to
that for the early-type galaxies having SBH detections (open squares and dotted
histogram). This latter sample is based on data from Table~II of Ferrarese \& 
Ford (2005), which reports SBH mass measurements from
resolved dynamical studies for 30 galaxies. Among this sample, there are 23
early-type galaxies with measured SBH masses (all based on stellar and/or gas dynamical
methods).  It is clear from Figure~\ref{fig25} 
that the two samples have very different distributions. With the exception
of M32 (with $M_B = -15.76$~mag and ${\cal M}_{\bullet} = 2.5\times10^6$ solar masses),
the SBH galaxies are all brighter than
$M_B \approx -18$, a cutoff that is thought to reflect the
formidable technical challenges involved in detecting smaller SBHs in fainter early-type
galaxies.

By contrast, the Type Ia galaxies have $M_B \gtrsim -18.9$~mag. Note that this does
{\it not} reflect the true upper boundary for nucleated galaxies,
since nuclei definitely exist in galaxies brighter than this --- Table~\ref{tab:data}
lists 14 galaxies with certain or suspected nuclei (i.e., Types Ib, Ic or Id)
having $M_B \lesssim -18.9$~mag ---
but the high surface brightness of the host galaxies do not allow a reliable
measurement of the nuclei luminosities or sizes. As we have argued in
\S\ref{sec:fund}, the more fundamental cutoff seems to occur at $M_B \sim -20.5$~mag
since we find no nucleated galaxies brighter than this.

Before moving on, we note that four of the galaxies with SBH masses in Table~II
of Ferrarese \& Ford (2005) appear in our survey. In two cases --- VCC1978 (NGC4649)
and VCC1231 (NGC4473) --- there is no evidence for a nucleus so we classify the
galaxies as Type II. In a third case,
VCC763 (NGC4374), the center of the galaxy is partly obscured by an AGN (Type~O)
but we see no evidence of a resolved stellar nucleus (\S4). The fourth and final
galaxy, VCC1664 (NGC4564), has a reported SBH mass of 5.6$\times10^7{\cal M}_{\odot}$ 
(Gebhardt \etal\ 2003). We classify this object as Type~Ic, meaning that
we see evidence for a resolved nucleus but are unable to measure its properties
owing to the high surface of the galaxy.

In the lower panel of Figure~\ref{fig25}, we compare the frequency functions
of SBHs and Type~Ia nuclei. Bulge masses for the SBH galaxies were found by
assuming a constant bulge color of $(B-V)=0.9$~mag and combining the 
magnitudes from Ferrarese \& Ford (2005) with the mass-to-light ratio
relation $\Upsilon_V = 6.3(L_V/10^{11})^{0.3}$ from Paper VII.
For the SBH sample, we find
\begin{equation}
\begin{array}{rrlll}
\langle \log_{10}({\cal M}_{\bullet} / {\cal M}_{gal})\rangle & = &
-2.61\pm0.07~{\rm dex} & \\
& = & \phantom{-}0.25\pm0.04~\% \\
\sigma(\log_{10}{\cal M}_{\bullet} / {\cal M}_{gal}) & = &  \phantom{-}0.45\pm0.09~{\rm dex} \\
\end{array}
\label{eq14}
\end{equation}
whereas for the nuclei, we find
\begin{equation}
\begin{array}{rrrrr}
\langle \log_{10}{\eta}\rangle & = & -2.49\pm0.09~{\rm dex} & {\rm (= 0.32\pm0.07\%)} \\
\sigma(\log_{10}{\eta}) & = & 0.59\pm0.10~{\rm dex} \\
\end{array}
\label{eq15}
\end{equation}
For comparison, Gaussian distributions with these parameters are shown in the lower panel
of Figure~\ref{fig25}.

In our view, the available evidence favors the view that the compact stellar nuclei
found in many of the fainter galaxies may indeed be the counterparts of SBHs in the
brighter galaxies.  If this speculation is correct, then it might be more
appropriate to think in terms of {\it Central Massive 
Objects} --- either SBHs or compact stellar nuclei --- that accompany the formation
of almost all early-type galaxies and contain a mean fraction $\approx$ 0.3\%
of the total bulge mass. We note that a similar conclusion has been reached
independently by Wehner \& Harris (2006).
Models for the formation of SBHs in massive galaxies would
then be confronted with the additional challenge of explaining the different
manifestations of Central Massive Object formation along the galaxy luminosity
function, with an apparent preference for SBH formation above 
$-20.5 \lesssim M_B \sim -18$~mag. These issues are explored in
more detail in Ferrarese \etal\ (2006b).

\section{Summary}

Using multi-color ACS imaging from the {\it Hubble Space Telescope}, we have 
carefully examined the central structure of the 100 early-type galaxies which
make up the ACS Virgo Cluster Survey. In this paper, we have focussed
on the compact nuclei which are commonly found at, or close to, the photocenters
of many of the galaxies. Our main conclusions are as follows:

\begin{itemize}

\item[1.] Nuclei in early-type galaxies are more common than previously
believed. Excluding the six galaxies for which the presence of a nucleus
could not be established, either because of dust obscuration or the
presence of an AGN, and counting the five galaxies with possible offset
nuclei as non-nucleated, we find the 
frequency of nucleation to fall in the range $66 \% \lesssim f_n \lesssim 82$\% for
early-type galaxies brighter than $M_B \approx -15$.

\item[2.] Nuclei are found in galaxies both above and below the canonical dividing
line for dwarfs and giants ($M_B \approx -17.6$). Half (21/42) of the galaxies brighter
than $M_B \approx -17.6$ are found to contain nuclei. This is almost
certainly a lower limit on the true frequency of nucleation because
of incompleteness and surface brightness selection effects in these bright galaxies.

\item[3.] On the other hand, galaxies brighter than $M_B \approx -20.5$ appear to
be distinct in that they do {\it not} contain nuclei --- at least, not those of the
kind expected from upward extrapolations of the scaling relations obeyed by nuclei
in fainter galaxies. Whether this means that nuclei 
never formed in these ``core-S\'{e}rsic" galaxies (see Paper VI and references therein), or 
were instead subsequently destroyed by violent mergers and core evacuation due to
coalescing supermassive black holes, is unclear. The absence of nuclei in galaxies
brighter than $M_B \approx -20.5$ adds to the mounting evidence for a fundamental
transition in the structural properties of early-type galaxies at this magnitude.

\item[4.] Few, if any, of the nuclei are found to be significantly offset from the
photocenters of their host galaxies. In only five galaxies do we measure offsets
$\delta{r_n} \gtrsim 0\farcs5$ or $\delta{r_n}/\langle{r_e}\rangle \gtrsim 0.04$.
In all fives cases, however, the available evidence (i.e., magnitudes, colors, 
half-light radii and surface brightness measurements) suggests that such ``nuclei"
are probably star clusters projected close to the galaxy photocenters.

\item[5.] The central nuclei are {\it resolved}, with a few notable exceptions: such as
the two AGN galaxies, M87 and M84, which have prominent non-thermal nuclei, and
a half dozen of the faintest galaxies with very compact, but presumably stellar, nuclei.
This observation rules out low-level AGN as an explanation
for the central luminosity excess in the vast majority of the galaxies. Excluding 
those galaxies with faint, unresolved nuclei, we find the half-light
radii of the nuclei to vary with luminosity according to the relation
$r_h \propto {\cal L}^{0.50\pm0.03}$.

\item[6.] A Gaussian distribution provides an adequate, though by no means unique, description
of the luminosity function for the nuclei. The peak of the luminosity function occurs at
$\langle{M_g}\rangle = -11.2\pm0.6$ and $\langle{M_z}\rangle = -12.2\pm0.6$, or
roughly 25$\times$ brighter than the peak of the globular cluster luminosity functions
in these same galaxies. We find the ratio of nucleus-to-galaxy luminosities to
be $\eta \approx $ 0.3\%, independent of galaxy luminosity but with significant
scatter.

\item[7.] A comparison of the nuclei to the nuclear star clusters in
late-type galaxies shows a remarkable similarity in luminosity, size
and overall frequency. This suggests that a quite generic formation mechanism
is required to explain the nuclei: one which is largely independent of local
or global environmental factors, such as the gas content and present-day morphology
of the host galaxy, or the density of neighboring galaxies.

\item[8.] In terms of color, luminosity and size, the UCDs from Paper~VII bear
a strong resemblance to the compact nuclei in a number of these galaxies,
leading credence to the ``threshing" scenario in which UCDs are assumed to be the
tidally stripped remains of nucleated galaxies. If this model is correct, then
we argue that the UCD progenitor galaxies would --- if they avoided ``threshing" --- 
now resemble galaxies with magnitudes in the range $-18.2 \lesssim M_B \lesssim -16.6$.
Simulations to test the viability of the threshing mechanism for such galaxies are advisable.

\item[9.] The colors of the nuclei are tighly correlated with their total luminosities,
but only weakly with those of their host galaxies. This may suggest that the history
of chemical enrichment in the nuclei was governed by local or internal factors.

\item[10.] The mean of the frequency function for the nucleus-to-galaxy luminosity ratio
in our nucleated galaxies, $\langle\log_{10}\eta\rangle = -2.49\pm0.09$~dex ($\sigma = 0.59\pm0.10$),
is indistinguishable from that of the SBH-to-bulge mass ratio,
$\langle \log_{10} {( {\cal M}_{\bullet} / {\cal M}_{gal} )} \rangle = -2.61\pm0.07$~dex
($\sigma = 0.45\pm0.09$),
calculated in 23 early-type galaxies with detected SBHs.

\item[11.] We argue that the compact stellar nuclei
found in many of our program galaxies may be the low-mass counterparts of
SBHs detected in the bright galaxies, a conclusion reached
independently by Wehner \& Harris (2006).
If this view is correct, then one should
think in terms of {\it Central Massive Objects (CMOs)} --- either SBHs or compact
stellar nuclei --- that accompany the formation of almost all early-type galaxies and
contain a mean fraction $\approx$ 0.3\% of the total bulge mass. In
this view, SBHs would be the dominant mode of CMO formation above $M_B \approx -20.5$.

\end{itemize}

As the nearest large collection of early-type galaxies, the Virgo cluster is 
likely to remain, for the forseeable future, at the center of efforts to understand the physical
processes that drive nucleus formation. Unfortunately, exploring the stellar
dynamics of the most compact nuclei --- and modeling the mass distribution within
the central few parsecs of the host galaxies --- requires integrated-light spectra with an angular 
resolution of $\sim$~0\farcs1 or better. Thus, the Virgo nuclei are obvious
targets for diffraction-limited, near-IR spectrographs on 8m-class
ground-based telescopes, particularly since the demise of the {\it Space
Telescope Imaging Spectrograph} on {\it HST}. For the time being, though, ACS imaging of the nuclei should
serve to guide models of their formation and evolution. This will be the 
subject of a future paper in this series, in which we will examine the
implications of these observations for theories of nucleus formation
(Merritt \etal\ 2006).

\acknowledgments

We thank Peter Stetson for assistance with the construction of the PSFs used
in this study. Support for program GO-9401 was provided through a grant from
the Space Telescope Science Institute, which is operated by the Association of
Universities for Research in Astronomy, Inc., under NASA contract
NAS5-26555.
P.C. acknowledges additional support provided by NASA LTSA grant NAG5-11714.
M.M. acknowledges additional financial support provided by the Sherman
M. Fairchild foundation. D.M. is supported by NSF grant AST-020631,
NASA grant NAG5-9046, and grant HST-AR-09519.01-A from STScI.
M.J.W. acknowledges support through NSF grant AST-0205960.
This research has made use of the NASA/IPAC Extragalactic Database (NED)
which is operated by the Jet Propulsion Laboratory, California Institute
of Technology, under contract with NASA.

\begin{appendix}

\section{Tests for Completeness, Resolvability and Bias}

The approach used to classify galaxies according to the presence or absence of
a central nucleus has been described in \S4. Briefly, the classification procedure
relies on both a visual inspection of the ACS images and the detection of
a central ``excess" in the brightness profile relative to the fitted S\'{e}rsic
or core-S\'{e}rsic galaxy model. The results are summarized in Table~\ref{tab:class}.
We find a total of 62 galaxies in which the presence of a nucleus could be
established with a high level of confidence (i.e., the Type~Ia and Ib galaxies).
Five more galaxies (Type~Ie) {\it may} contain an offset nucleus but,
as we have argued above, the weight of evidence favors the view that these
``nuclei" are actually globular clusters. Six other galaxies (Type~0) contain
either an AGN or dust at the photocenters, making the identification of a nucleus
difficult or impossible.

This leaves us with a sample of 100 -- 62 -- 5 -- 6 = 27 galaxies which
may be classified provisionally as non-nucleated. Of course,
the faintest, most extended nuclei will go undetected in any survey, especially
when superimposed on a bright galaxy background. It therefore seems
likely that at least some of these galaxies may, in fact, be nucleated. In this
Appendix, we attempt to elucidate the nature of these
galaxies with the aid of numerical simulations guided by our findings from \S4.  

For the 27 galaxies in question, Figure~\ref{fig26} plots residuals, over the
innermost 10\arcsec, between the {\it observed} brightness profile and the fitted
models shown in Figure~\ref{fig04}. Because a few of 
these galaxies contain multiple components (e.g., rings, bars or shells), or have
outer brightness profiles that are contaminated by the light of nearby giant galaxies,
the profiles were sometimes fit over a restricted range in radius. In two cases
where this outer fitting radius is $\le 10$\arcsec~(VCC778 = NGC4377 and
VCC575 = NGC4318), an upward arrow shows the adopted limit. Likewise, six galaxies
(VCC1664 = NGC4564, VCC944 = NGC4417, VCC1279 = NGC4478, VCC355 = NGC4262, 
VCC1025 = NGC4434 and VCC575) in which the presence of
a faint central nucleus was suspected on the basis of an upturn in the central
brightness profile, an upward arrow at 0\farcs2--0\farcs3 shows the
inner limit used to avoid biasing the fitted galaxy parameters. Note that in
most cases, the fitted S\'{e}rsic or core-S\'{e}rsic model provides a reasonably
accurate match to the central brightness profile, meaning that any nuclei which
may be hiding in these galaxies have had only a minor impact on the observed profiles.

Of course, two possibilities exist for any given galaxy: either it contains a
nucleus or it does not. To test the first possibility, we use the scaling relation
from \S4 which links the luminosity of the galaxy to that of its nucleus 
(Equation 6). Meanwhile, Equation 11 provides a link between galaxy luminosity
and nucleus size (half-light radius). For each of the 27 galaxies 
in Figure~\ref{fig26}, we then subtract a nucleus of the expected
size and luminosity based on the magnitude of the galaxy itself.
If the nucleus-subtracted profile of the galaxy is better represented by a 
Sersic or core-S\'{e}rsic model than the original profile, this would be (circumstantial)
evidence for the presence of a faint nucleus. 

The alternative possibility is that the galaxy is truly non-nucleated. In this
case, we can {\it add} a
simulated nucleus of the appropriate size and magnitude to see if it would be detectable
from the images and/or the surface brightness profile. Taken together, these two experiments
allow us to crudely estimate the overall completeness of our survey and to refine the
provisional classifications for these 27 galaxies. We caution, however, that
the approach of adding and subtracting nuclei not only assumes that the scaling relations
found in \S4 are valid for all galaxies in the survey, but it ignores the significant
scatter about the fitted relations.

With these caveats in mind, we present the results of this exercise in Figure~\ref{fig26}.
In each panel, the small blue squares show the residuals profile obtained after adding
a simulated nucleus to the image and recalculating the brightness profile. Small red squares
show the profile obtained if this nucleus is instead subtracted. For four galaxies
(VCC1692, 1664, 944 and 1025), the best-fit S\'{e}rsic/core-S\'{e}rsic model for
the subtracted profile provides some improvement over the original fit.
We therefore classify these four as galaxies as Type~Ic systems.

At the same time, we identify 12 other galaxies which can be classified
unambiguously as non-nucleated (Type~II). In these galaxies, the subtracted brightness profiles 
show strong inward gradients in their central regions: an obviously 
non-physical result. Interestingly, these 12 galaxies fall into two rather
distinct categories: (1) {\it bright giants} which have shallow ``cores" in the
central few arcseconds and thus are best fit with core-S\'{e}rsic models; and (2)
{\it faint dwarfs} which are best
fit with pure S\'{e}rsic models. The common feature linking these two
populations is the presence of a low-surface brightness core that facilitates
the detection of a central nucleus. For this reason, we can say with some confidence
that these 12 galaxies do {\it not} contain nuclei that follow the scaling
relations observed in \S4 for the Type~Ia galaxies.
The final 11 galaxies remain elusive since we can neither confirm
nor rule out the presence of a nucleus in these cases. We classify these objects as
possibly nucleated (Type~Id).

Figures~\ref{fig06}-\ref{fig07} clearly demonstrate that it is possible to detect
nuclei in galaxies that span a wide range in luminosity and central surface brightness.
But to what extent is our ability to detect nuclei --- and to measure their sizes and 
magnitudes ---
affected by the surface brightness of the underlying galaxy and their own
luminosity or size? Needless to say, a complete characterization of the biases and incompleteness 
affecting the nuclei requires {\it a priori} knowledge of their intrinsic properties:
information that we are
obviously lacking. Nevertheless, we may take a first step towards answering these questions
by adding simulated nuclei of known size and
magnitude to the center of a non-nucleated galaxy. For these experiments,
we focus on a single non-nucleated galaxy, VCC1833, which, as a S\'{e}rsic-law galaxy with a central
surface brightness of $\mu_g(1\arcsec) \approx 19.3$ and
$\mu_z(1\arcsec) \approx 18.1$~mag~arcsec$^{-2}$, is representative of the Type Ia galaxies
in our survey.

Nuclei that span a range in both magnitude and size were added to the
galaxy photocenter. Input magnitudes covered the intervals $16 \le g \le 25$
and $16 \le z \le 24$ in 1~mag increments; at each magnitude, nuclei were added
with half-light radii of 0\farcs00, 0\farcs02, 0\farcs03, 0\farcs04, 0\farcs1,
0\farcs05, 0\farcs1, 0\farcs15 and 0\farcs2. 
Simulations were carried out independently for the
F475W and F850LP images, and for each simulation, the surface brightness profile was
measured from the artificial image using the same procedure as for the real galaxy.
A nucleated S\'{e}rsic model was then fitted to the profile of the simulated
galaxy+nucleus and the best-fit parameters for the nucleus recorded.

The results of these simulations are shown in Figure~\ref{fig27}. The upper panel
of this figure shows the difference between the recovered and input half-light
radius, $\Delta{r_h}$, as a function of input radius. The lower panel plots the difference 
between the recovered and input magnitude, $\Delta{m}$, as a function of input magnitude.
In both panels, results are shown for the separate F475W and F850LP bandpasses (blue
and red squares, respectively). The symbol size is proportional to either input magnitude
(as in the upper panel, where larger symbols correspond to brighter nuclei) or input
radius (as in the lower panel, where larger symbols correspond to more compact nuclei).

There are several conclusions which may be drawn from this figure, although
sweeping claims must be avoided because the results of the simulations
will almost certainly depend on the central surface brightness of the galaxy,
the steepness of its brightness profile, etc, so the findings are 
not generalizable to the other program galaxies in any straightforward way.
With these caveats in mind, we note that nuclei brighter than $g \approx 19$~mag
or $z \approx 18$~mag in this particular galaxy
would be detected for any choice of $r_h$ in the range 0-0\farcs2. Conversely,
nuclei fainter than $g \approx 23$~mag or $z \approx 24$~mag would never be
detected. There appears to be no serious bias affecting the $r_h$ measurements
for nuclei with $r_h \le 0\farcs05$, at least for nuclei brighter than
$g \sim$ 20--21~mag. In this size regime --- a range which 
encompasses half of the nuclei in Table~\ref{tab:data} --- the input half-light
radii are recovered to a precision of $\sim$ 15\% or better. For larger nuclei,
with $r_h \gtrsim 0\farcs05$, there is a bias which ranges from $\lesssim$ 10\% 
for the brightest nuclei, to nearly a factor of two for the faintest detectable
nuclei, in the sense that the recovered nuclei are smaller. Unfortunately, the intrinsic
distribution of nuclei sizes is unknown, so it is not possible to apply an {\it a posteriori}
correction to the measured sizes. In any case, we note that the result from
\S\ref{sec:results} that would be most directly affected by a bias of this sort
is the observed scaling between nucleus and luminosity (Figure~\ref{fig17}),
where it was found ${\cal L} \propto r^{\beta}$ with $\beta = 0.50\pm0.03$. 
If we make the admittedly dubious assumption that the luminosity dependence of the
bias found in the case of VCC1833 is representative of the full sample of 
Type~Ia galaxies, then we would expect the exponent in Equation~11 to fall to
$\beta \sim 0.4$.

For the faintest nuclei, the simulations reveal that completeness is a function
of surface brightness, in the expected sense that, at fixed luminosity, more
compact nuclei (i.e., higher surface brightness) are more likely to be detected.
As the lower panel of Figure~\ref{fig27} shows, there is also a tendency to 
measure fainter magnitudes for the simulated nuclei, at least in this galaxy. Not
surprisingly, the importance of this bias depends sensitively on the input magnitude;
for the brightest nuclei, the bias is less than $0.1$~mag in both filters, 
irrespective of the input $r_h$. For the fainter nuclei, the bias can be as large as
$\sim$ 0.5~mag, and is slightly worse in the F850LP filter. To the
extent that the simulations for VCC1833 are applicable to other galaxies in
the survey, this means that the faintest Type~Ia nuclei may have
measured colors that are systematically too blue by $\sim$ 0.1~mag.

\section{Comparison with de Propris \etal\ (2005) and Strader \etal\ (2006)}

Two recent papers, de~Propris~et~al.~(2005) and Strader \etal\ (2006), have presented
magnitudes, colors and half-light radii for the nuclei belonging to a subset of our
program galaxies. Since the same observational material that forms the basis of
our analysis was used in each of these studies, it is of interest to compare the various 
measurements.

Based on the VCC classifications that were available when the ACS Virgo Cluster Survey
was begun, 25 of the 100 program galaxies were thought to contain nuclei
(see Table~1 of Paper~I). As we have shown in \S4, the actual number of nucleated
galaxies in our survey is about three times larger than this, although in a number
of cases the nuclei were too faint to determine reliable photometric or
structural parameters; in the final analysis, magnitudes, colors and sizes could
be measured for 51 (Type~Ia) nuclei.

We first consider the results of de~Propris~et~al.~(2005), who studied 18 of the
25 galaxies originally classified as nucleated dwarfs. These authors parameterized
the underlying galaxies as S\'{e}rsic models. After subtracting this S\'{e}rsic component,
colors and magnitudes for the nuclei were determined by summing the light within
a 1\arcsec~aperture, while half-light radii for the nuclei were measured with the
ISHAPE software package (Larsen 1999) for a circular Plummer profile and Tiny Tim
PSF.
We have transformed the de Propris \etal\ (2005) VEGAMAG photometry onto
the AB system using the zeropoints given in Table~11 of Sirianni \etal\
(2005). Their half-light radii were converted from parsecs to
arcseconds using their adopted Virgo distance of 15.3~Mpc. Extinction
corrections, which in both studies are based on the DIRBE maps of Schlegel
\etal\ (1998), were applied to our photometry as described in Paper~II.

The two upper panels of Figure~\ref{fig28} compare our magnitudes
with those of de~Propris~et~al. (2005) (filled circles), where the dashed lines show
the one-to-one relations. There is only fair agreement between the measured
magnitudes (the $rms$ scatter is $\approx$ 0.30~mag in both bands).
In the lower left panel of Figure~\ref{fig28}, we compare our two estimates for the nuclei colors
with those of de~Propris~et~al. (2005). Whether one uses integrated or aperture
colors, the agreement is fair at best ($rms$ scatter $\approx$ 0.17~mag in either case). 
As discussed in \S4.1, an internal comparison of our color measurements
shows a an $rms$ scatter of 0.059 mag between the integrated and aperture
colors. In any case, the scatter in the comparison with the de~Propris~et~al.
(2005) colors is largely driven by three galaxies --- VCC200, VCC1826 and
VCC2050 (IC3779) --- which de~Propris~et~al.
(2005) find to host exeptionally blue nuclei, $(g-z)_{AB} \lesssim 0.75$~mag. For
single burst stellar populations, such colors would require ages
$\lesssim 3$~Gyr for virtually any choice of metallicity (see Figure~6
of Paper~I). By contrast, we measure colors in the range 0.8--1.2 for
these three nuclei.

In addition, we find poor agreement ($rms$ scatter = 0\farcs056)
between the half-light radii measured in the two studies. In the lower 
right panel of Figure~\ref{fig28}, we plot the de~Propris~et~al.~(2005)
half-light radii against the mean of our
measurements in the $g$ and $z$ bandpasses. An internal comparison of
our $g-$ and $z$-band measurements shows good agreement, with an $rms$ scatter of
$\sim$ 0\farcs01 (\S4.1). Unfortunately, an internal comparison of the
de~Propris~et~al.~(2005) is not possible since they report a single value of
the half-light radius for each nucleus, and it is not clear
if this value refers to a measurement made in a single bandpass, or the average
of measurements in the $g$ and $z$ bandpasses.

Figure~\ref{fig28} also shows a comparison of our magnitudes, colors and 
half-light radii for 25 nuclei to those of Strader \etal\ (2006) (open squares).
The Strader \etal\ (2006) measurements were also determined using
the ISHAPE package (Larsen 1999), although these authors used an empirical
PSF and assumed a King model 
nucleus of fixed concentration index $c \equiv \log{(r_t/r_c)} = 1.477$. 
Although there is no discussion of how the contribution from the underlying galaxy 
was modeled in their analysis of the nuclei, 
the authors do state that photometry and size measurements for the nuclei
were carried out using procedures identical to the globular
clusters, in which the background is usually modeled as a constant
or a plane. However, near the photocenter where the nuclei are found, the galaxy
light is varying rapidly in both the radial and azimuthal directions, and since
the galaxy brightness profiles nearly always exhibit an inward
rise, this procedure will lead to overestimates of the nuclei 
luminosities and sizes.

From the upper panels of Figure~\ref{fig28}, we see
that the Strader \etal\ (2006) magnitudes are, on average, $\sim$ 0.4~mag
brighter than ours. In addition, the discrepancy rises to $\gtrsim$~1~mag 
for the faintest nuclei --- those which should be most prone to errors in
modeling the underlying galaxy light.
There is better agreement between the measured
colors from the two studies, as the lower left panel shows ($rms$ scatter
= 0.098 and 0.072~mag for the integrated and aperture colors, respectively).
At the same time, however, there is poor agreement between the measured half-light
radii ($rms$ scatter = 0\farcs055), where their radii
are $\sim$ 80\% larger than ours. Unfortunately, Strader \etal\ (2006) tabulate a single
value of the radius for each nucleus, so no internal comparison of their
size measurements is possible.

\end{appendix}

\clearpage

\tabletypesize{\tiny}
\LongTables
\begin{landscape}
\begin{deluxetable}{cccccccccccccccccl}
\tablecaption{Basic Data for Nuclei of Program Galaxies\label{tab:data}}
\tablecolumns{18}
\tablehead{
\colhead{ID}         & 
\colhead{VCC}        & 
\colhead{Other}      & 
\colhead{$B_T$}      & 
\colhead{E(B-V)}   & 
\colhead{$\mu_g(1\arcsec$)}       & 
\colhead{$\mu_z(1\arcsec$)}       &
\colhead{Class}    &
\colhead{Class}    &
\colhead{Mod}   &
\colhead{$g_{AB}$}    &
\colhead{$z_{AB}$}    &
\colhead{$(g-z)_{AB}$}    &
\colhead{$(g-z)_{AB}^a$}    &
\colhead{$r_{h,g}$} &
\colhead{$r_{h,z}$} &
\colhead{$\delta_n$} &
\colhead{Comments} \\ 
\colhead{} & 
\colhead{} & 
\colhead{} & 
\colhead{(mag)} & 
\colhead{(mag)} & 
\colhead{(mag $\Box\arcsec$)} &
\colhead{(mag $\Box\arcsec$)} &
\colhead{(VCC)}&
\colhead{(ACS)}&
\colhead{}&
\colhead{(mag)} &
\colhead{(mag)} &
\colhead{(mag)} &
\colhead{(mag)} &
\colhead{($\arcsec$)} &
\colhead{($\arcsec$)} &
\colhead{($\arcsec$)} &
\colhead{} \\
\colhead{(1)} &
\colhead{(2)} &
\colhead{(3)} &
\colhead{(4)} &
\colhead{(5)} &
\colhead{(6)} &
\colhead{(7)} &
\colhead{(8)}&
\colhead{(9)}&
\colhead{(10)}&
\colhead{(11)}&
\colhead{(12)} &
\colhead{(13)} &
\colhead{(14)} &
\colhead{(15)} &
\colhead{(16)} &
\colhead{(17)} &
\colhead{(18)}  
}
\startdata
  1 & 1226 & M49,~N4472 &  9.31 & 0.022 & 16.74 & 15.12 & N & II & cS & \nodata&  \nodata&  \nodata&  \nodata&  \nodata& \nodata & \nodata & Dust.~AGN?\tablenotemark{a}\\
  2 & 1316\tablenotemark{b} & M87,~N4486 &  9.58 & 0.023 & 17.58 & 15.95 & N &  0 & cS & \nodata&  \nodata&  \nodata&  \nodata&  \nodata& \nodata &  & Dust.~AGN.\tablenotemark{c}\\
  3 & 1978 & M60,~N4649 &  9.81 & 0.026 & 16.93 & 15.27 & N & II & cS & \nodata&  \nodata&  \nodata&  \nodata&  \nodata& \nodata & \nodata & AGN?\tablenotemark{a} \\
  4 &  881 & M86,~N4406 & 10.06 & 0.029 & 16.66 & 15.08 & N & II & cS & \nodata&  \nodata&  \nodata&  \nodata&  \nodata& \nodata & \nodata & Dust.\\
  5 &  798 & M85,~N4382 & 10.09 & 0.030 & 16.23 & 14.89 & N & II & cS & \nodata&  \nodata&  \nodata&  \nodata&  \nodata& \nodata & \nodata & Dust:.\\
  6 &  763 & M84,~N4374 & 10.26 & 0.041 & 16.47 & 14.88 & N &  0 & cS & \nodata&  \nodata&  \nodata&  \nodata&  \nodata& \nodata & & Dust.~AGN.\tablenotemark{c}\\
  7 &  731 & N4365 & 10.51 & 0.021 & 16.95 & 15.34 & N & Ib & cS & \nodata&  \nodata&  \nodata&  \nodata&  \nodata& \nodata & 0.014$\pm$0.008 & \\
  8 & 1535 & N4526 & 10.61 & 0.022 & \nodata & \nodata & N &  0 &  \nodata & \nodata&  \nodata&  \nodata&  \nodata&  \nodata& \nodata & \nodata & Dust.\\
  9 & 1903\tablenotemark{b} & M59,~N4621 & 10.76 & 0.032 & 16.44 & 14.80 & N & Id &  S & \nodata&  \nodata&  \nodata&  \nodata&  \nodata& \nodata & \nodata & \\
 10 & 1632 & M89,~N4552 & 10.78 & 0.041 & 16.34 & 14.68 & N & II & cS & \nodata&  \nodata&  \nodata&  \nodata&  \nodata& \nodata & \nodata & Dust.~AGN?\tablenotemark{a}\\
 11 & 1231 & N4473      & 11.10 & 0.028 & 16.45 & 14.87 & N & II &  S & \nodata&  \nodata&  \nodata&  \nodata&  \nodata& \nodata & \nodata & \\
 12 & 2095 & N4762      & 11.18 & 0.022 & 16.76 & 15.25 & N & Ib &  S & \nodata&  \nodata&  \nodata&  \nodata&  \nodata& \nodata & 0.009$\pm$0.009 & \\
 13 & 1154 & N4459      & 11.37 & 0.045 & 16.67 & 15.11 & N & Id &  S & \nodata&  \nodata&  \nodata&  \nodata&  \nodata& \nodata & \nodata & Dust.\\
 14 & 1062 & N4442      & 11.40 & 0.022 & 16.63 & 15.02 & N & Id &  S & \nodata&  \nodata&  \nodata&  \nodata&  \nodata& \nodata & \nodata & \\
 15 & 2092 & N4754      & 11.51 & 0.032 & 16.77 & 15.15 & N & Ib &  S & \nodata&  \nodata&  \nodata&  \nodata&  \nodata& \nodata & 0.027$\pm$0.008 & \\
 16 &  369 & N4267      & 11.80 & 0.047 & 16.98 & 15.36 & N & Ib &  S & \nodata&  \nodata&  \nodata&  \nodata&  \nodata& \nodata & 0.015$\pm$0.004 & \\
 17 &  759 & N4371      & 11.80 & 0.036 & 17.50 & 15.93 & N & II &  S & \nodata&  \nodata&  \nodata&  \nodata&  \nodata& \nodata & \nodata & Dust.\\
 18 & 1692 & N4570      & 11.82 & 0.022 & 16.65 & 15.06 & N & Ic &  S & \nodata&  \nodata&  \nodata&  \nodata&  \nodata& \nodata & \nodata & \\
 19 & 1030 & N4435      & 11.84 & 0.029 & \nodata & \nodata & N &  0 &  \nodata & \nodata&  \nodata&  \nodata&  \nodata&  \nodata& \nodata & \nodata & Dust.\\
 20 & 2000\tablenotemark{b} & N4660  & 11.94 & 0.034 & 16.47 & 14.89 & N & Id &  S & \nodata&  \nodata&  \nodata&  \nodata&  \nodata& \nodata & \nodata & \\
 21 &  685 & N4350      & 11.99 & 0.028 & 16.57 & 14.98 & N &  0 &  S & \nodata&  \nodata&  \nodata&  \nodata&  \nodata& \nodata & \nodata & Dust.\\
 22 & 1664 & N4564      & 12.02 & 0.033 & 16.90 & 15.27 & N & Ic &  S & \nodata&  \nodata&  \nodata&  \nodata&  \nodata& \nodata & \nodata & \\
 23 &  654 & N4340      & 12.03 & 0.026 & 17.45 & 15.95 & N & Ib &  S & \nodata&  \nodata&  \nodata&  \nodata&  \nodata& \nodata & 0.025$\pm$0.007 & \\
 24 &  944 & N4417      & 12.08 & 0.025 & 16.93 & 15.42 & N & Ic &  S & \nodata&  \nodata&  \nodata&  \nodata&  \nodata& \nodata & \nodata & \\
 25 & 1938 & N4638      & 12.11 & 0.026 & 16.97 & 15.43 & N & Ib &  S & \nodata&  \nodata&  \nodata&  \nodata&  \nodata& \nodata & 0.017$\pm$0.069 & \\
 26 & 1279 & N4478      & 12.15 & 0.024 & 17.40 & 15.90 & N & Id &  S & \nodata&  \nodata&  \nodata&  \nodata&  \nodata& \nodata & \nodata & \\
 27 & 1720 & N4578      & 12.29 & 0.021 & 17.66 & 16.13 & N & Ia &  S & 18.40 & 16.82 & 1.57 & 1.57 & 0.085 & 0.085 & 0.007$\pm$0.005 & \\
 28 &  355 & N4262      & 12.41 & 0.036 & 16.75 & 15.15 & N & Id &  S & \nodata&  \nodata&  \nodata&  \nodata&  \nodata& \nodata & \nodata & Dust.\\
 29 & 1619 & N4550      & 12.50 & 0.040 & 17.69  & 16.25 & N & Ia &  S & 17.13 & 15.58 & 1.55 & 1.58 & 0.323 & 0.323 & 0.058$\pm$0.028 & Dust.~AGN.\tablenotemark{d}\\
 30 & 1883 & N4612      & 12.57 & 0.025 & 17.20 & 15.86 & N & Ia &  S & 18.74 & 17.60 & 1.14 & 1.11 & (0.024) & (0.024) & 0.013$\pm$0.006 & \\
 31 & 1242 & N4474      & 12.60 & 0.042 & 17.59 & 16.09 & N & Ia &  S & 19.84 & 18.14 & 1.70 & 1.68 & 0.035 & 0.035 & 0.017$\pm$0.007 & \\
 32 &  784 & N4379      & 12.67 & 0.024 & 17.53 & 16.05 & N & Ia &  S & 18.34 & 16.68 & 1.66 & 1.67 & 0.161 & 0.161 & 0.018$\pm$0.006 & \\
 33 & 1537 & N4528      & 12.70 & 0.046 & 17.30 & 15.82 & N & Id &  S & \nodata&  \nodata&  \nodata&  \nodata&  \nodata& \nodata & \nodata & \\
 34 &  778 & N4377      & 12.72 & 0.038 & 16.94 & 15.44 & N & Id &  S & \nodata&  \nodata&  \nodata&  \nodata&  \nodata& \nodata & \nodata & \\
 35 & 1321 & N4489      & 12.84 & 0.028 & 17.92 & 16.52 & N & Id &  S & \nodata&  \nodata&  \nodata&  \nodata&  \nodata& \nodata & \nodata & \\
 36 &  828 & N4387      & 12.84 & 0.033 & 18.01 & 16.50 & N & Ia &  S & 18.53 & 16.96 & 1.57 & 1.59 & 0.208 & 0.208 & 0.011$\pm$0.005 & \\
 37 & 1250 & N4476      & 12.91 & 0.028 & 17.81 & 16.58 & N & Ia & cS & 19.73 & 18.19 & 1.55 & 1.52 & 0.026 & 0.026 & 0.083$\pm$0.054 & Dust.\\
 38 & 1630 & N4551      & 12.91 & 0.039 & 18.00 & 16.46 & N & Ia &  S & 17.39 & 15.72 & 1.68 & 1.71 & 0.501 & 0.501 & 0.014$\pm$0.007 & \\
 39 & 1146 & N4458      & 12.93 & 0.023 & 18.03 & 16.59 & N & Ia &  S & 15.37 & 13.95 & 1.42 & 1.47 & 0.780 & 0.780 & 0.006$\pm$0.008 & \\
 40 & 1025 & N4434      & 13.06 & 0.022 & 17.48 & 16.00 & N & Ic &  S & \nodata&  \nodata&  \nodata&  \nodata&  \nodata& \nodata & \nodata & \\
 41 & 1303 & N4483      & 13.10 & 0.020 & 18.00 & 16.50 & N & Ib &  S & \nodata&  \nodata&  \nodata&  \nodata& \nodata& \nodata & 0.016$\pm$0.008 & \\
 42 & 1913 & N4623      & 13.22 & 0.022 & 18.51 & 17.02 & N & Ia &  S & 17.55 & 15.95 & 1.60 & 1.64 & 0.597 & 0.597 & 0.011$\pm$0.010 & \\
 43 & 1327 & N4486A     & 13.26 & 0.023 & 17.49 & 15.90 & N & II &  S & \nodata&  \nodata&  \nodata&  \nodata&  \nodata& \nodata & \nodata & Dust.\\
 44 & 1125 & N4452      & 13.30 & 0.030 & 18.57 & 17.15 & N & Ia &  S & 20.48 & 19.51 & 0.97 & 0.950 & 0.060 & 0.060 & 0.021$\pm$0.029 & \\
 45 & 1475 & N4515      & 13.36 & 0.031 & 17.85 & 16.46 & N & Ib &  S & \nodata&  \nodata&  \nodata&  \nodata& \nodata& \nodata & 0.017$\pm$0.008 & \\
 46 & 1178 & N4464      & 13.37 & 0.022 & 17.47 & 15.99 & N & Ib &  S & \nodata&  \nodata&  \nodata&  \nodata& \nodata& \nodata & 0.014$\pm$0.008 & \\
 47 & 1283 & N4479      & 13.45 & 0.029 & 19.11 & 17.62 & N & Ia &  S & 20.65 & 19.07 & 1.58 & 1.55 & 0.053 & 0.053 & 0.034$\pm$0.021 & \\
 48 & 1261 & N4482      & 13.56 & 0.029 & 19.71 & 18.49 & Y & Ia &  S & 19.50 & 18.28 & 1.22 & 1.25 & 0.041 & 0.036 & 0.028$\pm$0.018 & \\
 49 &  698 & N4352      & 13.60 & 0.026 & 18.43 & 17.03 & N & Ia &  S & 19.93 & 18.61 & 1.32 & 1.29 & 0.041 & 0.041 & 0.041$\pm$0.016 & \\
 50 & 1422 & I3468      & 13.64 & 0.031 & 20.04 & 18.76 & Y & Ia &  S & 20.22 & 19.00 & 1.22 & 1.26 & 0.038 & 0.035 & 0.038$\pm$0.018 & Dust:.\\
 51 & 2048 & I3773      & 13.81 & 0.032 & 19.39 & 18.17 & N & Ia &  S & 21.45 & 20.30 & 1.15 & 1.14 & 0.037 & 0.031 & 0.076$\pm$0.033 & \\
 52 & 1871 & I3653      & 13.86 & 0.030 & 18.89 & 17.42 & N & Ia &  S & 18.73 & 17.48 & 1.25 & 1.35 & 0.125 & 0.108 & 0.030$\pm$0.013 & \\
 53 &    9 & I3019      & 13.93 & 0.039 & 22.11 & 20.95 & Y & Ie &  S & 22.01 & 21.22 & 0.79 & 0.88 & 0.040 & 0.034 & 1.907$\pm$0.070 & dIrr/dE.\\
 54 &  575 & N4318      & 14.14 & 0.025 & 18.38 & 16.96 & N & Id &  S & \nodata&  \nodata&  \nodata&  \nodata&  \nodata& \nodata & \nodata & \\
 55 & 1910 & I809       & 14.17 & 0.031 & 20.05 & 18.64 & Y & Ia &  S & 19.82 & 18.63 & 1.19 & 1.19 & 0.038 & 0.038 & 0.005$\pm$0.018 & \\
 56 & 1049 & U7580      & 14.20 & 0.022 & 19.74 & 18.73 & N & II &  S & \nodata&  \nodata&  \nodata&  \nodata&  \nodata& \nodata & \nodata & \\
 57 &  856 & I3328      & 14.25 & 0.024 & 20.67 & 19.46 & Y & Ia &  S & 18.97 & 17.85 & 1.12 & 1.17 & 0.163 & 0.153 & 0.073$\pm$0.047 & \\
 58 &  140 & I3065      & 14.30 & 0.037 & 19.60 & 18.37 & N & Ia &  S & 22.09 & 21.19 & 0.91 & 0.93 & 0.030 & (0.022) & 0.045$\pm$0.023 & \\
 59 & 1355 & I3442      & 14.31 & 0.034 & 21.84 & 20.59 & Y & Ia &  S & 21.10 & 20.07 & 1.03 & 1.03 & 0.043 & 0.038 & 0.030$\pm$0.069 & \\
 60 & 1087 & I3381      & 14.31 & 0.027 & 20.60 & 19.27 & Y & Ia &  S & 20.22 & 18.89 & 1.33 & 1.33 & 0.027 & 0.027 & 0.012$\pm$0.032 & \\
 61 & 1297 & N4486B     & 14.33 & 0.021 & 17.49 & 15.93 & N & Id &  S & \nodata&  \nodata&  \nodata&  \nodata&  \nodata& \nodata & \nodata & \\
 62 & 1861 & I3652      & 14.37 & 0.029 & 20.76 & 19.43 & Y & Ia &  S & 20.11 & 19.06 & 1.04 & 1.19 & 0.137 & 0.119 & 0.061$\pm$0.053 & \\
 63 &  543 & U7436      & 14.39 & 0.031 & 20.56 & 19.29 & N & Ia & cS\tablenotemark{e} & 22.56 & 21.20 & 1.36 & 1.19 & 0.157 & 0.196 & 0.056$\pm$0.034 & \\
 64 & 1431 & I3470      & 14.51 & 0.051 & 19.82 & 18.40 & Y & Ia &  S & 19.66 & 18.54 & 1.13 & 1.17 & 0.238 & 0.233 & 0.073$\pm$0.027 & \\
 65 & 1528 & I3501      & 14.51 & 0.028 & 19.51 & 18.18 & N & Ia & cS\tablenotemark{e} & 22.27 & 21.31 & 0.96 & 0.89 & (0.015) & (0.018) & 0.016$\pm$0.016 & \\
 66 & 1695 & I3586      & 14.53 & 0.045 & 20.34 & 19.22 & N & Ia & cS\tablenotemark{e} & 22.61 & 21.23 & 1.38 & 1.35 & (0.022) & (0.024) & 0.433$\pm$0.196 & \\
 67 & 1833 &            & 14.54 & 0.036 & 19.28 & 18.09 & N & II &  S & \nodata&  \nodata&  \nodata&  \nodata&  \nodata& \nodata & \nodata & \\
 68 &  437 & U7399A     & 14.54 & 0.029 & 20.75 & 19.45 & Y & Ia &  S & 20.00 & 19.01 & 1.00 & 1.03 & 0.089 & 0.083 & 0.032$\pm$0.029 & \\
 69 & 2019 & I3735      & 14.55 & 0.022 & 20.83 & 19.63 & Y & Ia &  S & 20.31 & 19.20 & 1.12 & 1.18 & 0.037 & 0.029 & 0.183$\pm$0.077 & \\
 70 &   33 & I3032      & 14.67 & 0.037 & 20.79 & 19.75 & Y & Ia &  S & 22.18 & 21.27 & 0.91 & 0.91 & 0.033 & 0.032 & 0.046$\pm$0.056 & \\
 71 &  200 &            & 14.69 & 0.030 & 20.56 & 19.31 & Y & Ia &  S & 22.86 & 21.94 & 0.92 & 1.06 & 0.053 & 0.038 & 0.068$\pm$0.043 & \\
 72 &  571 &            & 14.74 & 0.022 & 20.12 & 19.11 & N &  0 &  S & \nodata&  \nodata&  \nodata&  \nodata&  \nodata& \nodata & \nodata & Dust.\\
 73 &   21 & I3025      & 14.75 & 0.021 & 20.42 & 18.79 & N & Ie &  S & 20.64 & 20.36 & 0.28 & 0.32 & 0.033 & 0.029 & 0.759$\pm$0.070 & dIrr/dE\\
 74 & 1488 & I3487      & 14.76 & 0.021 & 20.24 & 19.51 & N & Ia &  S & 23.71 & 22.99 & 0.72 & 0.69 & 0.025 & 0.025 & 0.038$\pm$0.071 & \\
 75 & 1779 & I3612      & 14.83 & 0.028 & 20.45 & 19.62 & N & Ie &  S & 22.41 & 21.58 & 0.83 & 0.85 & 0.021 & 0.024 & 0.542$\pm$0.191 & Dust. dIrr/dE\\
 76 & 1895 & U7854      & 14.91 & 0.017 & 20.54 & 19.42 & N & Ia &  S & 23.48 & 22.61 & 0.88 & 0.89 & (0.023) & (0.021) & 0.198$\pm$0.043 & \\
 77 & 1499 & I3492      & 14.94 & 0.030 & 19.55 & 19.12 & N & II &  S & \nodata&  \nodata&  \nodata&  \nodata&  \nodata& \nodata & \nodata & dIrr/dE\\
 78 & 1545 & I3509      & 14.96 & 0.042 & 19.98 & 18.53 & N & Ia &  S & 21.93 & 20.88 & 1.05 & 1.16 & 0.050 & 0.037 & 0.027$\pm$0.017 & \\
 79 & 1192 & N4467      & 15.04 & 0.023 & 19.05 & 17.53 & N & Ia &  S & 19.09 & 17.90 & 1.19 & 1.22 & 0.120 & 0.121 & 0.007$\pm$0.013 & \\
 80 & 1857 & I3647      & 15.07 & 0.025 & 22.72 & 21.72 & Y & Ie &  S & 21.11 & 20.27 & 0.84 & 0.86 & 0.041 & 0.041 & 6.986$\pm$0.347 & \\
 81 & 1075 & I3383      & 15.08 & 0.027 & 21.43 & 21.40 & Y & Ia &  S & 21.07 & 20.11 & 0.96 & 0.97 & 0.040 & 0.039 & 0.012$\pm$0.070 & \\
 82 & 1948 &            & 15.10 & 0.025 & 21.59 & 20.57 & N & Ie &  S & 24.30 & 23.21 & 1.09 & 1.04 & 0.041 & 0.054 & 1.426$\pm$0.057 & \\
 83 & 1627 &            & 15.16 & 0.039 & 19.04 & 17.59 & N & Ia &  S & 18.83 & 17.39 & 1.44 & 1.46 & 0.197 & 0.197 & 0.024$\pm$0.017 & \\
 84 & 1440 & I798       & 15.20 & 0.028 & 19.25 & 17.96 & N & Ia &  S & 19.70 & 18.38 & 1.32 & 1.33 & 0.063 & 0.056 & 0.015$\pm$0.014 & \\
 85 &  230 & I3101      & 15.20 & 0.028 & 21.14 & 19.91 & Y & Ia &  S & 20.31 & 19.22 & 1.09 & 1.08 & 0.038 & 0.033 & 0.058$\pm$0.032 & \\
 86 & 2050 & I3779      & 15.20 & 0.023 & 21.02 & 19.82 & Y & Ia &  S & 22.38 & 21.41 & 0.97 & 1.08 & 0.073 & 0.066 & 0.037$\pm$0.039 & \\
 87 & 1993 &            & 15.30 & 0.025 & 20.61 & 19.34 & N & Ib &  S & \nodata&  \nodata&  \nodata&  \nodata& \nodata &\nodata & 0.043$\pm$0.052 & \\
 88 &  751 & I3292      & 15.30 & 0.032 & 19.92 & 18.58 & N & Ia &  S & 21.22 & 20.17 & 1.05 & 1.22 & 0.046 & 0.035 & 0.030$\pm$0.024 & \\
 89 & 1828 & I3635      & 15.33 & 0.037 & 21.40 & 20.12 & Y & Ia &  S & 21.50 & 20.50 & 1.00 & 1.04 & 0.060 & 0.057 & 0.018$\pm$0.070 & \\
 90 &  538 & N4309A     & 15.40 & 0.020 & 20.14 & 18.98 & N & Ia &  S & 21.27 & 20.15 & 1.13 & 1.13 & 0.033 & 0.030 & 0.028$\pm$0.016 & \\
 91 & 1407 & I3461      & 15.49 & 0.032 & 20.75 & 19.50 & Y & Ia &  S & 20.39 & 19.40 & 0.98 & 1.08 & 0.145 & 0.127 & 0.060$\pm$0.029 & \\
 92 & 1886 &            & 15.49 & 0.033 & 22.08 & 21.12 & Y & Ia &  S & 22.05 & 21.04 & 1.01 & 0.97 & 0.036 & 0.041 & 0.022$\pm$0.070 & \\
 93 & 1199 &            & 15.50 & 0.022 & 19.56 & 17.97 & N & Ia &  S & 19.75 & 18.39 & 1.36 & 1.45 & 0.075 & 0.063 & 0.040$\pm$0.036 & \\
 94 & 1743 & I3602      & 15.50 & 0.019 & 21.45 & 20.31 & N & Ib &  S & \nodata&  \nodata&  \nodata&  \nodata& \nodata &\nodata & 0.015$\pm$0.078 & \\
 95 & 1539 &            & 15.68 & 0.032 & 22.01 & 20.86 & Y & Ia &  S & 20.93 & 19.81 & 1.11 & 1.02 & 0.231 & 0.265 & 0.143$\pm$0.063 & \\
 96 & 1185 &            & 15.68 & 0.023 & 21.89 & 20.62 & Y & Ia &  S & 20.86 & 19.91 & 0.95 & 1.00 & 0.057 & 0.050 & 0.004$\pm$0.069 & \\
 97 & 1826 & I3633      & 15.70 & 0.017 & 20.52 & 19.34 & Y & Ia &  S & 20.10 & 18.91 & 1.19 & 1.17 & (0.024) & (0.025) & 0.043$\pm$0.028 & \\
 98 & 1512 &            & 15.73 & 0.050 & 20.27 & 19.42 & N & II & cS & \nodata&  \nodata&  \nodata&  \nodata&  \nodata& \nodata & \nodata & dIrr/dE\\
 99 & 1489 & I3490      & 15.89 & 0.034 & 22.03 & 20.98 & Y & Ia &  S & 22.38 & 21.51 & 0.87 & 0.83 & 0.051 & 0.058 & 0.021$\pm$0.070 & \\
100 & 1661 &            & 15.97 & 0.020 & 22.57 & 21.34 & Y & Ia &  S & 20.30 & 19.25 & 1.05 & 1.02 & 0.079 & 0.082 & 0.027$\pm$0.069 & \\
\enddata
\tablenotetext{~}{Key to Columns:\\
(1) identification number;\\
(2) Virgo Cluster Catalog (VCC) number (Binggeli, Sandage \& Tammann (1987);\\
(3) alternative names in the Messier, NGC, UGC or IC catalogs;\\
(4) total blue magnitude from VCC;\\
(5) extinction from Schlegel, Finkbeiner \& Davis (1998);\\
(6-7) $g$- and $z$-band surface brightness measured at a geometric radius of 0\farcs1\\
(8) nuclear classification in VCC;\\
(9) nuclear classification (see Table~\ref{tab:class});\\
(10) model used fit the galaxy brightness profile: (cS) = core-S\'ersic; (S) = S\'ersic;\\
(11-12) $g$- and $z$-band magnitudes for nucleus;\\
(13) integrated color of nucleus;\\
(14) nucleus color within a 4-pixel aperture;\\
(15-16) King model half-light radii in the $g$- and $z$-bands. Radii in parantheses indicate that the nucleus is formally unresolved in our images;\\
(17) offset from galaxy photocenter in arcseconds;\\
(18) comments on galaxy morphology, AGN and dust properties. Magnitudes and colors in this table have
not been corrected for extinction.}
\tablenotetext{a}{Low-level radio emission detected by Wrobel (1991) and/or Ho \etal\ (1997).}
\tablenotetext{b}{Central $\lesssim$0.3\arcsec saturated in F475W images.}
\tablenotetext{c}{Fr I radio galaxy (e.g., Ho 1999; Chiaberge \etal\ 1999).}
\tablenotetext{d}{LINER galaxy according to Ho \etal\ (1997).}
\tablenotetext{e}{Galaxy classified as S\'ersic in Paper VI.}
\end{deluxetable}
\clearpage
\end{landscape}
\tabletypesize{\normalsize}

\begin{deluxetable}{lr}
\tablecaption{Nuclear Classifications for the ACS Virgo Cluster Survey.\label{tab:class}}
\tablewidth{0pt}
\tablehead{
\colhead{Classification} &
\colhead{Number} 
}
\startdata
No Classification Possible (Type 0)                &    \\
\hskip0.2truein {\it Dust}                         &  4 \\
\hskip0.2truein {\it AGN}                          &  2 \\
\hskip0.2truein Total                              & {\bf 6} \\
& \\
Nucleated (Type I)                                 & \\
\hskip0.2truein {\it fitted nucleus} (Ia)          & 51 \\
\hskip0.2truein {\it no fit possible} (Ib)         & 11 \\
\hskip0.2truein Total                              & {\bf 62} \\
& \\
Non-Nucleated (Type II)                            & {\bf 12} \\
& \\
Uncertain (Type I)                                 &    \\
\hskip0.2truein {\it likely nucleated} (Ic)        &  4 \\
\hskip0.2truein {\it possibly nucleated} (Id)      & 11 \\
\hskip0.2truein {\it possible offset nucleus} (Ie) &  5 \\
\hskip0.2truein Total                              & {\bf 20} \\
& \\
All Galaxies & {\bf 100} \\
\enddata
\end{deluxetable}

\clearpage

\begin{landscape}
\begin{deluxetable}{cccccccc}
\tablecaption{Comparison of Nuclear Classifications from Previous {\it HST} Studies.\label{tab:comp}}
\tabletypesize{\scriptsize}
\tablewidth{0pt}
\tablehead{
\colhead{VCC} &
\colhead{Other} &
\colhead{$M_B$} &
\colhead{Rest \etal\ (2001)} &
\colhead{Ravindranath \etal\ (2001)} &
\colhead{Lauer \etal\ (2005)} &
\colhead{Lotz \etal\ (2004)} &
\colhead{ACSVCS} \\
\colhead{} &
\colhead{} &
\colhead{(mag)} &
\colhead{} &
\colhead{} &
\colhead{} &
\colhead{} &
\colhead{}
}
\startdata
1226 & NGC4472   & -21.87 & \nodata   & no      & yes     & \nodata & no (II)       \\
1978 & NGC4649   & -21.39 & \nodata   & \nodata & no      & \nodata & no (II)       \\
 881 & NGC4406   & -21.15 & \nodata   & no      & yes     & \nodata & no (II)       \\
 798 & NGC4382   & -21.13 & \nodata   & \nodata & \nodata & \nodata & no (II)       \\
 763 & NGC4374   & -21.01 & \nodata   & yes     & \nodata & \nodata & AGN (O)       \\
 731 & NGC4365   & -20.67 & no        & \nodata & yes     & \nodata & yes (Ib)      \\
1632 & NGC4552   & -20.49 & \nodata   & \nodata & yes     & \nodata & no (II)       \\
1903 & NGC4621   & -20.47 & no        & \nodata & no      & \nodata & possibly (Id) \\
1231 & NGC4473   & -20.11 & \nodata   & \nodata & no      & \nodata & no (II)       \\
2000 & NGC4660   & -19.30 & \nodata   & \nodata & no      & \nodata & possibly (Id) \\
1664 & NGC4564   & -19.21 & no        & \nodata & \nodata & \nodata & probably (Ic) \\
 944 & NGC4417   & -19.12 & \nodata   & no      & \nodata & \nodata & probably (Ic) \\
1279 & NGC4478   & -19.04 & no        & \nodata & yes     & \nodata & possibly (Id) \\
1242 & NGC4474   & -18.67 & yes (II)  & \nodata & \nodata & \nodata & yes (Ia)      \\
1146 & NGC4458   & -18.26 & \nodata   & \nodata & no      & \nodata & yes (Ia)      \\
1261 & NGC4482   & -17.65 & yes (III) & \nodata & \nodata & \nodata & yes (Ia)      \\
   9 & IC3019    & -17.33 & \nodata   & \nodata & \nodata & no      & possible offset (Ie) \\
1297 & NGC4486B  & -16.85 & \nodata   & \nodata & no      & \nodata & possibly (Id) \\
 543 & UGC7436   & -16.93 & \nodata   & \nodata & \nodata & no      & yes (Ia)      \\
1948 & IC3693    & -16.10 & \nodata   & \nodata & \nodata & no      & possible offset (Ie) \\
\enddata
\tablenotetext{~}{Note: Classifications from Rest \etal\ (2001) vary from
I (weakly nucleated) to III (strongly nucleated).}
\end{deluxetable}
\clearpage
\end{landscape}

\begin{deluxetable}{lccccc}
\tablecaption{Luminosity Function Parameters.\label{tab:lf}}
\tablewidth{0pt}
\tablehead{
\colhead{} &
\colhead{} &
\colhead{$A_n$} &
\colhead{$\overline{m}_n^0$} &
\colhead{$\sigma_n$} &
\colhead{$\overline{M}_n^0$\tablenotemark{a}} \\
\colhead{} &
\colhead{} &
\colhead{} &
\colhead{(mag)} &
\colhead{(mag)} &
\colhead{(mag)}
}
\startdata
$g$ & & 12.3$\pm$1.1 & 20.35$\pm$0.18 & 1.47$\pm$0.16 & --10.74$\pm$0.23 \\
$z$ & & 11.5$\pm$0.9 & 19.12$\pm$0.19 & 1.77$\pm$0.15 & --11.97$\pm$0.24 \\
\enddata
\tablenotetext{a}{Adopted Virgo distance: 16.52$\pm$0.22 (random) $\pm$1.14 (systematic) Mpc.}
\end{deluxetable}


\begin{deluxetable}{lcc}
\tablecaption{Global Properties of Nuclei and Ultra Compact Dwarfs.\label{tab:global}}
\tablewidth{0pt}
\tablehead{
\colhead{Parameter} &
\colhead{Nuclei} &
\colhead{UCDs}
}
\startdata
$r_h$ (pc) & $\le$2--62                                           & 25.9$\pm$9.1 \\
           & 4.2$\pm$3.8 (med.)                            &  \\
$\langle\mu_{h,g}\rangle$ (mag arcsec$^{-2}$) & 16.5$\pm$1.5 & 19.2$\pm$0.5\\
$\langle\mu_{h,z}\rangle$ (mag arcsec$^{-2}$) & 15.2$\pm$1.6 & 18.5$\pm$0.6\\
$\langle M_g\rangle$ (mag) & $-$10.9$\pm$1.7 & $-$11.2$\pm$0.6\\
$\langle M_z \rangle$ (mag) & $-$12.0$\pm$1.9 & $-$12.2$\pm$0.6\\
$\langle(g-z)_{AB}\rangle$ (mag) & 0.67--1.61 & 1.03$\pm$0.06\\
           & 1.13$\pm$0.25                             &  \\
\enddata
\end{deluxetable}


\begin{deluxetable}{lcc}
\tablecaption{Mean Metallicities for Galactic Nuclei.\label{tab:met}}
\tablewidth{0pt}
\tablehead{
\colhead{$(g-z)_0$-[Fe/H]} &
\colhead{$\tau$} &
\colhead{$\langle$[Fe/H]$\rangle$} \\
\colhead{} &
\colhead{(Gyr)} &
\colhead{(dex)}
}
\startdata
Bruzual \& Charlot (2003) &      1 & $+$0.82$\pm$0.52 \\
                          &      2 & $+$0.04$\pm$0.64\\
                          &      5 & $-$0.54$\pm$0.65\\
                          &     10 & $-$0.90$\pm$0.71\\
Peng \etal\ (2006)\tablenotemark{a}        &\nodata & $-$0.88$\pm$0.79\\
\enddata
\tablenotetext{a}{Broken linear relation based on $gz$ photometry and spectroscopic
metallicities for 95 globular clusters in the Milky Way, M49 and M87 (Peng \etal\ 
2006 = Paper IX).}
\end{deluxetable}

\clearpage

\clearpage

\clearpage

\clearpage

\clearpage
\begin{figure}
\figurenum{4}
\plotone{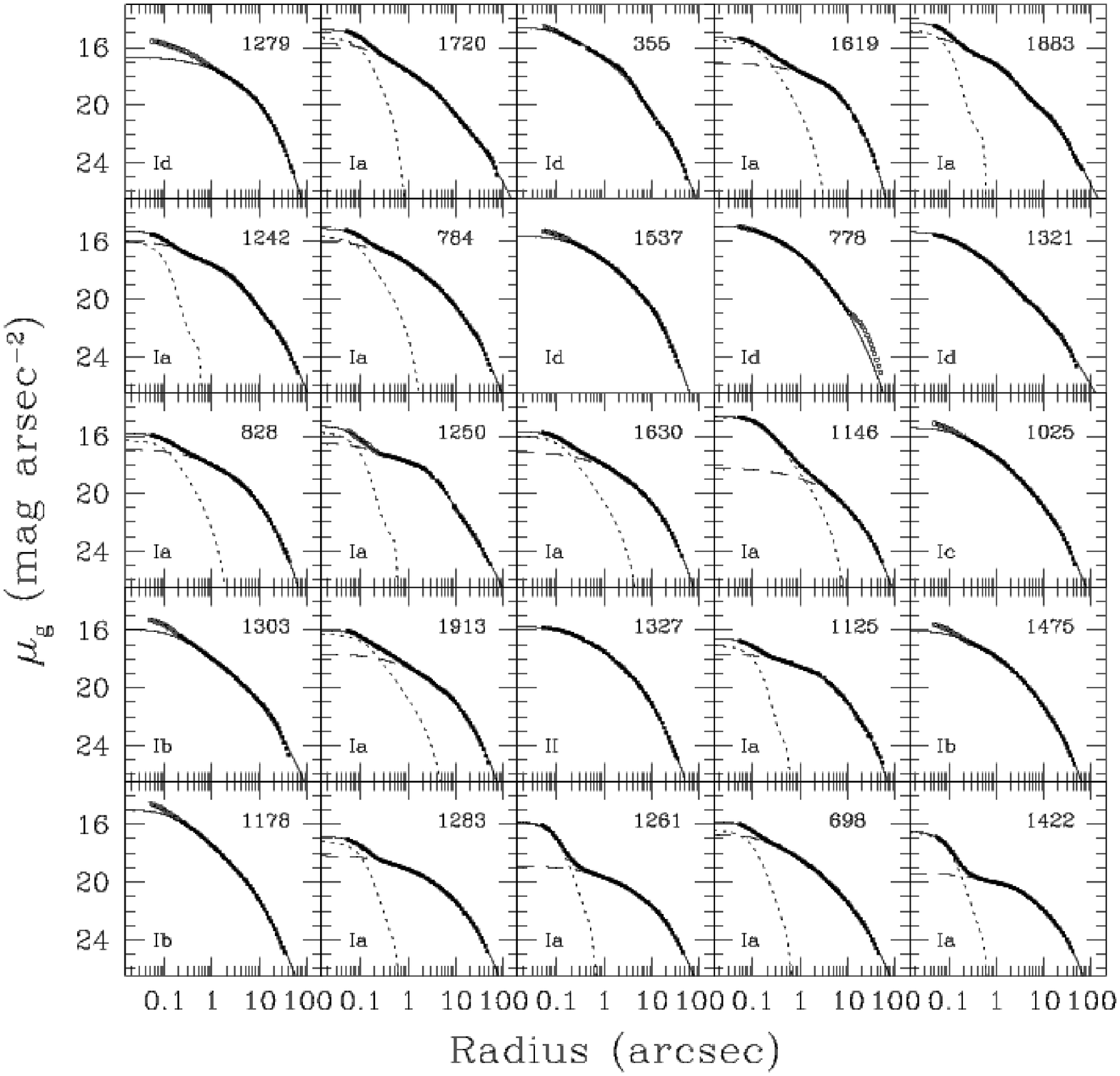}
\caption{Azimuthally-averaged $g$-band surface brightness profiles for the
galaxies from the ACS Virgo Cluster Survey (see Figure~\ref{fig03}).
The galaxies are rank ordered by blue magnitude, which decreases from left to
right and from top to bottom. Classifications from Table~\ref{tab:data} are reported in
the lower left corner of each panel. The classification scheme itself is
summarized in Table~\ref{tab:class}. In each panel, the solid curve shows the best-fit
galaxy model; open symbols show datapoints that were excluded during the
fitting of the models. For Type Ia galaxies, dashed and dotted curves show
the respective models for the galaxy and the nucleus. The fitted models for
galaxies with possible offset nuclei (the Type~Ie galaxies) do not include
components for the nuclei.
For VCC1316, VCC1903 and VCC2000, the crosses show regions affected by
saturation of the F475W images; these regions were excluded when fitting
models to the brightness profiles.
The brightness profiles for two galaxies (VCC1192 and VCC1199)
flatten outside of $\approx$ 10$^{\prime}$ due to contamination from the
halo of VCC1226 (M49), which is $\lesssim 5^{\prime}$ away in both cases.
These outer points have been excluded in the fits.}
\end{figure}

\clearpage
\begin{figure}
\figurenum{4}
\plotone{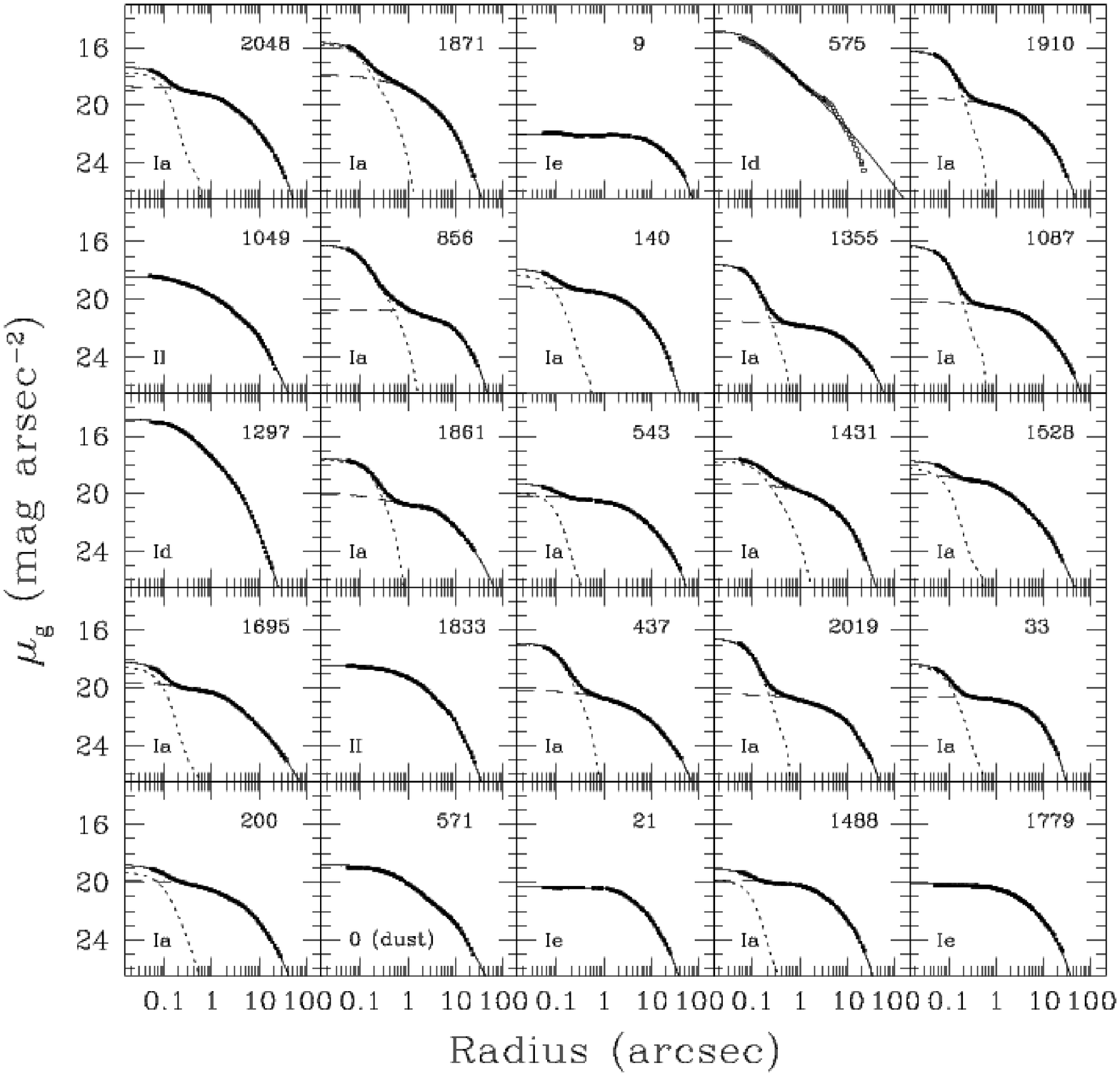}
\caption{Azimuthally-averaged $g$-band surface brightness profiles for the
galaxies from the ACS Virgo Cluster Survey (see Figure~\ref{fig03}).
The galaxies are rank ordered by blue magnitude, which decreases from left to
right and from top to bottom. Classifications from Table~\ref{tab:data} are reported in
the lower left corner of each panel. The classification scheme itself is
summarized in Table~\ref{tab:class}. In each panel, the solid curve shows the best-fit
galaxy model; open symbols show datapoints that were excluded during the
fitting of the models. For Type Ia galaxies, dashed and dotted curves show
the respective models for the galaxy and the nucleus. The fitted models for
galaxies with possible offset nuclei (the Type~Ie galaxies) do not include
components for the nuclei.
For VCC1316, VCC1903 and VCC2000, the crosses show regions affected by
saturation of the F475W images; these regions were excluded when fitting
models to the brightness profiles.
The brightness profiles for two galaxies (VCC1192 and VCC1199)
flatten outside of $\approx$ 10$^{\prime}$ due to contamination from the
halo of VCC1226 (M49), which is $\lesssim 5^{\prime}$ away in both cases.
These outer points have been excluded in the fits.}
\end{figure}

\clearpage
\begin{figure}
\figurenum{4}
\plotone{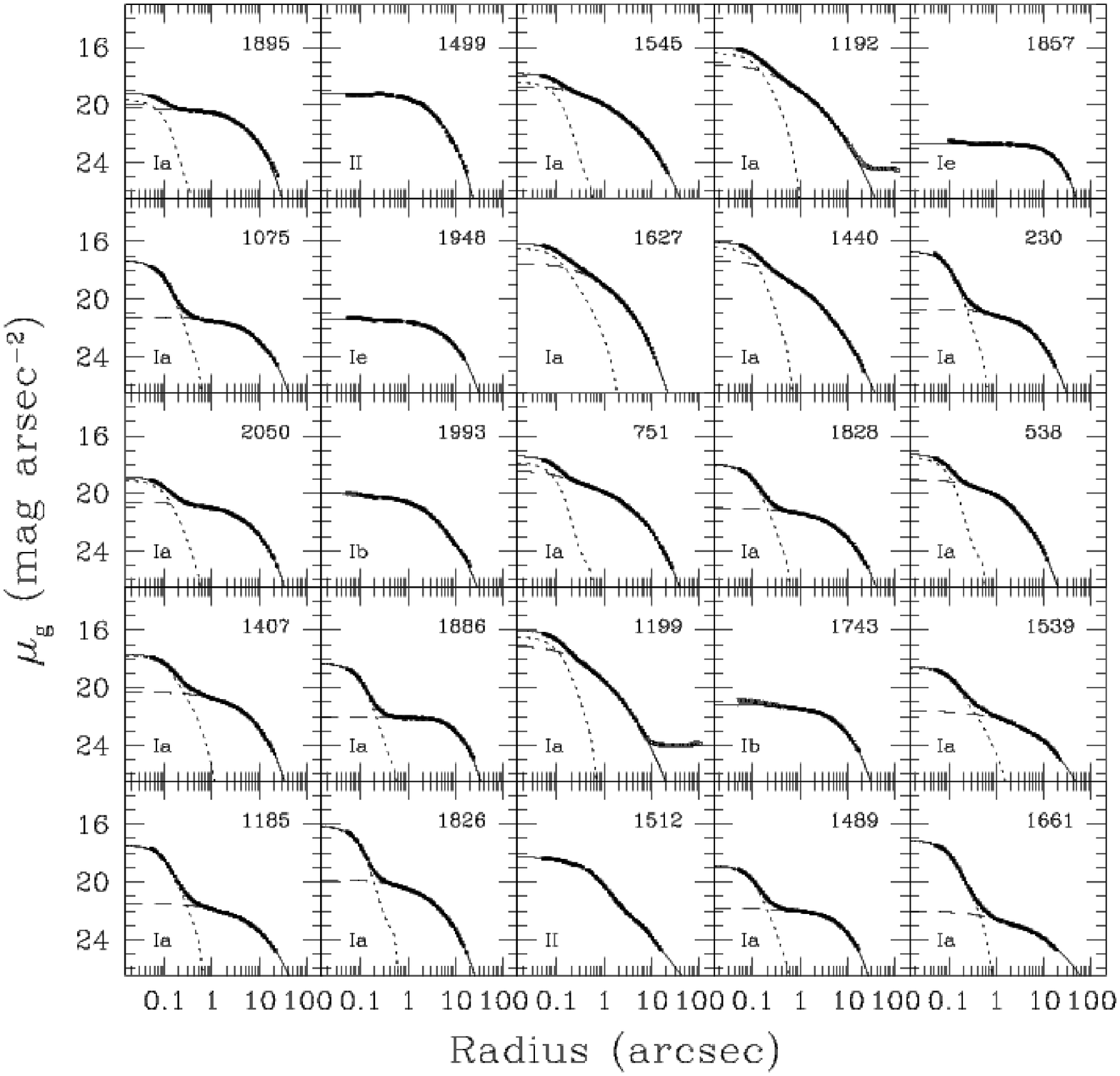}
\caption{Azimuthally-averaged $g$-band surface brightness profiles for the
galaxies from the ACS Virgo Cluster Survey (see Figure~\ref{fig03}).
The galaxies are rank ordered by blue magnitude, which decreases from left to
right and from top to bottom. Classifications from Table~\ref{tab:data} are reported in
the lower left corner of each panel. The classification scheme itself is
summarized in Table~\ref{tab:class}. In each panel, the solid curve shows the best-fit
galaxy model; open symbols show datapoints that were excluded during the
fitting of the models. For Type Ia galaxies, dashed and dotted curves show
the respective models for the galaxy and the nucleus. The fitted models for
galaxies with possible offset nuclei (the Type~Ie galaxies) do not include
components for the nuclei.
For VCC1316, VCC1903 and VCC2000, the crosses show regions affected by
saturation of the F475W images; these regions were excluded when fitting
models to the brightness profiles.
The brightness profiles for two galaxies (VCC1192 and VCC1199)
flatten outside of $\approx$ 10$^{\prime}$ due to contamination from the
halo of VCC1226 (M49), which is $\lesssim 5^{\prime}$ away in both cases.
These outer points have been excluded in the fits.}
\end{figure}

\clearpage
\begin{figure}
\figurenum{5}
\plotone{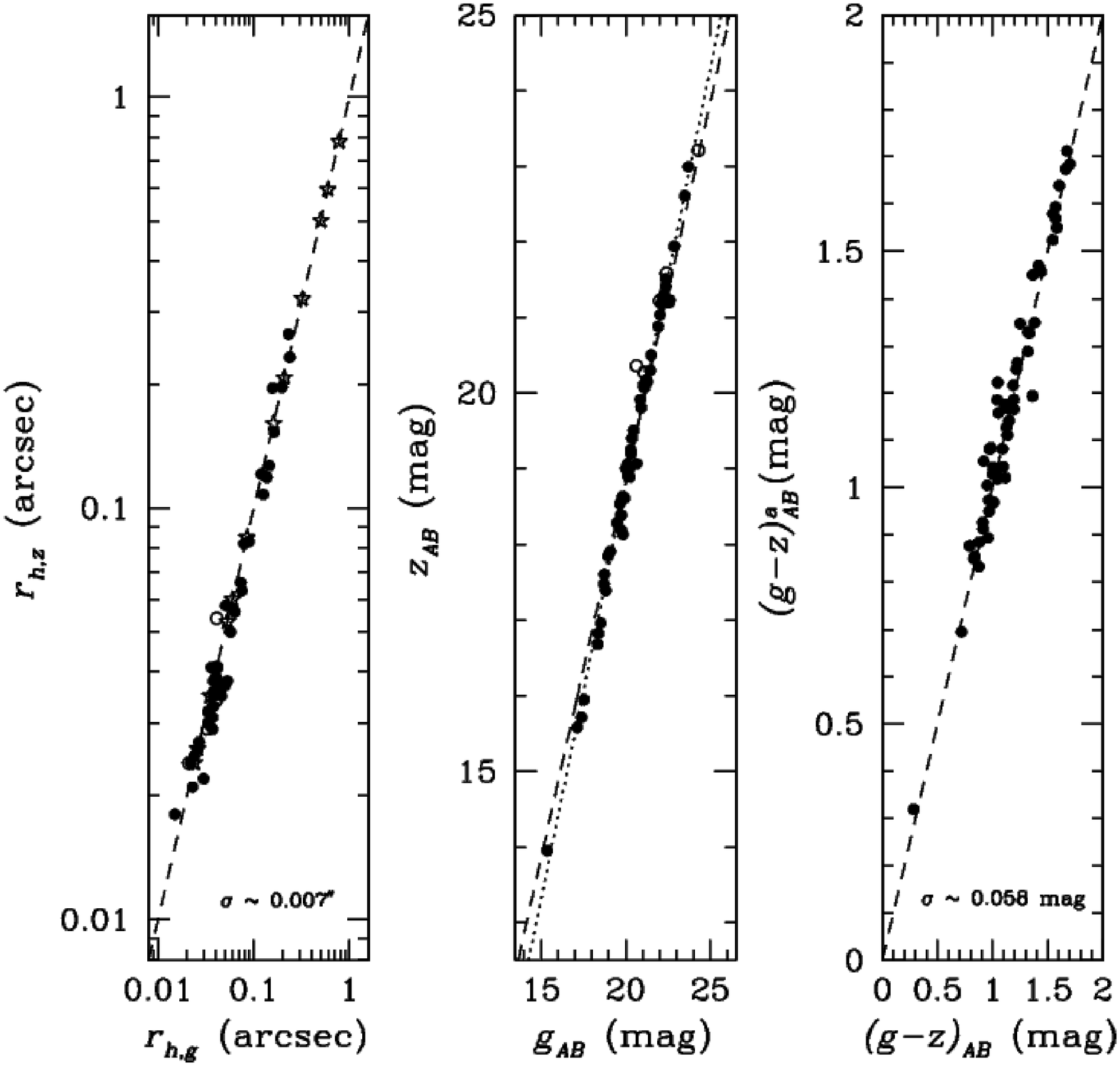}
\caption{{\it (Left Panel)}: Comparison of the King model half-light radii,
$r_h$, measured for the nuclei in the F850LP ($z$) and F475W ($g$) filters (filled
circles). The 51 nuclei with reliable photometric and structural measurements are 
shown by the filled circles and open stars. The latter symbols indicate the 11 galaxies
with $B_T \le 13.5$, for which the half-light radii were constrained to be the
same in the two bandpasses. The five galaxies with possible offset nuclei are
shown as open circles. The dashed line shows the one-to-one relation.
{\it (Middle Panel)}: Comparison of the measured $z$- and $g$-band magnitudes
for the nuclei. The dotted
line shows the least-squares line of best fit. The dashed line shows the
relation corresponding to the mean color of the nuclei: $\langle(g-z)\rangle = 1.15$~mag.
{\it (Right Panel)}: Comparison of the nuclei colors obtained by
direct integration of the best-fit model, $(g-z)_{AB}$, with those from 4-pixel aperture
measurements, $(g-z)^a_{AB}$. The dashed line shows the one-to-one relation.
\label{fig05}}
\end{figure}

\clearpage
\begin{figure}
\figurenum{6}
\plotone{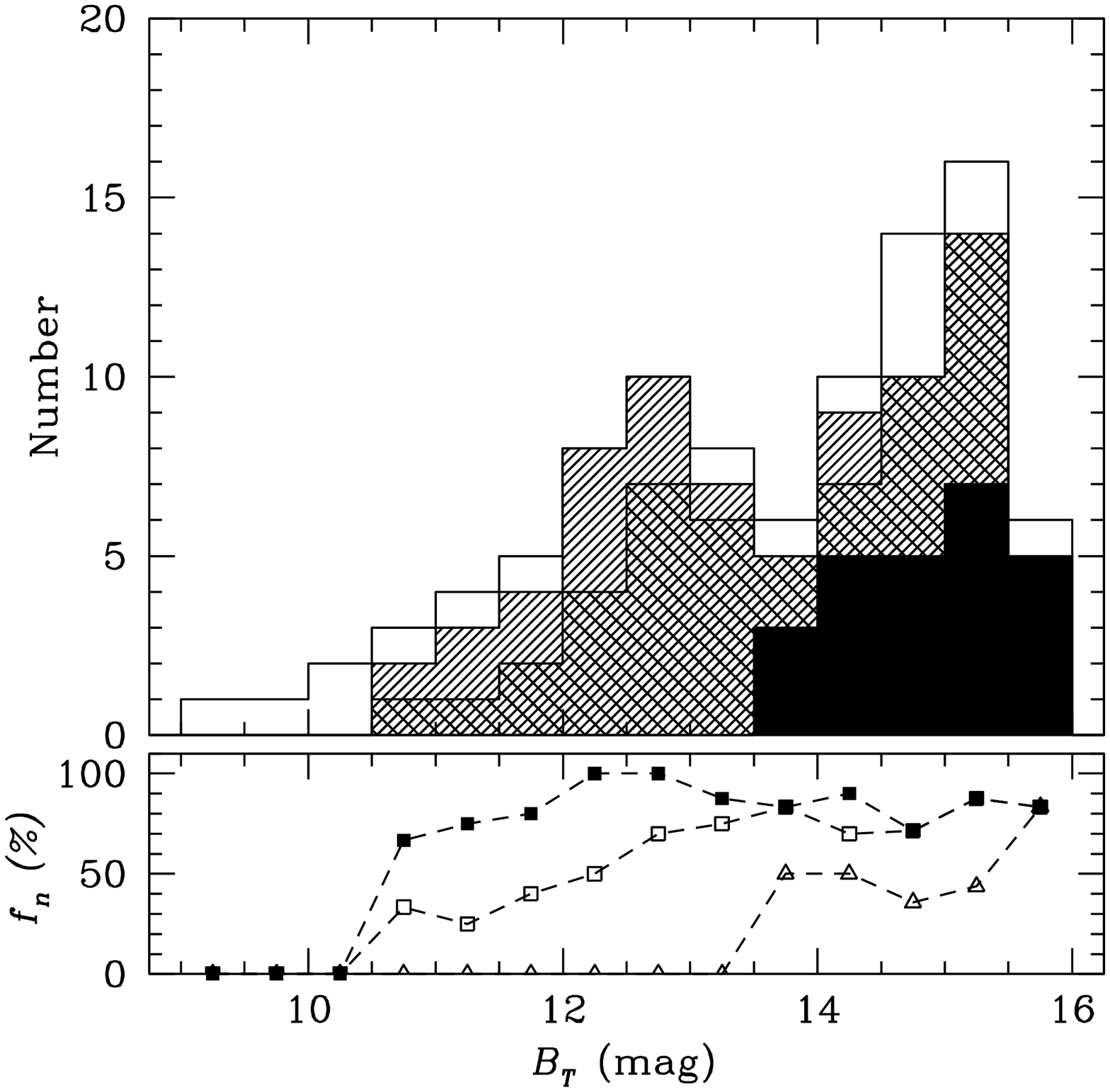}
\caption{{\it (Upper Panel)} Luminosity distribution of the 94 program galaxies for
which a classification as nucleated or non-nucleated is possible from our ACS images
(open histogram).  The double-hatched histogram shows the luminosity distribution for the
62 galaxies which we classify unambiguously as nucleated (i.e., Types Ia and Ib).
The hatched histogram shows this same sample plus the 15 galaxies which
{\it may} have central nuclei (i.e., Types Ia and Ib, plus Types Ic and Id).
The solid histogram shows the 25 galaxies in our survey which were classified as 
nucleated in the Virgo Cluster Catalog (Binggeli, Sandage \& Tammann 1985).
{\it (Lower Panel)} Percentage of nucleated galaxies, $f_n$, as a
function of blue magnitude. Open and filled squares show the frequency of
nucleation for Types Ia and Ib (62 galaxies) and Types Ia, Ib, Ic and Id (77
galaxies). Open triangles show results found using the classifications of Binggeli, 
Sandage \& Tammann (1985).
\label{fig06}}
\end{figure}

\clearpage
\begin{figure}
\figurenum{7}
\plotone{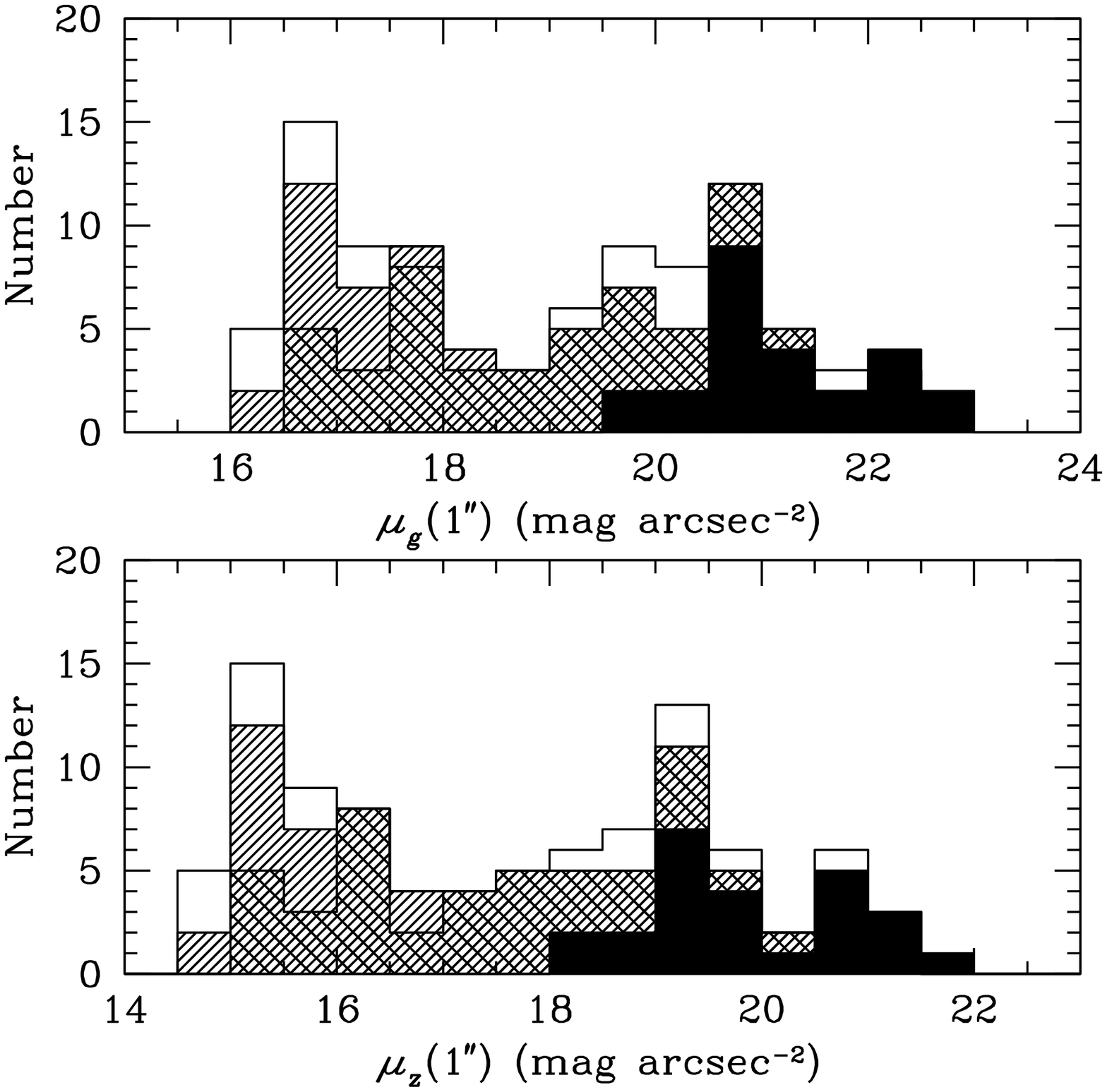}
\caption{{\it (Upper Panel)} Distribution of $g$-band surface brightnesses, measured
at a geometric mean radius of 1\arcsec, for the 94 galaxies from the ACS Virgo Cluster 
Survey for which a classification as nucleated or non-nucleated is possible (open 
histogram).  The hatched histogram shows the luminosity distribution for the
62 galaxies which we classify unambiguously as nucleated (i.e., Types Ia and Ib).
The double-hatched histogram shows this same sample plus the 15 galaxies which
{\it may} have central nuclei (i.e., Types Ia, Ib, Ic or Id).
The 25 ACS Virgo Cluster Survey galaxies classified as nucleated in Binggeli,
Sandage \& Tammann (1985) are shown by the filled histogram.
{\it (Lower Panel)} Same as above, except for the $z$ band.
\label{fig07}}
\end{figure}

\clearpage
\begin{figure}
\figurenum{8}
\plotone{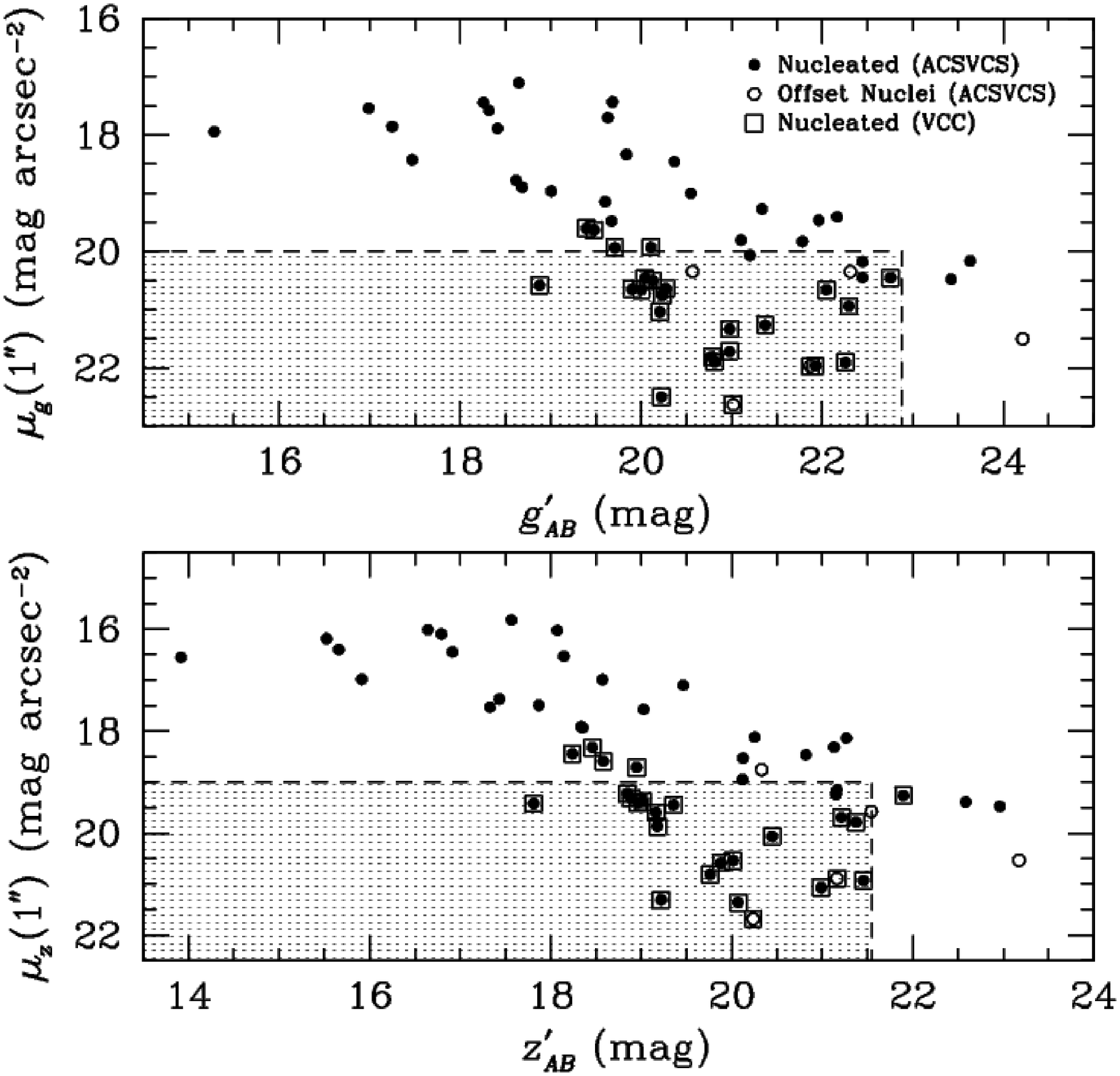}
\caption{{\it (Upper Panel)}: Galaxy $g$-band surface brightness, measured at a geometric
mean radius of 1$\arcsec$, plotted against the dereddened magnitude of the nucleus.
Filled circles represent the 51 galaxies in our survey which show unambiguous evidence
for a nucleus at or near their photocenters, and for which we are able to measure
reliable photometric and structural parameters (i.e., Type Ia galaxies). Open
circles show the five galaxies with possible offset nuclei (i.e., Type Ie galaxies).
Open squares show the 25 galaxies classified as nucleated by
Binggeli, Sandage \& Tammann (1985). The shaded region
shows the approximate region of the magnitude vs. surface brightness plane 
where selection effects were relatively unimportant in the
survey of Binggeli, Tammann \& Sandage (1987).
{\it (Lower Panel)}: Same as above, except for the $z$ band.
\label{fig08}}
\end{figure}

\clearpage
\begin{figure}
\figurenum{9}
\plotone{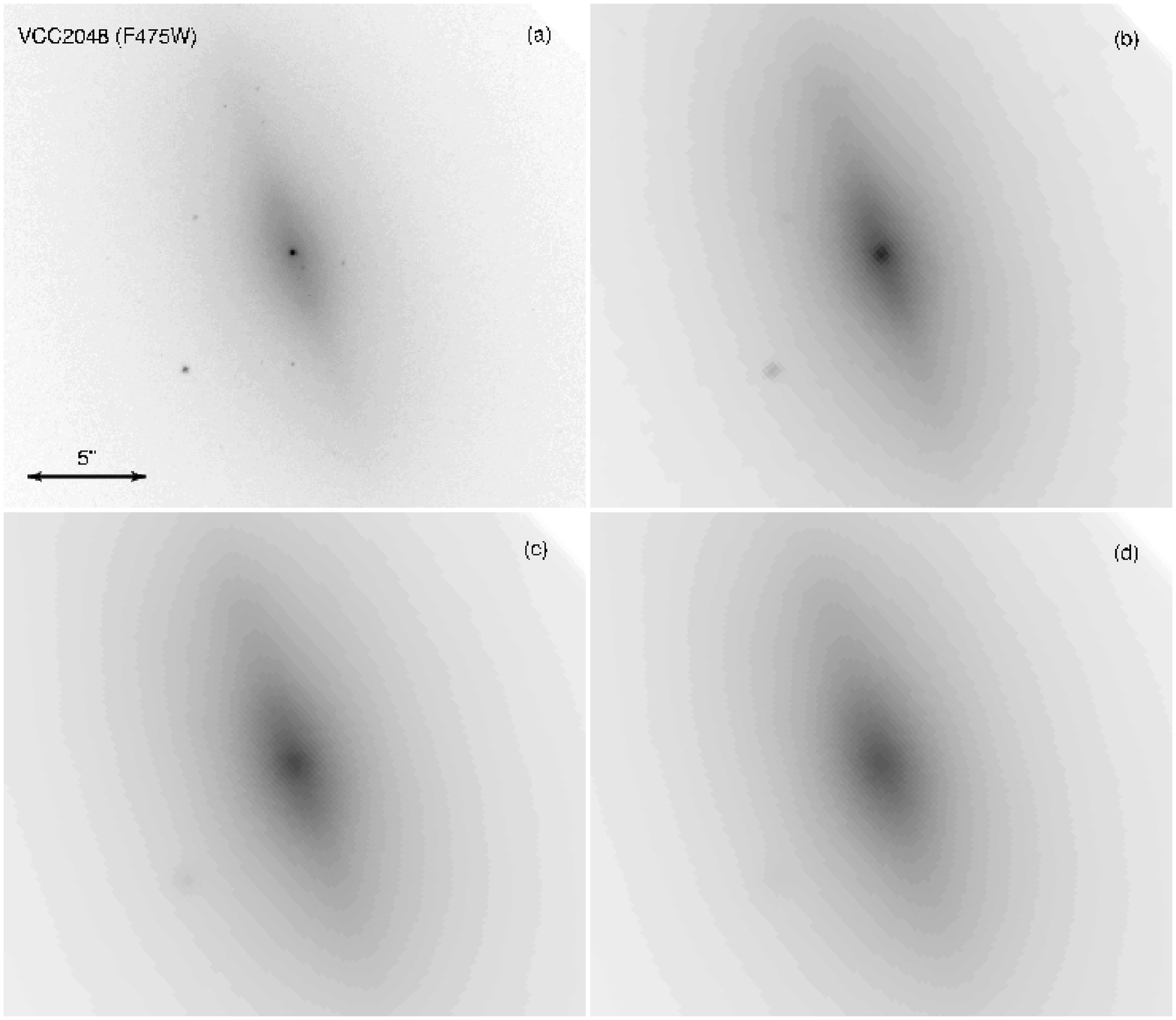}
\caption{{\it (Panel a)}: A magnified view of the F475W image of VCC2048 (IC3773). 
This is the brightest dwarf galaxy in our sample according to the morphological
classifications of Binggeli, Sandage and Tamman (1985). It is
classified as non-nucleated in the Virgo Cluster Catalog, with type d:S0(9).
{\it (Panels b--d)}: The same image, after binning 4$\times$4 pixels
and convolving with Gaussians of FWHM = 0\farcs5, 0\farcs9
and 1\farcs4, respectively.
\label{fig09}}
\end{figure}

\clearpage
\begin{figure}
\figurenum{10}
\plottwo{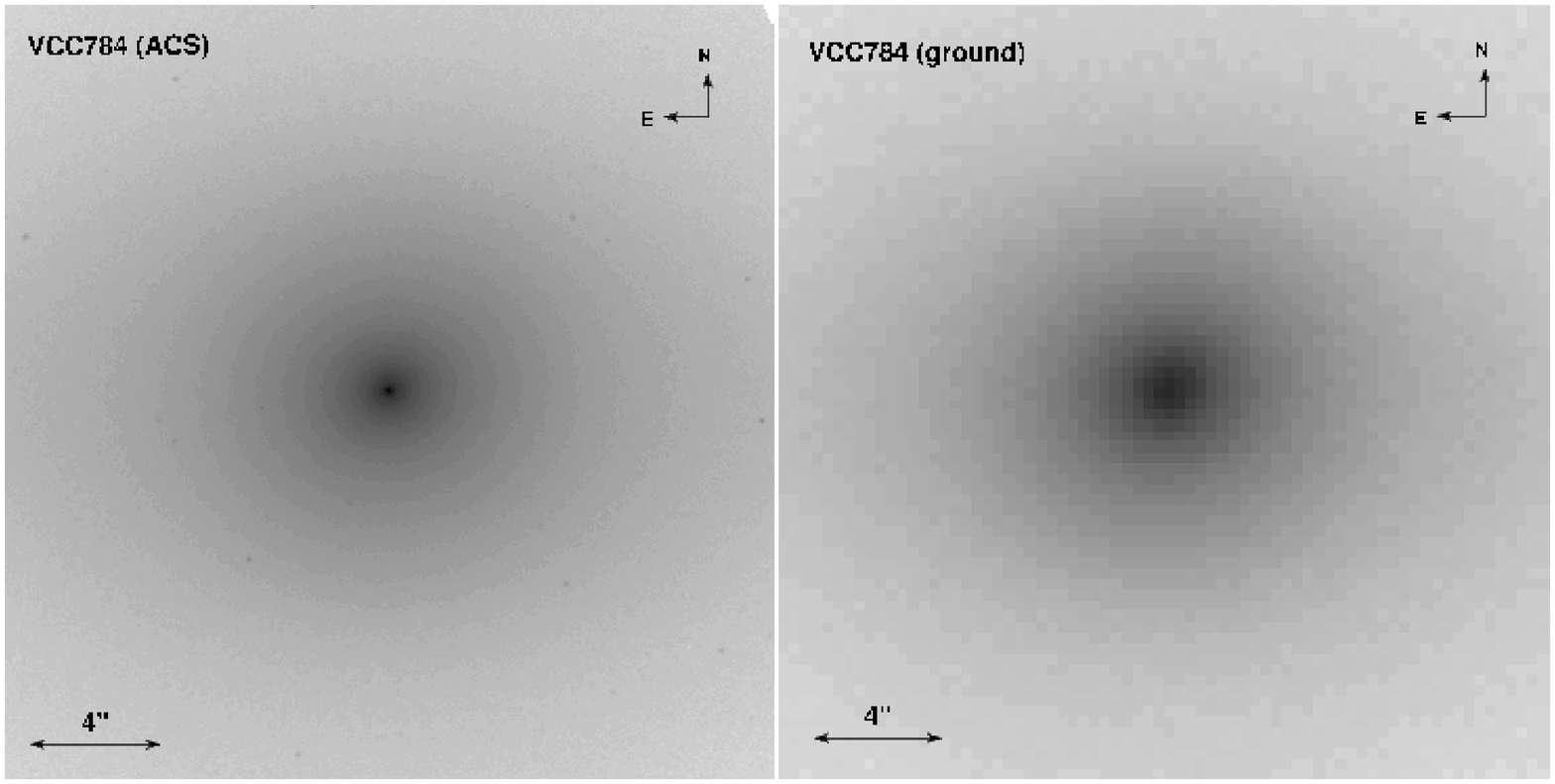}{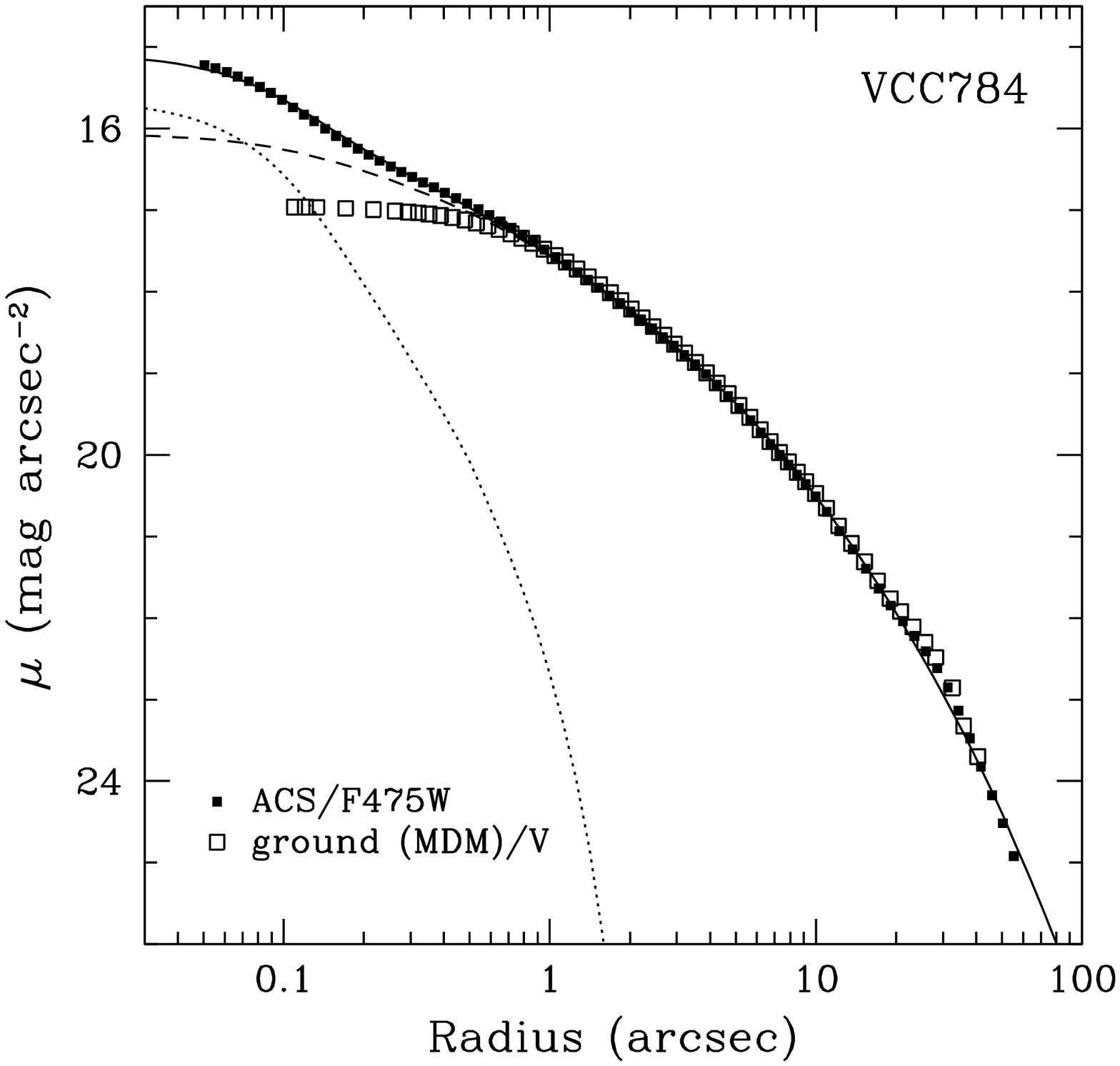}
\caption{{\it (First Panel)} F475W ($g$) image for VCC784 (NGC4379), one of the brightest Type Ia
galaxies in our survey. 
{\it (Second Panel)} Ground-based $V$-band image of VCC784, taken with the 2.4m
Hiltner telescope (FWHM = 1\farcs14).
{\it (Third Panel)} Comparison of surface brightness profiles measured for VCC784
using the ACS (filled squares) and ground-based images (open squares). To aid in
the comparison, the $V$-band profile has been matched to the ACS profile at a
radius of 1\arcsec. The best-fit two-component model for the galaxy and nucleus 
based on the ACS profile are shown by the dashed and dotted curves. The solid
curve shows the combined profile.
\label{fig10}}
\end{figure}

\clearpage
\begin{figure}
\figurenum{11}
\plotone{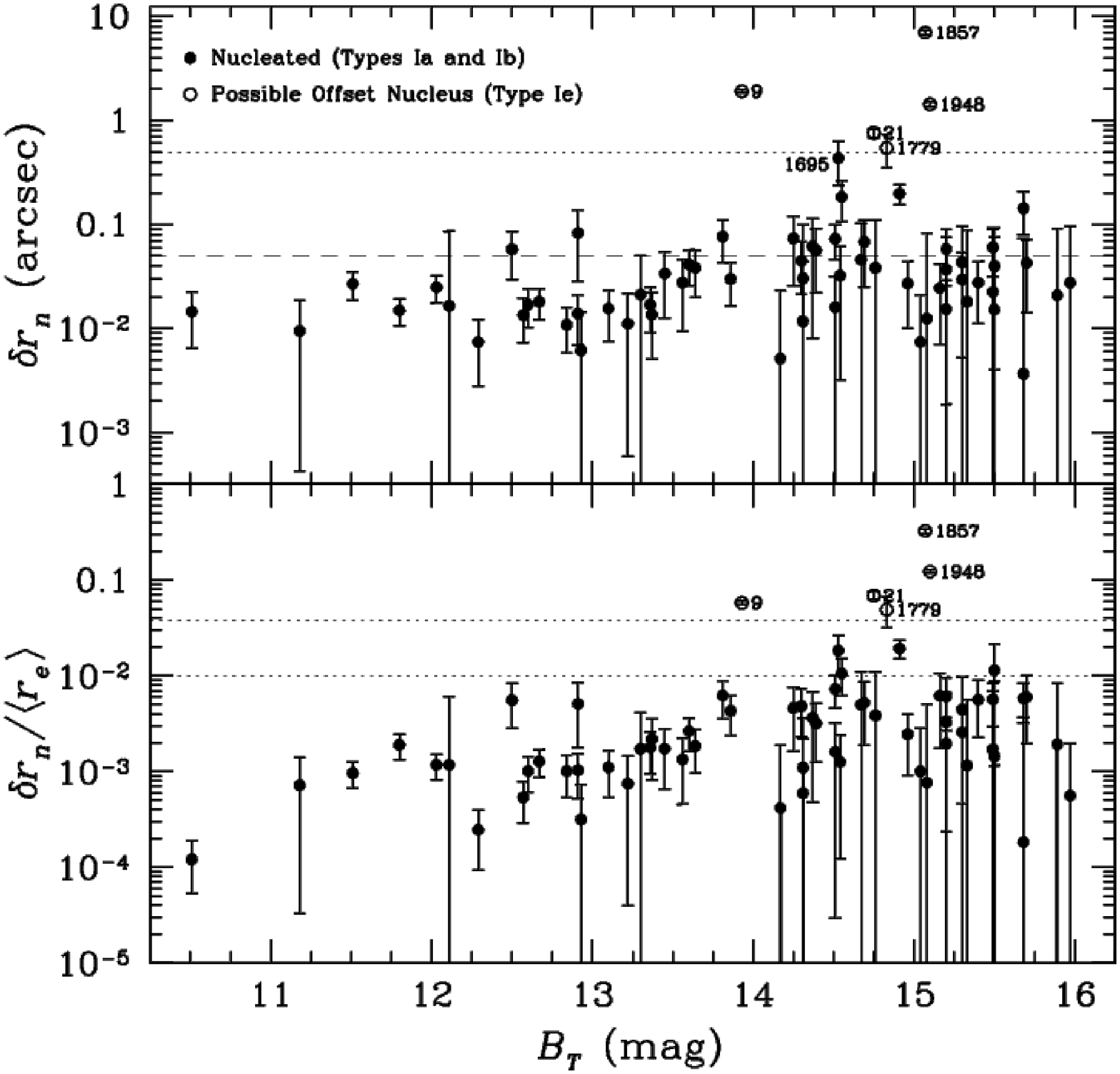}
\caption{{\it (Upper Panel)}: Projected offset, $\delta{r_n}$, between the
position of the galaxy center and that of the nucleus, plotted as a function of
galaxy magnitude. Filled circles show 62 galaxies of Types Ia and Ib. Open 
circles indicate the five galaxies in our sample which may have offset nuclei
(Type~Ie). The dotted line shows an offset of $\delta{r_n} =$ 0\farcs5, or
ten ACS/WFC pixels; a single pixel is represented by the lower, dashed line.
{\it (Lower Panel)} Ratio of the offset to the mean effective radius,
$\langle r_{e} \rangle$, of the host galaxy in the two bands, plotted as a function
of galaxy magnitude. The symbols are the same as in the previous panel. The
upper dotted line in this case is drawn at
$\delta{r_n} / \langle r_e \rangle = 0\farcs5 / 13\farcs13 \approx 0.038$,
where $\langle r_e \rangle = 13\farcs13$ is the mean effective radius of
Type Ia galaxies. The lower dotted line is drawn at a fractional offset of
1\% the effective radius. A total of nine galaxies lie above
this lower line.
\label{fig11}}
\end{figure}

\clearpage
\begin{figure}
\figurenum{12}
\plotone{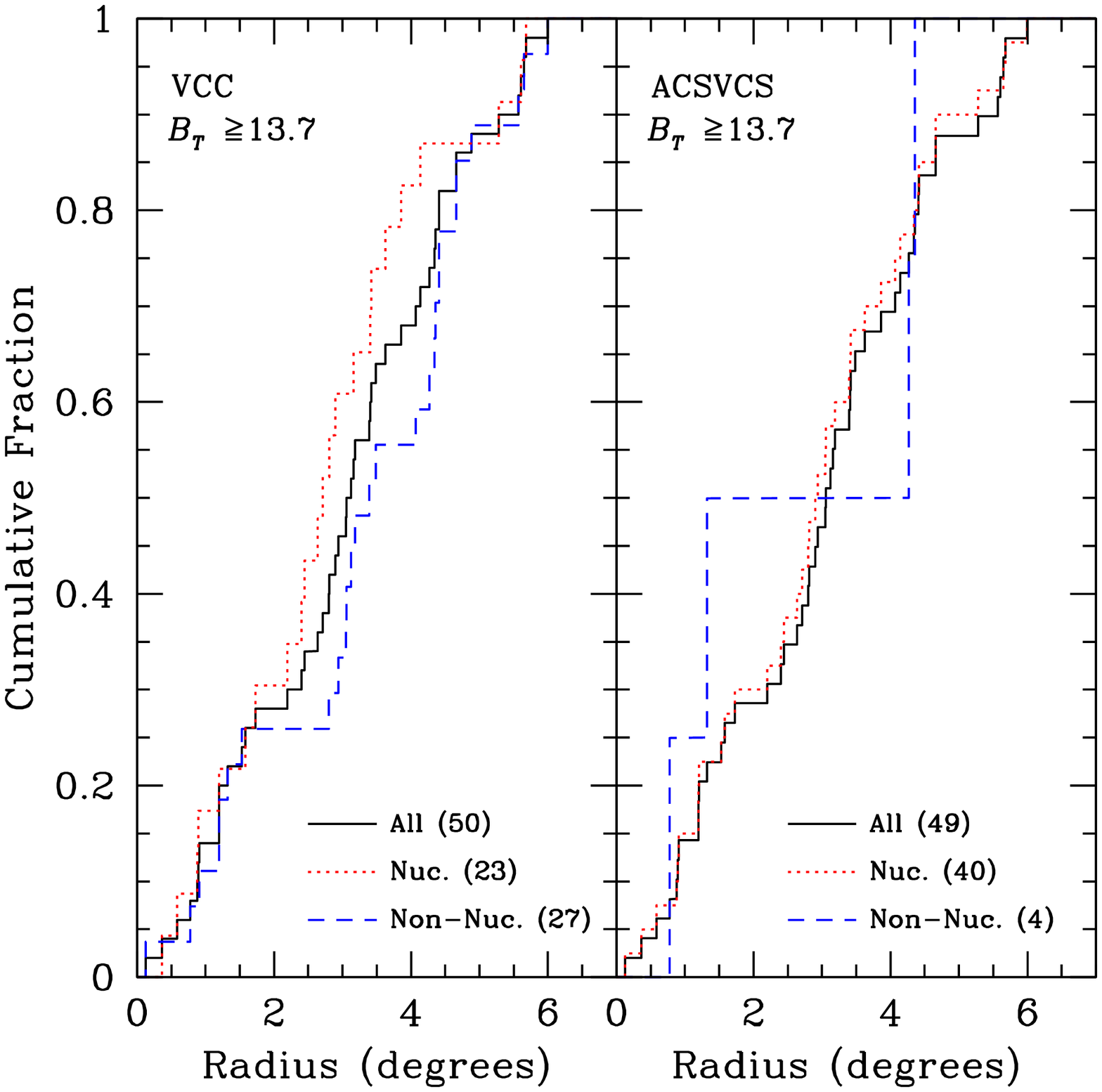}
\caption{{\it (Left Panel)}: Cumulative distribution of Virgocentric radii for
the 50 galaxies from the ACS Virgo Cluster Survey fainter than $B_{T}=13.7$. 
Using the morphological classifications from Binggeli, Sandage \& Tammann (1985), this
sample is sub-divided into 23 nucleated and 27 non-nucleated galaxies.
The dotted and dashed curves show the cumulative distributions for these
two samples.
{\it (Right Panel)} Cumulative distribution of Virgocentric radii for
the 49 galaxies from the ACS Virgo Cluster Survey fainter than $B_{T}=13.7$
for which a classification as nucleated or non-nucleated is possible
from our ACS images. (One galaxy in this magnitude range, VCC571, cannot
be classified due to the
presence of dust.) This sample is shown sub-divided into 40 nucleated
(Type Ia and Ib) and 4 non-nucleated (Type II) galaxies (dotted and
dashed curves, respectively). The five galaxies
with possible offset nuclei have been excluded from the analysis.
\label{fig12}}
\end{figure}

\clearpage
\begin{figure}
\figurenum{13}
\plotone{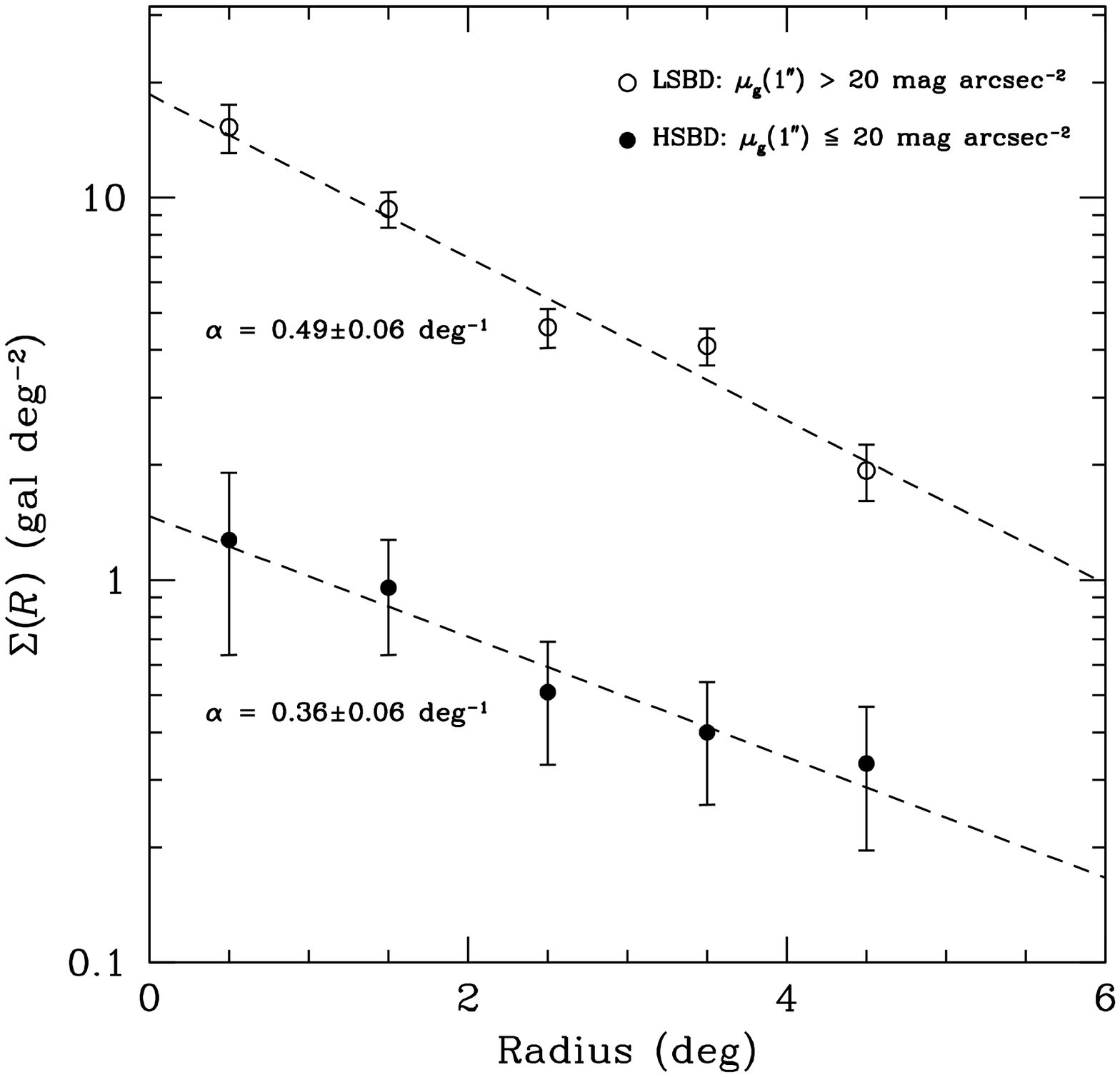}
\caption{Surface density profile for early-type, low-surface-brightness dwarf
(LSBD) galaxies in the Virgo Cluster (open circles). These galaxies have 
blue magnitudes in the range $14.55 < B_T \le 18.0$, which correspond to
central $g$-band surface brightnesses of
$\mu_g(1\arcsec) \gtrsim 20$~mag~arcsec$^{-2}$.
High-surface-brightness dwarfs (HSBDs) are defined as early-type galaxies
with $13.7 \le B_T \le 14.55$ and $\mu_g(1\arcsec) \lesssim 20$~mag~arcsec$^{-2}$
(filled circles).
Nuclei in the HSBDs were preferentially missed in the survey of Binggeli,
Tamman \& Sandage (1987), which may explain their observation that the
non-nucleated dwarfs in Virgo have a more extended distribution within
the cluster than the nucleated dwarfs.
\label{fig13}}
\end{figure}

\clearpage
\begin{figure}
\figurenum{14}
\plotone{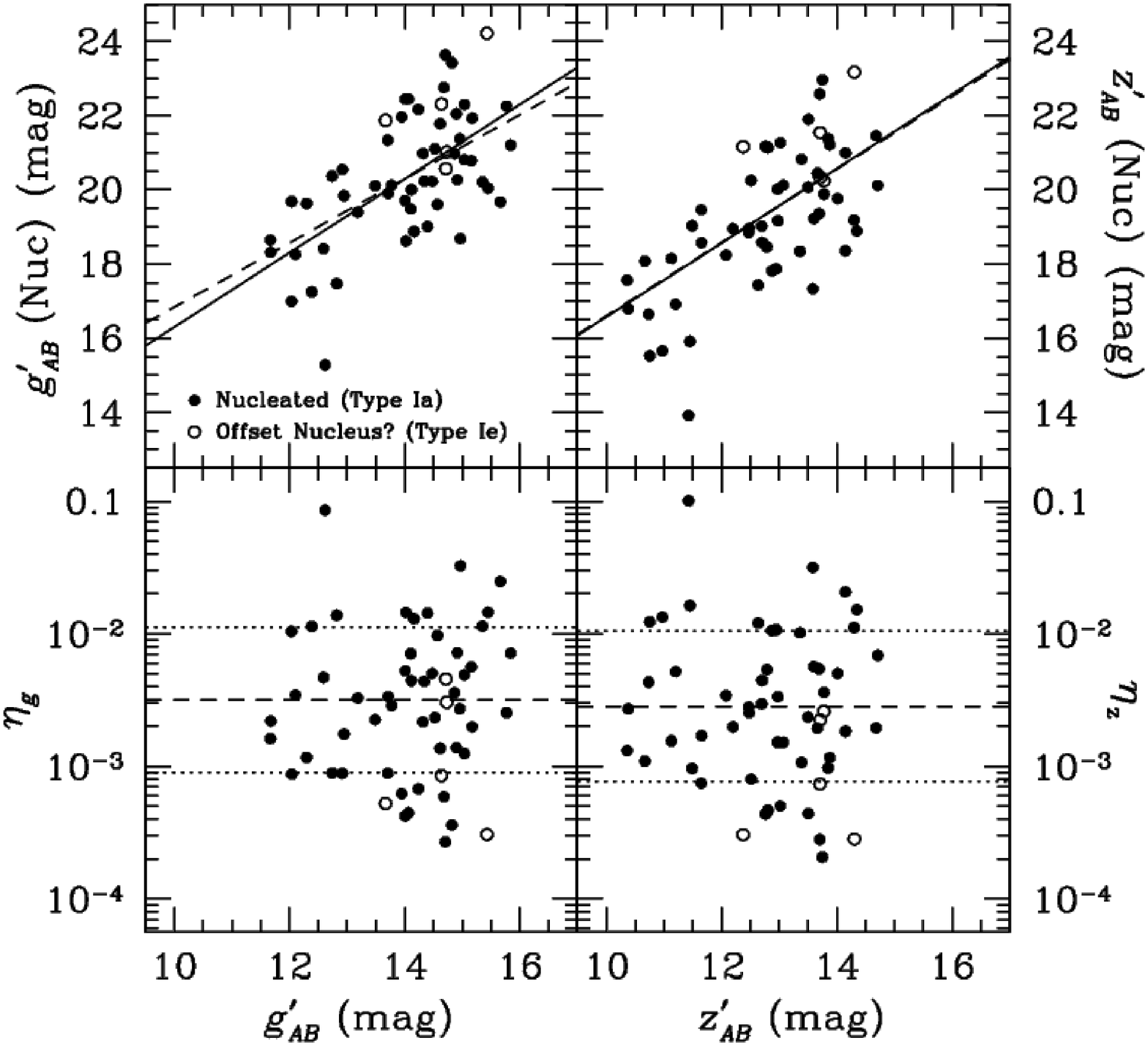}
\caption{{\it (Upper Panels)}: Nucleus magnitude plotted against
against that of the host galaxy. Results for the $g$ and
$z$ bands are shown in the left and right panels, respectively. Filled circles show
the 51 galaxies of Type Ia. Open circles indicate the five
galaxies in our sample which may have offset nuclei (Type Ie).
The dashed line shows the best-fit linear relation while the best-fit
relation with unity slope is shown by the solid line.
{\it (Lower Panels)}: Ratio of nucleus luminosity to that of the host galaxy, 
$\eta$, plotted against galaxy magnitude.  Results for the $g$ and
$z$ bands are shown in the left and right panels, respectively.
Filled circles show the 51 galaxies of Types Ia. Open
circles indicate the five galaxies in our sample which may
have offset nuclei (Type Ie). The dashed and dotted lines
show the mean and $\pm$1$\sigma$ limits for
the sample of Type~Ia galaxies.
\label{fig14}}
\end{figure}

\clearpage
\begin{figure}
\figurenum{15}
\plotone{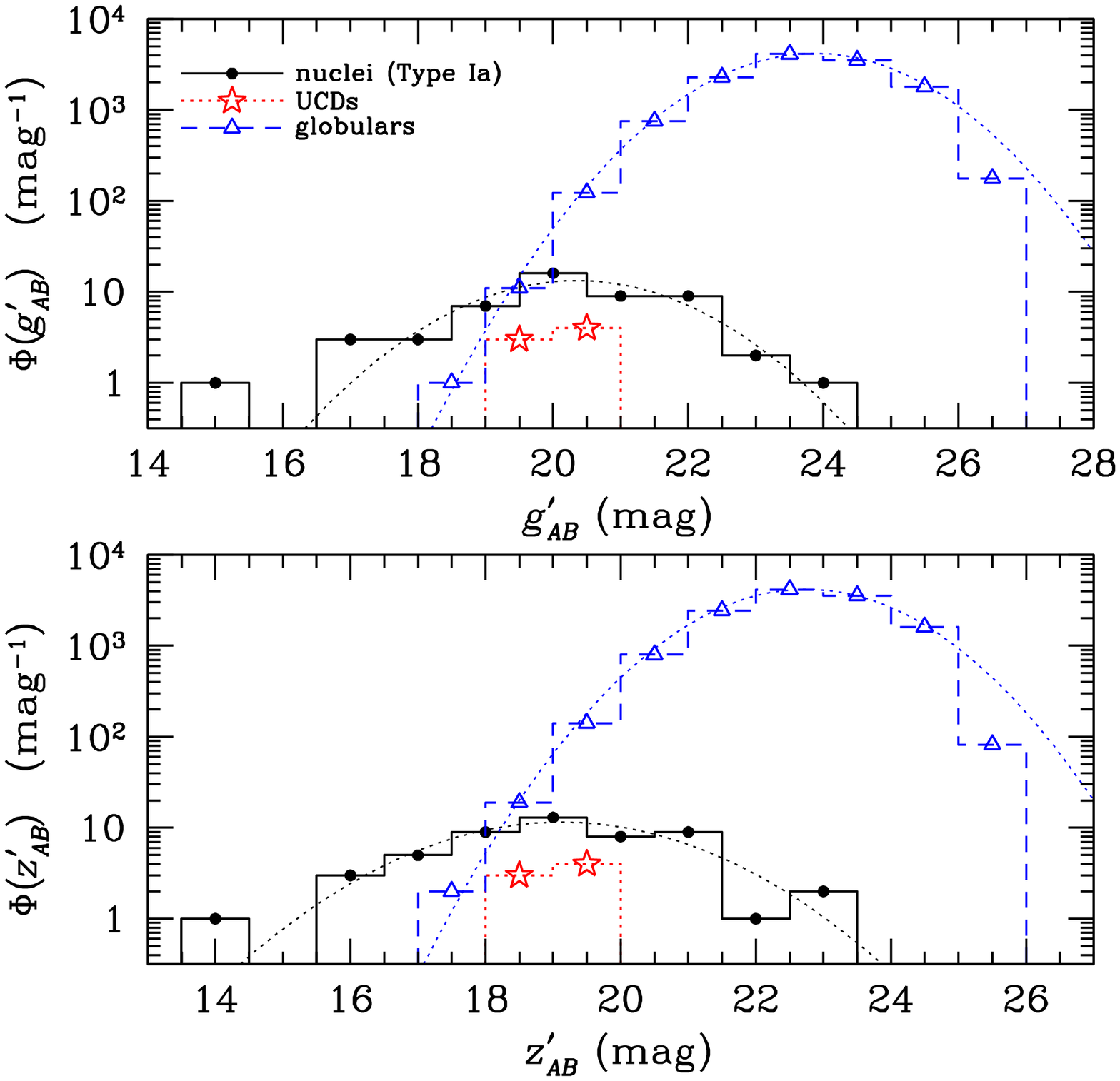}
\caption{{\it (Upper Panel)}: Luminosity function, measured in the $g$ band,
for the nuclei of the 51 galaxies classified as Type Ia. For comparison, we 
also show Virgo globular cluster candidates from Jord\'an \etal\ (2006) and
seven probable ultra-compact dwarf galaxies (UCDs) from
Ha\c{s}egan \etal~(2005; 2006).
{\it (Lower Panel)}: Same as above, except for the $z$ band.
\label{fig15}}
\end{figure}

\clearpage
\begin{figure}
\figurenum{16}
\plotone{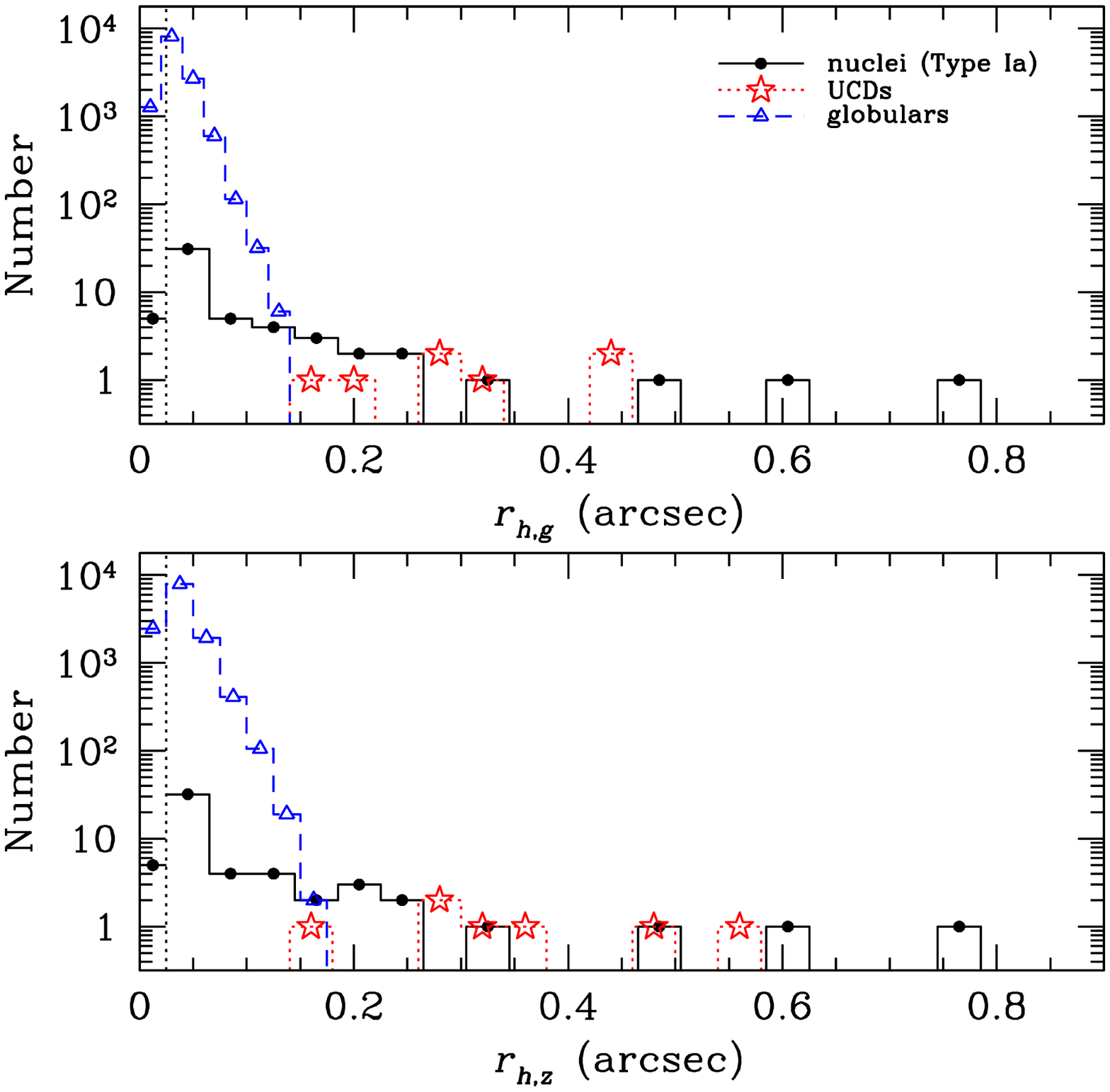}
\caption{{\it (Upper Panel)}: Distribution of half-light radii, measured in 
the $g$-band, for the nuclei of the 51 galaxies classified as Type Ia. Also
shown are seven probable ultra-compact dwarf galaxies (UCDs)
from Ha\c{s}egan \etal~(2005; 2006) and a sample
of Virgo globular cluster candidates from Jord\'an \etal\ (2006). The vertical
dotted line shows the approximate resolution limit for the ACS images drizzled
using a {\it Gaussian} interpolation kernel.
{\it (Lower Panel)}: Same as above, except for the $z$ band.
\label{fig16}}
\end{figure}

\clearpage
\begin{figure}
\figurenum{17}
\plotone{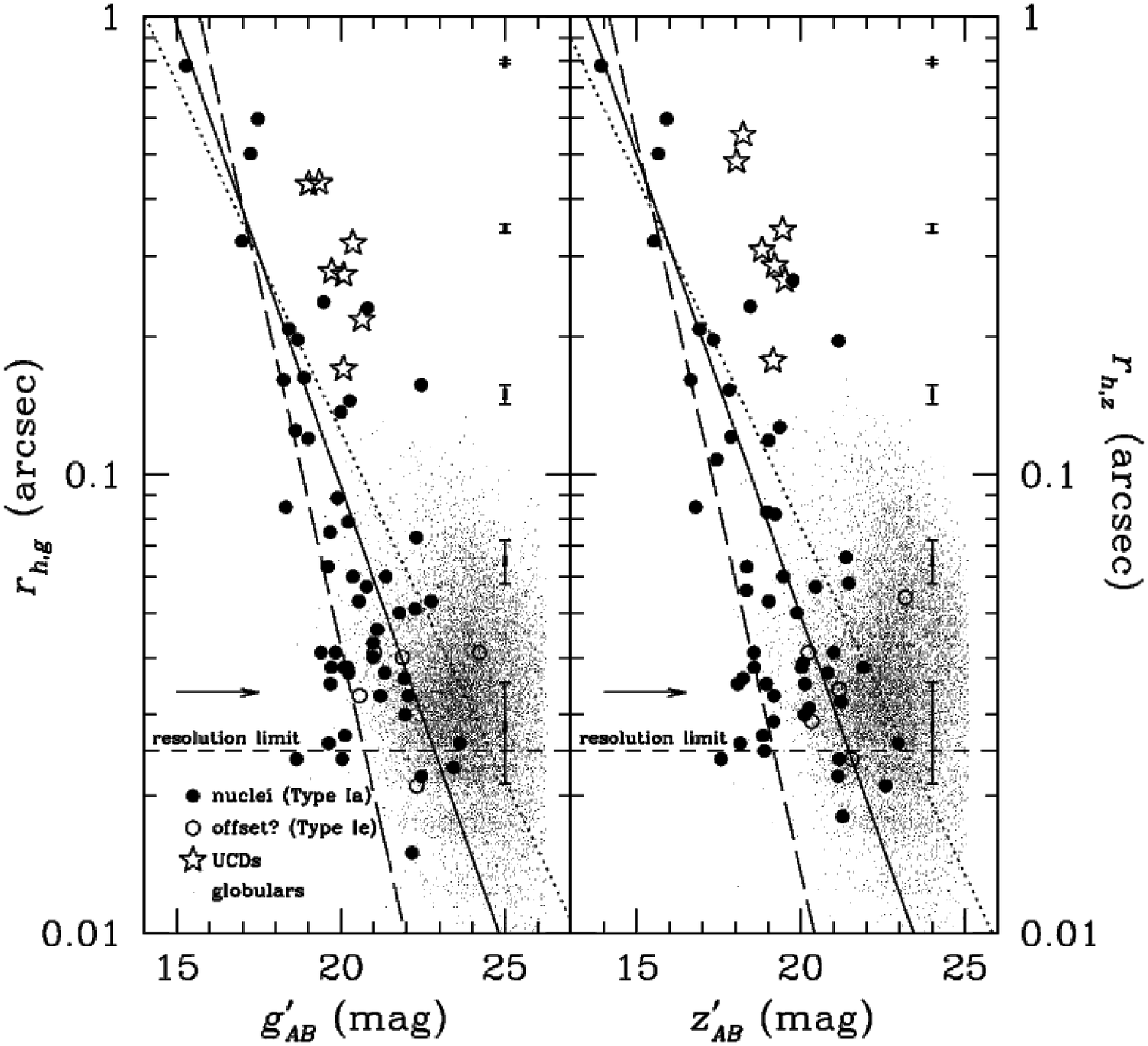}
\caption{{\it (Left Panel)}: The size-magnitude relation, in the $g$ band, for
the nuclei 
of the 51 galaxies classified as Type Ia (filled circles). Open circles
indicate the five galaxies which may have offset nuclei (Type Ie). Typical
errorbars for the nuclei are shown in the right had side of the panel.
Also shown are seven probable ultra-compact
dwarf galaxies (UCDs) from Ha\c{s}egan \etal~ (2005; 2006) and a sample of 
globular clusters from Jord\'an \etal\ (2006). The arrow shows
the ``universal" half-light radius of $\langle r_h  \rangle = 0\farcs033 \approx 2.7$~pc for
globular clusters in Virgo (Jord\'an \etal\ 2005), while the dashed line
shows a conservative estimate for the resolution limit of those ACS images 
which were drizzled using a Gaussian interpolation kernel (rather than
a Lanczos kernel).
The diagonal lines shows relations of the form $r_h \propto {\cal L}^{\beta}$.
The long-dashed line shows the extrapolation of the size-luminosity
relation for giant galaxies, from Equation (11) of Ha\c{s}egan \etal~(2005).
See text for details.
{\it (Right Panel)}: Same as the left panel, except for the $z$ band.
\label{fig17}}
\end{figure}

\clearpage
\begin{figure}
\figurenum{18}
\plotone{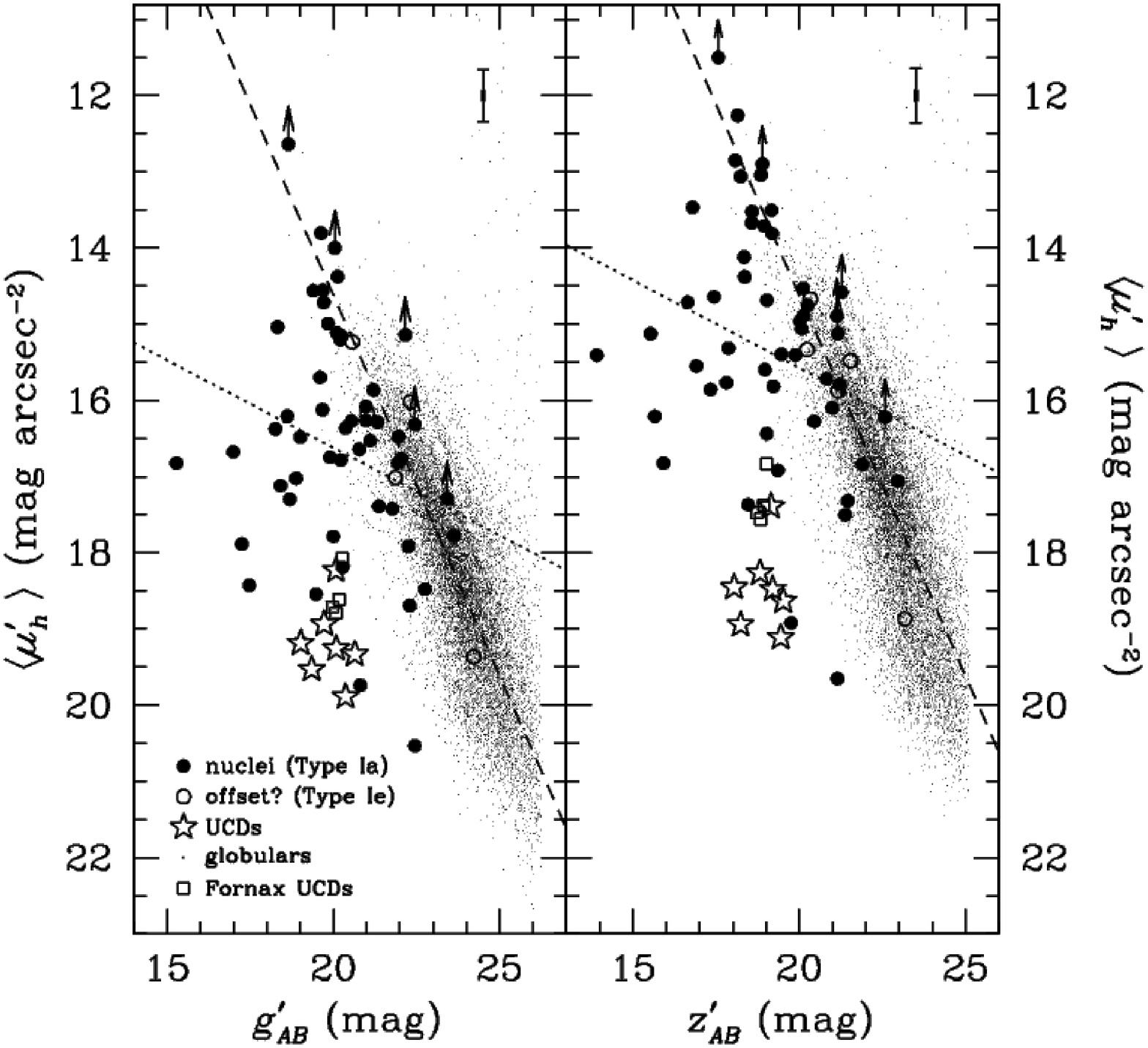}
\caption{{\it (Left Panel)}: Average surface brightness within the
half-light radius plotted against magnitude for the nuclei of the 51 galaxies
classified as Type Ia (filled circles). Open circles indicate the five galaxies 
which may have offset nuclei (Type Ie). Both the magnitude and
surface brightness measurements refer to the $g$ band. Also shown are seven
probable ultra-compact dwarf galaxies (UCDs) from Ha\c{s}egan \etal~(2005; 2006),
a sample of Virgo globular clusters candidates from Jord\'an \etal\ (2006) and
four UCDs in the Fornax Cluster studied by de Propris et al. (2005). The dashed
line shows the relation expected for globular clusters if they have a 
``universal" half-light radius of
$\langle r_h  \rangle = 0\farcs033 \approx 2.7$~pc (Jord\'an et~al. 2005). 
The dotted line shows the predicted scaling relation for nuclei formed by the
mergers of globular clusters (Bekki \etal\ 2004). Arrows indicate lower
limits on the surface brightness for unresolved nuclei.
{\it (Right Panel)}: Same as the left panel, except for the $z$ band.
\label{fig18}}
\end{figure}

\clearpage
\begin{figure}
\figurenum{19}
\plotone{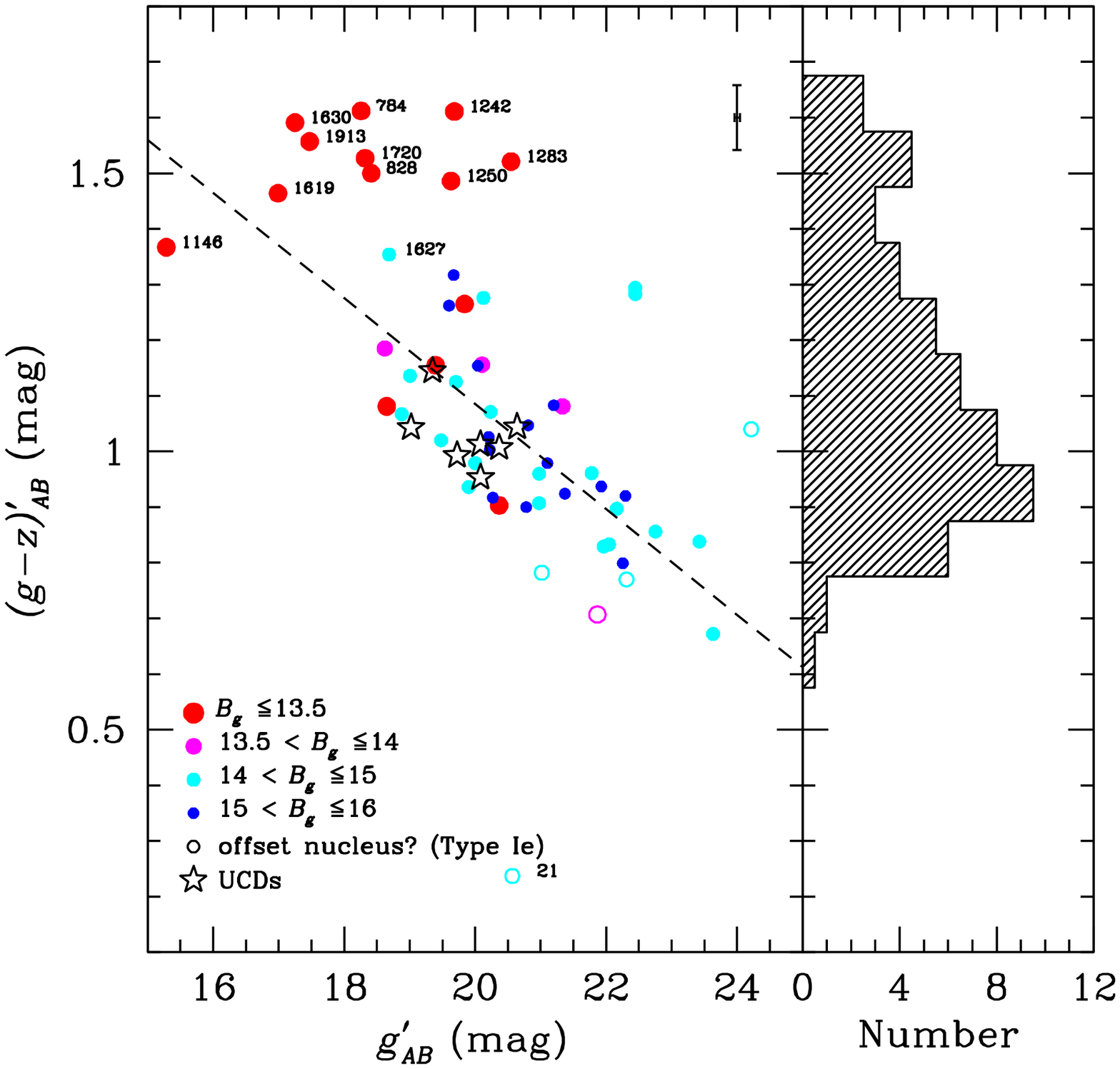}
\caption{{\it (Left Panel)} Color-magnitude diagram for the nuclei of the 51 galaxies
classified as Type~Ia (filled circles). Open circles indicate the five 
galaxies which may have offset nuclei (Type~Ie).
Magnitudes and colors have been corrected for extinction and
reddening. The size of the symbol for the nuclei is proportional to the
magnitude of the host galaxy. Galaxies with unusually blue or red nuclei
are labelled. For comparison, we show seven probable ultra-compact
dwarf galaxies from Ha\c{s}egan \etal~(2005; 2006). The short-dashed
line shows the best-fit relation for the nuclei of galaxies fainter
than $B_T = 13.5$ (see \S\ref{sec:col}). The dashed line shows the
the best-fit relation for the nuclei of galaxies fainter
than $B_T = 13.5$ (see \S\ref{sec:col}).
{\it (Right Panel)} Histogram of de-reddened nuclei colors.
\label{fig19}}
\end{figure}

\clearpage
\begin{figure}
\figurenum{20}
\plotone{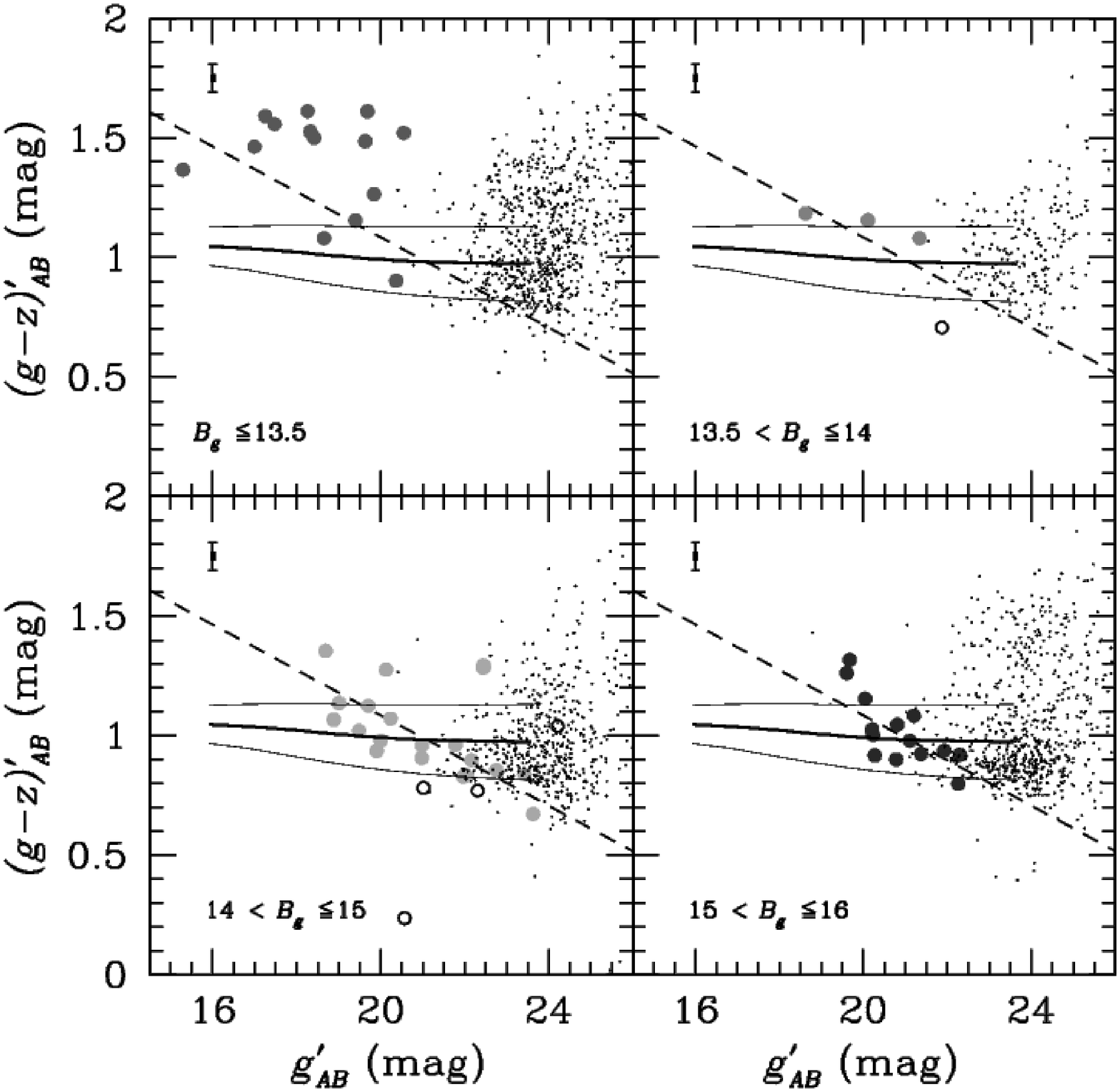}
\caption{Comparison of the color magnitude diagrams for the nuclei of the
51 galaxies classified as Type~Ia (filled circles) with those of the globular
clusters belonging to our program galaxies (points).
Possible offset nuclei are shown as open circles (i.e., Type~Ie galaxies).
The samples have been divided into four panels based on the blue luminosity
of the host galaxy. The dashed line in each panel shows
the best-fit relation for the nuclei of galaxies fainter
than $B_T = 13.5$ (see \S\ref{sec:col}). The heavy solid curve shows the
color magnitude relation predicted by Monte-Carlo experiments in which
the nuclei are assembled from globular cluster mergers; the thin solid
curves show the 95\% confidence limits (see \S\ref{sec:discussion} for details).
\label{fig20}}
\end{figure}

\clearpage
\begin{figure}
\figurenum{21}
\plotone{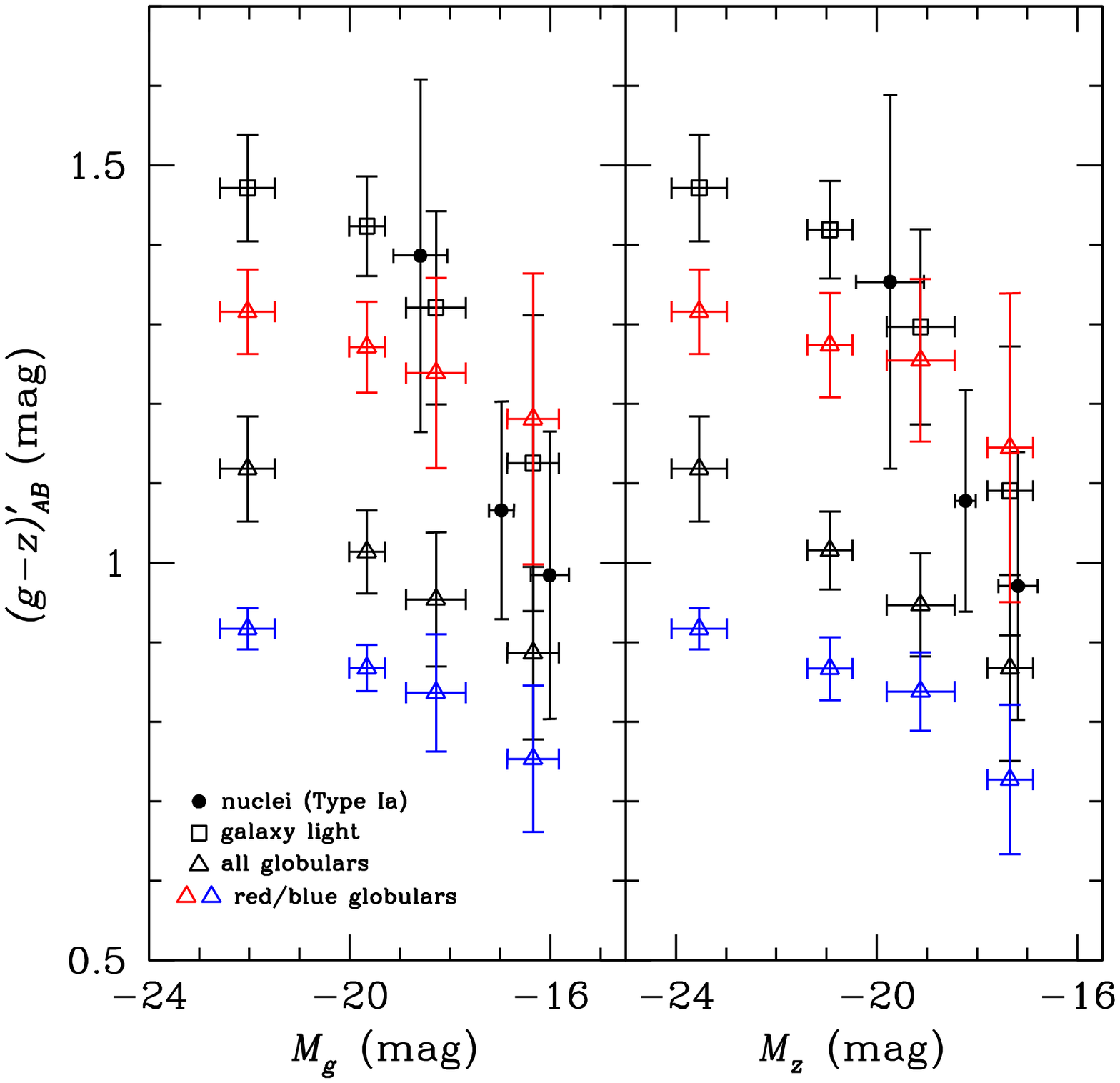}
\caption{{\it (Left Panel)} Comparison of the $M_g$ vs. $(g-z)_{AB}^{\prime}$ color
magnitude relation for nuclei (filled circles), globular clusters
(open triangles) and galaxies (open squares) from the ACS Virgo Cluster Survey.
The galaxies have been divided into four bins in absolute magnitude.
The globular clusters have been further divided into red and blue subcomponents
as described in Peng \etal\ (2006a).
The nuclei of the 51 Type~Ia galaxies have been divided into
three magnitude bins containing roughly equal numbers of nuclei.
{\it (Right Panel)} $M_z$ vs. $(g-z)_{AB}^{\prime}$ color
magnitude relation for nuclei, globular clusters and galaxies.
\label{fig21}}
\end{figure}

\clearpage
\begin{figure}
\figurenum{22}
\plotone{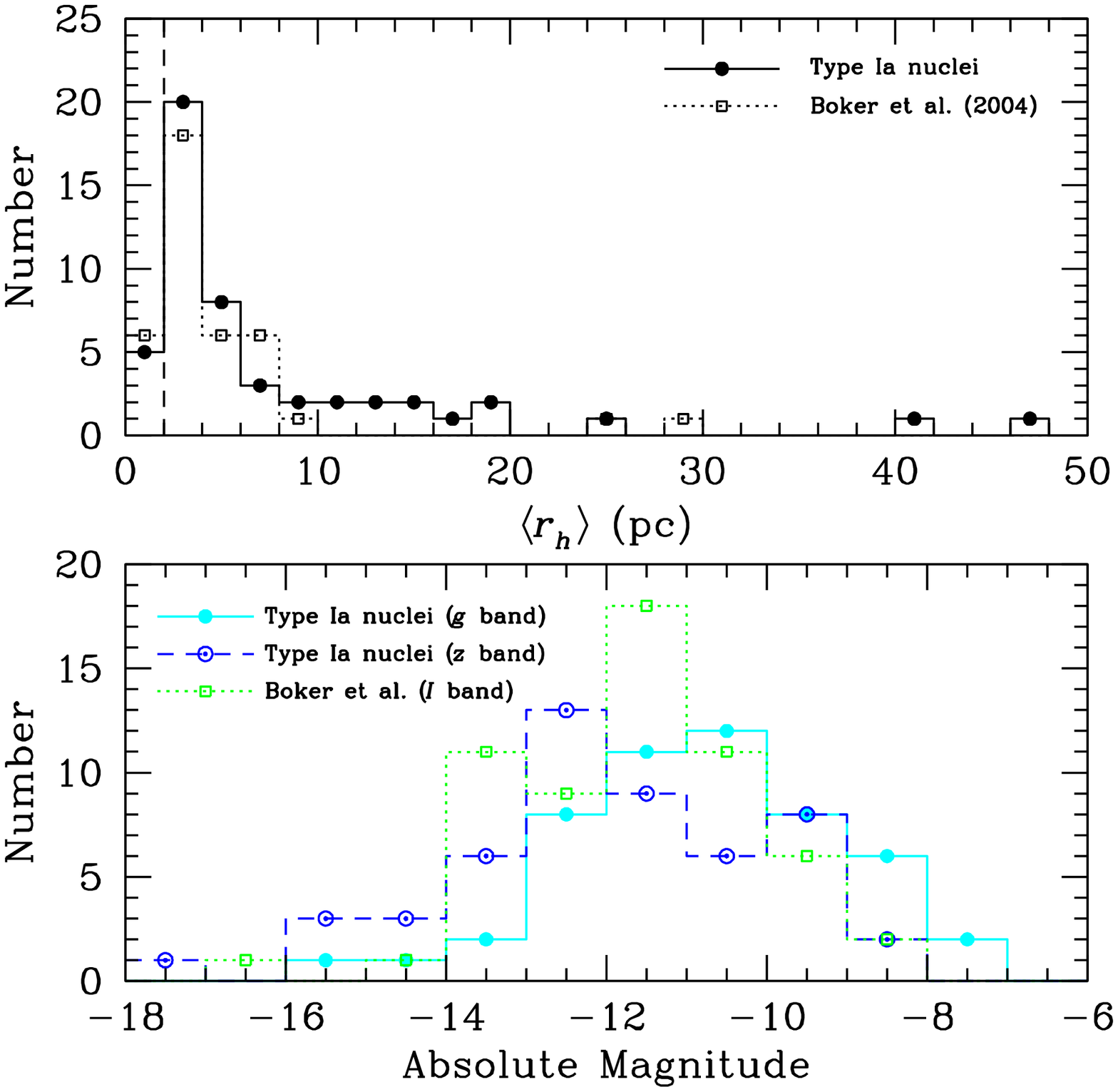}
\caption{{\it (Upper Panel)} Distribution of half-light radii for the nuclei of early-type
galaxies from the ACS Virgo Cluster Survey, compared to those of late-type galaxies
from B\"oker \etal\ (2004). The plotted half-light radii for the early-type galaxies 
are averages of the measurements in the $g$ and $z$ bands.
{\it (Lower Panel)} Distribution of absolute magnitudes for the nuclei of early-type
galaxies from the ACS Virgo Cluster Survey, compared to those of late-type galaxies
from B\"oker \etal\ (2004). Results in both the $g$ and $z$ bands are shown 
for the early-type galaxies, while those for the late-type galaxies refer to the
$I$ band.
\label{fig22}}
\end{figure}

\clearpage
\begin{figure}
\figurenum{23}
\plotone{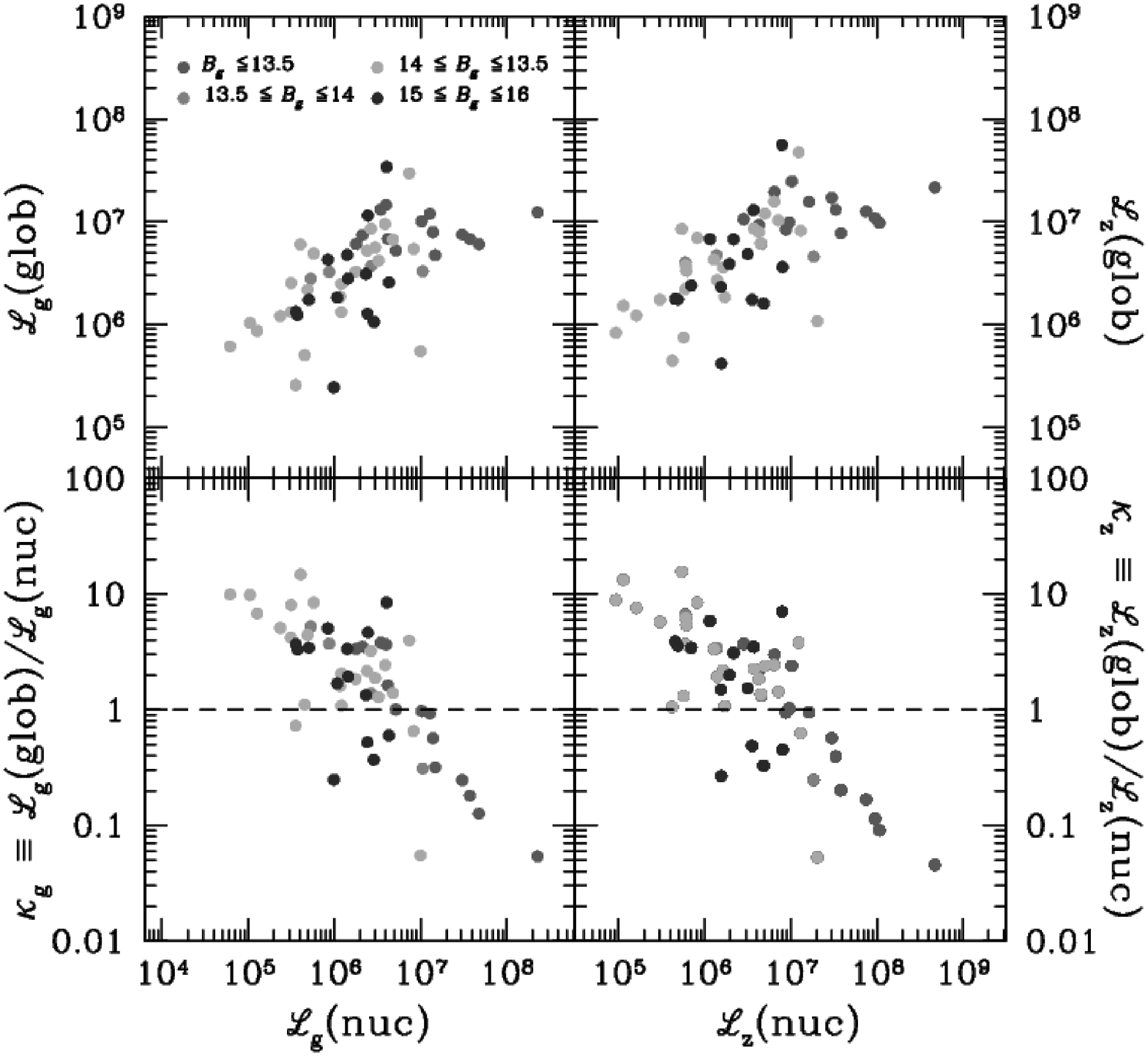}
\caption{{\it (Upper Left Panel)} Total $g$-band luminosity of globular clusters
belonging to 51 nucleated (Type~Ia) galaxies, plotted against $g$-band nucleus
luminosity. Symbols are color coded according to the blue magnitude of host
galaxy. 
{\it (Lower Left Panel)} Ratio of total $g$-band luminosity in globular clusters
to that contained in the nucleus. The dashed line corresponds to $\kappa_g =1$.
{\it (Right Panels)} As before, except for the $z$ band.
\label{fig23}}
\end{figure}

\clearpage
\begin{figure}
\figurenum{24}
\plotone{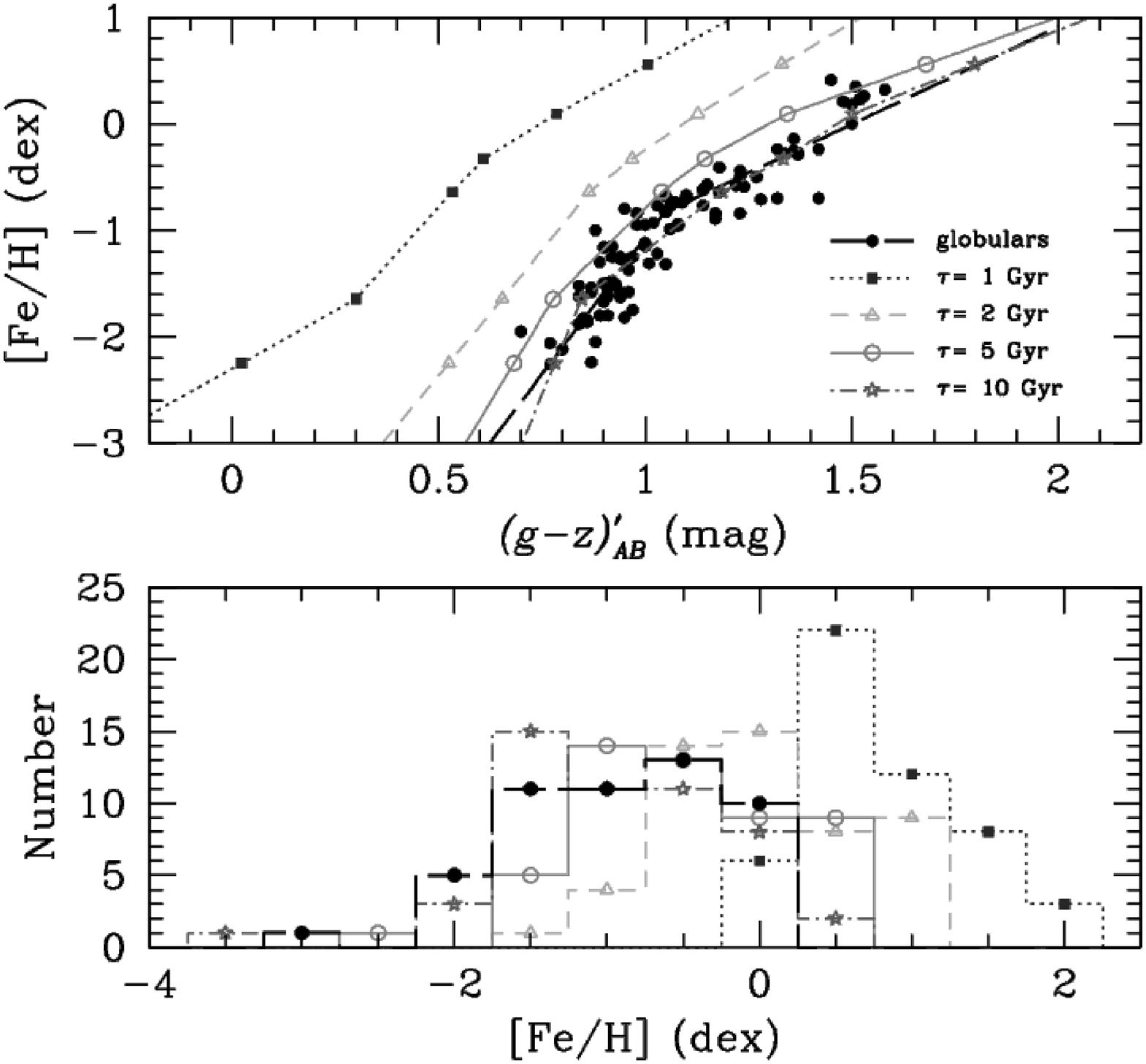}
\caption{{\it (Upper Panel)} Color-metallicity relations for old stellar populations.
The long-dashed (broken linear) relation shows the empirical metallicity relation
of Peng \etal\ (2006 = Paper IX) which is based on 95 globular clusters (asterisks)
in the Milky Way, M49 (NGC4472 = VCC1226) and M87 (NGC4486 = VCC1316). The remaining
relations show theoretical predictions based on the models of Bruzual \&
Charlot (2003) which assume simple stellar populations with a Chabrier (2003) 
initial mass function and ages of $\tau =$ 1, 2, 5 and 10 Gyr. The curves
show linear interpolations of the models. 
{\it (Lower Panel)} Histograms of metallicities for the 51 Type Ia nuclei in our
sample, based on the color metallicity relations shown in the upper panel. The
symbols are the same as above.
\label{fig24}}
\end{figure}

\clearpage
\begin{figure}
\figurenum{25}
\plotone{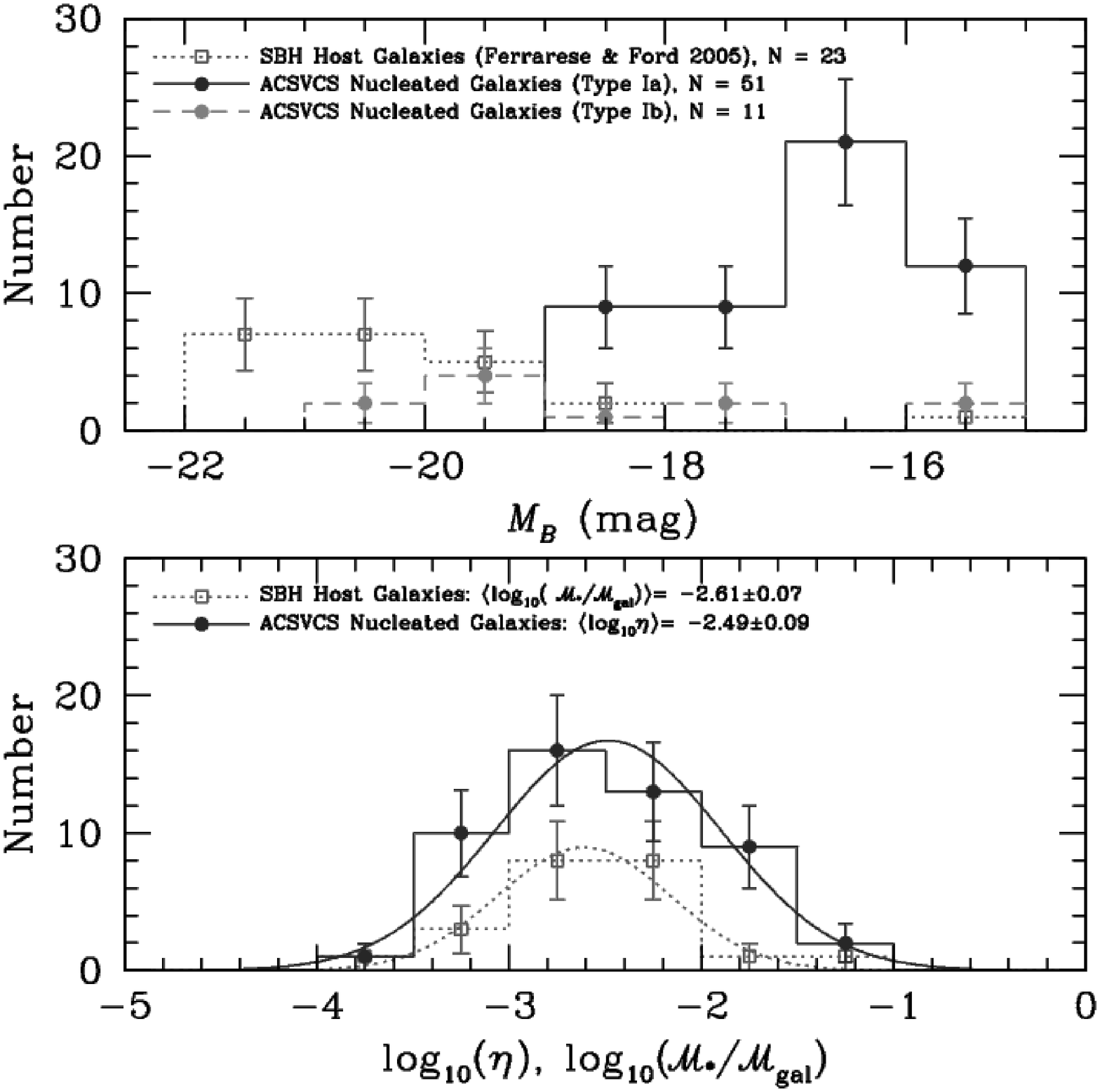}
\caption{{\it (Upper Panel)} Distribution of absolute blue magnitudes for
23 early-type galaxies with supermassive black holes (SBHs) taken from the
compilation of Ferrarese \& Ford (2005) (open squares). The distribution of
51 nucleated (Type~Ia) galaxies in the ACS Virgo Cluster Survey is shown by
the solid histogram; 11 galaxies with Type~Ib nuclei are shown by the dashed
histogram.
{\it (Lower Panel)} Distribution of the mass fraction, $\eta$, for early-type galaxies containing
supermassive black holes (SBHs) with that for nucleated galaxies (Type Ia) in the ACS
Virgo Cluster Survey. In the former case, ${\cal M}_{\bullet}/{\cal M}_{gal}$
measures the dynamical mass of the SBH relative to host galaxy's bulge mass.
For the nucleated galaxies, $\eta$ is the ratio of the nucleus and galaxy 
luminosities, averaged in the $g$ and $z$ bandpasses.
The smooth curves show the best-fit Gaussians.
\label{fig25}}
\end{figure}

\clearpage
\begin{figure}
\figurenum{26}
\plotone{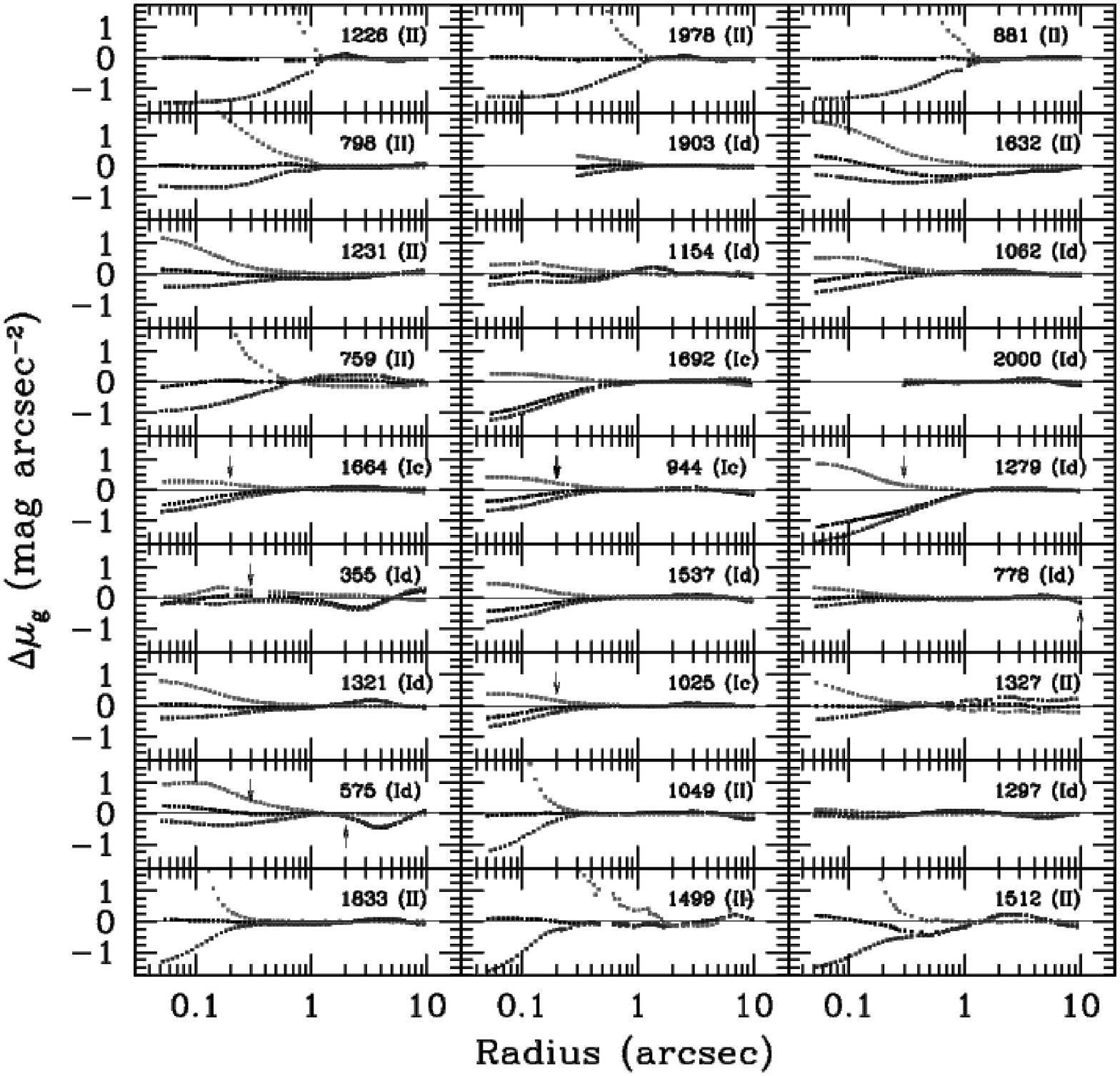}
\caption{The 27 galaxies from the ACS Virgo Cluster Survey which
are classified as non-nucleated, or having uncertain classifications, in
Tables~\ref{tab:data} and \ref{tab:class}. Galaxies are ordered according to blue luminosity, which
decreases left to right and top to bottom. Galaxy names and nuclear
classifications from Table~\ref{tab:class} are given in the upper right corner of each
panel.  In each panel, we plot the difference, $\Delta{\mu_g}$,
between the fitted model surface
brightness profile and: (1) the observed profile (black squares); (2) the
profile obtained after {\it adding} a central nucleus to the observed profile
(blue squares); and (3) the profile obtained after {\it subtracting} a
central nucleus from the observed profile (red squares). In the interests
of clarity, residuals are shown for the inner 10\arcsec~ only.
For VCC1903 (M59 = NGC4621) and VCC2000 (NGC4660), no data are plotted with 
$\lesssim$ 0\farcs3 since their $g$-band profiles are saturated inside
this point. 
\label{fig26}}
\end{figure}

\clearpage
\begin{figure}
\figurenum{27}
\plotone{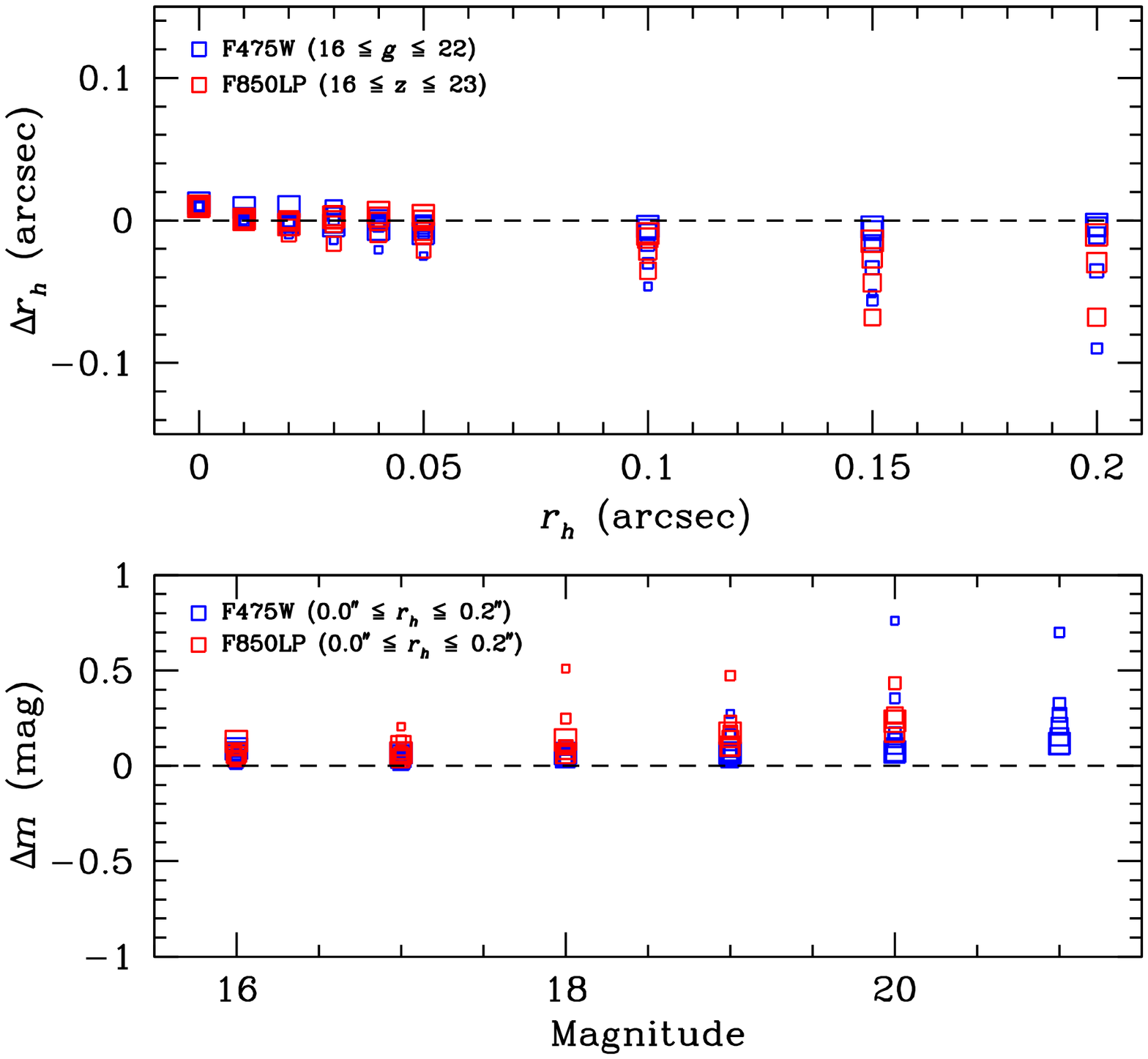}
\caption{{\it (Upper Panel)} Difference between the recovered and input half-light radii,
${\Delta}r_h$, for simulated nuclei added to VCC1833, the faintest
non-nucleated galaxy in our survey (excluding two dIrr/dE transition objects). The
central surface brightness of this galaxy is $\mu_g(1\arcsec) \approx$ 19.3 and
$\mu_z(1\arcsec) \approx$ 18.1 mag~arcsec$^{-2}$,
near the average for our sample galaxies. The blue and red squares show the 
respective results for the F475W and F850LP images, where the symbol size
is proportional to the input magnitude (in the sense that brighter nuclei
are plotted with larger symbols).
{\it (Lower Panel)} Difference between the recovered and input magnitude
${\Delta}m$, for simulated nuclei added to VCC1833. In this case, the
symbol size is proportional to input radius: i.e., the most compact nuclei are
plotted with the largest symbols.
\label{fig27}}
\end{figure}

\clearpage
\begin{figure}
\figurenum{28}
\plotone{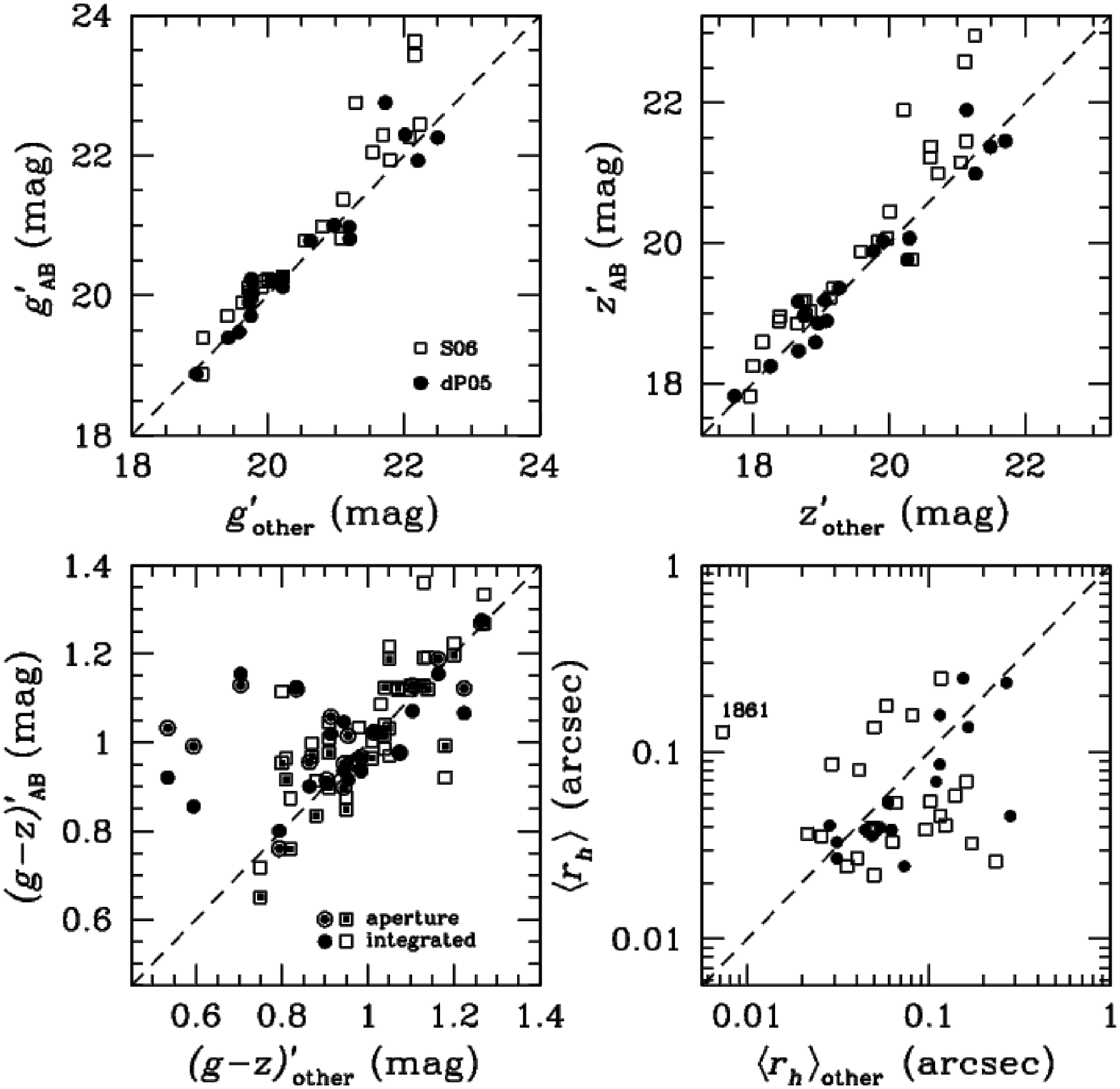}
\caption{Comparison of photometric and structural parameters for 
nuclei in common with the studies of de Propris \etal\ (2005; dP06) and
Strader \etal\ (2006; S06).
{\it (Upper Left Panel)} Comparison of $g$-band magnitudes, on the AB system,
for 18 and 25 nuclei from dP06 and S06 (open squares and filled circles,
respectively).
{\it (Upper Right Panel)} Comparison of $z$-band magnitudes, on the AB system.
{\it (Lower Left Panel)}  Integrated $(g-z)$ colors, on the AB system
(filled circles and open squares). Circled symbols show our {\it aperture}
measurements for the nuclei colors.
{\it (Lower Right Panel)} Comparison of measured half-light radii in
arcseconds. Our measurements refer to unweighted averages from the two bandpasses.
VCC1861 (IC3652) has been labelled since the half-light radius 
reported by Strader \etal\ (2006), $r_h$ = 0\farcs007, is 
$\sim 4\times$ below the ACS resolution limit.
\label{fig28}}
\end{figure}


\begin{thebibliography}{}
\bibitem[]{} Babul, A., \& Rees, M.J. 1992, MNRAS, 255, 346
\bibitem[]{} Bassino, L.P., Muzzio, J.C., \& Rabolli, M.  1994, \apj, 431, 634
\bibitem[Bekki \etal\ 2001]{} Bekki, K., Couch, W.J., \& Drinkwater, M.J. 2001, \apj, 552, L105
\bibitem[]{} Bekki, K., Couch, W.J., Drinkwater, M.J., \& Shioya,
Y. 2004, ApJ, 610, L13
\bibitem[]{} Ben\'itez, N., \etal\ 2002, Bull.. Am. Astron. Soc., 34, 1236
\bibitem[]{} Binggeli, B., Barazza, F., \& Jerjen, H. 2000, A\&A, 359, 447
\bibitem[]{} Binggeli, B., \& Cameron, L.M. 1991, A\&A, 252, 27
\bibitem[]{} Binggeli, B., \& Cameron, L.M. 1993, A\&ASS, 98, 297
\bibitem[]{} Binggeli, B., Sandage, A., \& Tammann, G.A. 1985, AJ, 90, 1681 (BST85)
\bibitem[]{} Binggeli, B., Tammann, G.A., \& Sandage, A. 1987, AJ, 94, 251 (BTS87)
\bibitem[]{} B\"oker, T., Laine, S., van der Marel, R.P., Sarzi, M., Rix,
H.-W., Ho, L., \& Shields, J.C. 2002, \aj, 123, 1389
\bibitem[]{} B\"oker, T., Sarzi, M., McLaughlin, D.E., van der Marel, R.P.,
Rix, H.-W., Ho, L.C., \& Shields, J.C. 2004, \aj, 127, 105
\bibitem[]{} Bower, R.G., Lucey, J.R., \& Ellis, R.S. 1992, \mnras, 254, 601
\bibitem[]{} Butler, D. J., \& Mart\'inez-Delgado, D. 2005, \aj, 129, 2217
\bibitem[]{} Byun, Y.I., \etal\ 1996, \aj, 111, 1889
\bibitem[]{} Caldwell, N. 1983, AJ, 88, 808
\bibitem[]{} Caldwell, N. 1987, AJ, 94, 1116
\bibitem[]{} Caldwell, N., \& Bothun, G. D. 1987, AJ, 94, 1126
\bibitem[]{} Caldwell, N., Armandroff, T.E., Seitzer, P., \& Da Costa, G.S.
1992, \aj, 103, 840
\bibitem[]{} Caon, N., Capaccioli, M., \& D'Onofrio, M. 1993, \mnras, 265, 1013
\bibitem[]{} Capuzzo-Dolcetta, R. 1993, \apj, 415, 616
\bibitem[]{} Capuzzo-Dolcetta, R., \& Tesseri, A. 1999, \mnras, 308, 961
\bibitem[]{} Carollo, C.M., Franx, M., Illingworth, G.D., \& Forbes, D.A.
1997, \apj, 481, 710
\bibitem[]{} Carollo, C.M., Stiavelli, M., \& Mack, J. 1998, \aj, 116, 68
\bibitem[]{} Chabrier, G. 2003, \pasp, 115, 763
\bibitem[]{} Chandrasekhar, S. 1943, \apj, 97, 255
\bibitem[]{} Chiaberge, M., Capetti, A., \& Celotti, A. 1999, \aap, 349, 77
\bibitem[]{} Coleman, M., Da Costa, G.S., Bland-Hawthorn, J., Mart\'inez-Delgado,
Freeman, K.C. \& Malin, D. 2004, \aj, 127, 832
\bibitem[]{} C\^ot\'e, P., Blakeslee, J.P., Ferrarese, L., Jord\'an, A.,
Mei, S., Merritt, D., Milosavljevi\'c, M., Peng, E.W., Tonry, J.L., \& West, M.J.
2004, \apjs, 153, 223 (Paper I)
\bibitem[]{} Crane, P., \etal\ 1993, \aj, 106, 1371
\bibitem[]{} Davies, J.I., \& Phillipps, S. 1988, \mnras, 233, 553
\bibitem[]{} de Propris, R., Phillipps, S., Drinkwater, M.J., Gregg, M.D., 
Jones, J.B., Evstigneeva, E., \& Bekki, K. 2005, \apj, 623, 105
\bibitem[]{} De Rijcke, S., \& Debattista, V.P. 2004, ApJ, 603, L25
\bibitem[]{} de Vaucouleurs, G. 1948, Ann. d'Astrophys., 11, 247
\bibitem[]{} Djorgovski, S., Bendinelli, O., Parmeggiani, G., \& Zavatti, F.
1992, in Morphological and Physical Classification of Galaxies, ed. 
G. Longo, M. Capaccioli, \& G. Busarello (Kluwer, Dordrecht), p. 439
\bibitem[]{} Dressler, A. 1980, \apj, 236, 351
\bibitem[]{} Drinkwater, M.J., Phillipps, S., Gregg, M.D., Parker, Q.A.,
Smith, R.M., Davies, J.I., Jones, J.B., Sadler, E.M. 1999, \apj, 511, L97
\bibitem[]{} Drinkwater, M. J., Phillipps, S., Jones, J. B., Gregg, M. D.,
 Deady, J. H., Davies, J. I., Parker, Q. A., Sadler, E. M., \& Smith, R. M. 2000, \aap, 355, 900
\bibitem[]{} Durrell, P.R. 1997, \aj, 113, 531
\bibitem[]{} Ebisuzaki, T., Makino, J., \& Okumura, S.K. 1991, Nature, 354, 212
\bibitem[]{} Fabbiano, G., Gioia, I.M., \& Trinchieri, G. 1989, \apj, 347, 127
\bibitem[]{} Faber, S., \etal\ 1997, \aj, 114, 1771
\bibitem[]{} Ferrarese, L., \& Merritt, D. 2000, 539, L9
\bibitem[]{} Ferrarese, L., \& Ford, H.C. 2005, SSRv, 116, 523
\bibitem[]{} Ferrarese, L., C\^ot\'e, P., Jord\'an, A., Peng, E.W., Blakeslee, J.P., Piatek, S.,
Mei, S., Merritt, D., Milosavljevi\'c, M., Tonry, J.L., \& West, M.J. 2006a, \apj, in press, 
(astro-ph/0602297) (Paper VI)
\bibitem[]{} Ferrarese, L., C\^ot\'e, P., Dalla Bont\`a, E., Peng, E.W., Merritt, D., Jord\'an, A.,
Blakeslee, J.P., Ha\c{s}egan, M., Mei, S., Piatek, S., Tonry, J.L., \& West, M.J. 2006b, \apj, submitted
\bibitem[]{} Ferguson, H.C. 1989, \aj, 98, 367
\bibitem[]{} Ferguson, H.C., \& Sandage, A. 1989, \apj, 346, L53
\bibitem[]{} Ford, H.C., \etal\ 1998, Proc. SPIE, 3356, 234
\bibitem[]{} Gebhardt, K., \etal\ 2000, \apj, 539, L13
\bibitem[]{} Graham, A.W., \& Guzm\'an, R. 2003, \aj, 125, 2936
\bibitem[]{} Graham, A.W., Erwin, P., Trujillo, I., \& Asensio Ramos, A. 2003b, AJ, 125, 2951
\bibitem[]{} Graham, A.W. 2004, \apj, 613, 33
\bibitem[]{} Graham, A.W., \& Driver, S.P. 2005, PASA, 22, 118
\bibitem[]{} Grant, N.I., Kuipers, J.A., \& Phillipps, S. 2005, \mnras, 363, 1019
\bibitem[]{} Ha\c{s}egan, M., Jord\'an, A., C\^ot\'e, P., Djorgovski, S.G.,
McLaughlin, D.E., Blakeslee, J.P., Mei, S., West, M.J., Peng, E.W., Ferrarese, L.,
Milosavljevi\'c, M., Tonry, J.L., \& Merritt, D. \ 2005, ApJ, 627, 203 (Paper VII)
\bibitem[]{} Ha\c{s}egan, M., \etal\ 2006, in preparation
\bibitem[]{} Haehnelt, M.G., Natarajan, P., \& Rees, M.J. 1998, \mnras, 300, 817
\bibitem[]{} Hilker, M., Infante, L., Vieira, G., Kissler-Patig, M., \&
Richtler, T. 1999, \aaps, 134, 75
\bibitem[]{} Ho, L.C., Filippenko, A.V., \& Sargent, W.L.W. 1997, \apjs, 112, 315
\bibitem[]{} Ho, L.C. 1999, \apj, 516, 672
\bibitem[]{} Jedrzejewski, R.I. 1987, \mnras, 226, 747
\bibitem[]{} Jones, J.B., Drinkwater, M.J., Jurek, R., Phillipps, S., Gregg, M.D., 
Bekki, K., Couch, W.J., Karick, A., Parker, Q.A., \& Smith, R.M. 2006, \aj, 131, 312
\bibitem[]{} Jord\'an, A., Blakeslee, J.P., Peng, E., Mei, S., C\^ot\'e, P.,
Ferrarese, L., Merritt, D., Milosavljevi\'c, M., Tonry, J.L., \& West, M.J. 
2004a, \apjs, 154, 509 (Paper II)
\bibitem[]{} Jord\'an, A., C\^ot\'e, P., Ferrarese, L., Blakeslee, J.P., Mei, S., Merritt, D.,
Milosavljevi\'{c}, M., Peng, E.W., Tonry, J.L., \& West, M.J. 2005b, \apj, 613, 279 (Paper~III)
\bibitem[]{} Jord\'an, A., C\^ot\'e, P., Blakeslee, J.P., Ferrarese, L., McLaughlin, D.E.,
Mei, S., Peng, E.W., Tonry, J.L., Merritt, D., Milosavljevi\'c, M., Sarazin, C.L.,
Sivakoff, G.R., West, M.J. 2005, \apj, 634, 1002 (Paper~X)
\bibitem[]{} Jord\'an, A., \etal\ 2006, in preparation
\bibitem[]{} Jerjen, H., \& Binggeli, B. 1997, in ASP Conf. Ser. 116, The Nature
of Elliptical Galaxies, ed. M. Arnaboldi, G. S. Da Costa, \&
P. Saha (San Francisco: ASP), 239
\bibitem[]{} Julian, W.H., Kooiman, B.L., \& Sanders, W.L. 1997, \pasp, 109, 297
\bibitem[]{} King, I.R. 1966, AJ, 71, 64 
\bibitem[]{} Kleyna, J.T., Wilkinson, M.I., Gilmore, G., \& Evans, N.W.
2003, \apj, 588, L21
\bibitem[]{} Kormendy, J. 1985, ApJ, 295, 73
\bibitem[]{} Kormendy, J., \& Richstone, D 1995, \araa, 33, 581
\bibitem[]{} Larsen, S. 1999, \aaps, 139, 393
\bibitem[]{} Lauer, T.R., \etal\ 1995, \aj, 110, 2622
\bibitem[]{} Lauer, T.R., \etal\ 2005, \aj, 129, 2138
\bibitem[]{} Layden, A.C., \& Sarajedini, A. 2000, \aj, 119, 1760
\bibitem[]{} Lotz, J.M., Telford, R., Ferguson, H.C., Miller, B.W.,
Stiavelli, M., \& Mack, J. 2001, ApJ, 552, 572
\bibitem[]{} Lotz, J.M., Miller, B.W., \& Ferguson, H.C. 2004, \apj, 613, 262
\bibitem[]{} Matthews, L.D., \etal\ 1999, AJ, 118, 208
\bibitem[]{} McConnachie, A.W., \& Irwin, M.J. 2006, \mnras, 365, 1263
\bibitem[]{} Mei, S., Blakeslee, J.P., Tonry, J.L., Jord\'an, A., Peng, E.W.,
C\^ot\'e, P., Ferrarese, L., Merritt, D., Milosavljevi\'{c}, M., \& West, M.J. 
2005a, \apjs, 156, 113 (Paper~IV)
\bibitem[]{} Mei, S., Blakeslee, J.P., Tonry,
J.L., Jord\'an, A., Peng, E.W., C\^ot\'e, P., Ferrarese, L., West, M.J.,
Merritt, D., \& Milosavljevic, M. 2005b, \apj, 625, 121 (Paper V)
\bibitem[]{} Merritt, D., \& Ferrarese, L. 2001, \mnras, 320, 30
\bibitem[]{} Merritt, D., Milosavljevi\'{c}, M., Favata, M.,
Hughes, S.A., \& Holz, D.E. 2004, ApJ, 607, L9
\bibitem[]{} Merritt, D., Navarro, J.F., Ludlow, A., \& Jenkins, A.
2005, \apj, 624, L85
\bibitem[]{} Merritt, D., \etal\ 2006, in preparation
\bibitem[]{} Michie, R.W. 1963, \mnras, 125, 127
\bibitem[]{} Mieske, S., Hilker, M., Infante, L., \& Jord\'an, A. 2006, \aj, in press (astro-ph/0512474)
\bibitem[]{} Mihos, J.C., \& Hernquist, L. 1994, \apj, 437, L47
\bibitem[]{} Miller, R.H., \& Smith, B.F. 1992, \apj, 393, 508
\bibitem[]{} Milosavljevi\'c, M., \& Merritt, D. 2001, \apj, 563, 34
\bibitem[]{} Monaco, L., Bellazini, M., Ferraro, F.R., \& Pancino, E. 
2005, \mnras, 356, 1396
\bibitem[]{} Mori, M., \& Burkert, A. 2000, \apj, 538, 559
\bibitem[]{} Moore, B., Lake, G., \& Katz, N. 1998, \apj, 495, 139
\bibitem[]{} Navarro, J.F., Hayashi, E., Power, C., Jenkins, A.R., Frenk, C.S.,
White, S.D.M., Springel, V., Stadel, J., \& Quinn, T.R 2004, \mnras, 349, 1039
\bibitem[]{} Nelder, J.A., \& Mead, R. 1965, Comput.J., 7, 308
\bibitem[]{} Oh, K.S., \& Lin, D.N.C. 2000, \apj, 543, 620
\bibitem[]{} Oemler, A. 1974, \apj, 194, 1
\bibitem[]{} Palma, C., Majewski, S.R., Siegel, M.H., Patterson, R.J.,
Ostheimer, J.C., \& Link, R. 2003, \aj, 125, 1352
\bibitem[]{} Peng, E.W., Jord\'an, A., C\^ot\'e, P., Blakeslee, J.P., Ferrarese, L.,
Mei, S., West, M.J., Merritt, D.,Milosavljevic, M., \& Tonry, J.L. 2006a, \apj, 639, 95 (Paper~IX)
\bibitem[]{} Peng, E.W.,  C\^ot\'e, P., Jord\'an, A., Blakeslee, J.P., Ferrarese, L.,
Mei, S., West, M.J., Merritt, D.,Milosavljevic, M., \& Tonry, J.L. 2006b, \apj, in press (Paper XI)
\bibitem[]{} Phillips, A.C., Illingworth, G.D., MacKenty, J.W., \&
Franx, M. 1996, \aj, 111, 1566
\bibitem[]{} Phillipps, S., Drinkwater, M., Gregg, M., \& Jones,
J. 2001, ApJ, 560, 201
\bibitem[]{} Postman, M., \etal\ 2005, \apj, 623, 721
\bibitem[]{} Quinlan, G.D., \& Hernquist, L. 1997, NewA, 2, 533
\bibitem[]{} Rangarajan, F.V.N., White, D.A., Ebeling, H., \& Fabian, A.C. 1995, \mnras, 277, 1047 
\bibitem[]{} Ravindranath, S., Ho, L.C., Peng, C.Y., Filippenko, A.V., \&
Sargent, W.L.W. 2001, \aj, 122, 653
\bibitem[]{} Reaves, G. 1983, \apjs, 53, 375
\bibitem[]{} Rest, A., van den Bosch, F.C., Jaffe, W., Tran, H., Tsvetanov, Z.,
Ford, H.C., Davies, J., \& Schafer, J. 2001, \aj, 121, 2431
\bibitem[]{} Sandage, A., Binggeli, B., \& Tammann, G.A. 1985, AJ,
90, 1759
\bibitem[]{} S\'ersic, J.-L. 1968, Atlas de Galaxias Australes (C\'ordoba:
Obs. Astron., Univ. Nac. C\'ordova) 
\bibitem[]{} Schlegel, D.J., Finkbeiner, D.P., \& Davis, M. 1998,
ApJ, 500, 525 
\bibitem[]{} Silk, J., Wyse, R.F.G., \& Shields, G.A. 1987, \apj, 332, L59
\bibitem[]{} Silk, J., \& Rees, M.J. 1998, \aap, 331, L1
\bibitem[]{} Sirianni, M., Jee, M.J., Ben\'itez, N., Blakeslee, J.P., Martel, A.R.,
Meurer, G., Clampin, M., De Marchi, G., Ford, H.C., Gilliland, R., Hartig, G.F.,
Illingworth, G.D.,  Mack, J., \& McCann, W.J. 2005, \pasp, 117, 1049
\bibitem[]{} Smith, G.P., Treu, T., Ellis, R.S., Moran, S.M., \&
Dressler, A. 2004, \apj, 620, 78
\bibitem[]{} Stiavelli, M., Miller, B.W., Ferguson, H.C., Mack, J.,
Whitmore, B.C., \& Lotz, J.M. 2001, \aj, 121, 1385
\bibitem[]{} Strader, J., Brodie, J.P., Spitler, L., \& Beasley, M.A. 2006, \aj,
submitted (astro-ph/0508001)
\bibitem[]{} Tonry, J.L., Dressler, A., Blakeslee, J.P., Ajhar, E.A.,
Fletcher, A.B., Luppino, G.A., Metzger, M.R., \& Moore, C.B.
2001, ApJ, 546, 681
\bibitem[]{} Tremaine, S.D., Ostriker, J.P., \& Spitzer, L.,
Jr. 1975, ApJ, 196, 407
\bibitem[]{} Tremaine, S.D. 1976, \apj, 203, 72
\bibitem[]{} Trujillo, I., Erwin, P., Asensio Ramos, A., \& Graham,
A.W. 2004, AJ, 127
\bibitem[]{} Valluri, M., Ferrarese, L., Merritt, D., \&
Joseph, C.L. 2005, \apj, 628, 137
\bibitem[]{} van den Bergh, S. 1986, AJ, 91, 271
\bibitem[]{} Walcher, C.J., van der Marel, R.P., McLaughlin, D., Rix, H.-W., 
B\"oker, T., H\"aring, N., Ho, L.C., Sarzi, M., \& Shields, J.C 2005, \apj, 618, 237
\bibitem[]{} Weedman, D.W. 1983, \apj, 266, 479
\bibitem[]{} Wehner, E.H., \& Harris, W.E. 2006, \apj, submitted
\bibitem[]{} West, M.J., C\^ot\'e, P., Jones, C., Forman, W., \& Marzke, R.O. 
1995, \apj, 453, L77.
\bibitem[]{} Wrobel, J.M. 1991, \aj, 101, 127
\bibitem[]{} Wyiethe, J.S.B., \& Loeb, A. 2002, \apj, 581, 886
\bibitem[]{} Zang, T.A., \& Hohl, F. 1978, ApJ, 226, 521
\end{thebibliography}
\end{document}